\documentclass[twocolumn,pre,aps,floatfix,superscriptaddress]{revtex4-1}

\usepackage{color}
\usepackage{siunitx}

\usepackage{graphicx}

\usepackage{amsthm}
\usepackage{amsmath}
\usepackage{amssymb}
\usepackage{enumerate}

\newcommand{\ket}[1]{\vert #1 \rangle}
\newcommand{\eket}[1]{\bigl \vert #1 \bigr \rangle}
\newcommand{\bra}[1]{\langle #1 \vert}
\newcommand{\ebra}[1]{\bigl \langle #1 \bigr \vert}
\newcommand{\eexp}[1]{\bigl \langle #1 \bigr \rangle}
\newcommand{\figref}[1]{Fig.~\ref{#1}}
\renewcommand{\vec}[1]{\boldsymbol{#1}}
\newcommand{\R}{\boldsymbol{R}}
\newcommand{\V}{\mathcal{V}}
\newcommand{\T}{\mathcal{T}}
\newcommand{\rv}{\boldsymbol{r}}
\newcommand{\ren}{R\'{e}nyi }
\newcommand{\omint}{\omega_{\mathrm{int}}}
\newcommand{\dt}{\tau}

\begin{document}

\title{A path integral Monte Carlo method for R\'{e}nyi entanglement entropies}

\author{C. M. Herdman}
\email{Christopher.Herdman@uvm.edu}
\affiliation{Department of Physics, University of Vermont, Burlington, VT 05405, USA}

\author{Stephen Inglis}
\affiliation{Department of Physics, Arnold Sommerfeld Center for Theoretical Physics, and Center for Nanoscience, Ludwig-Maximilians-Universit\"at M\"unchen, Theresienstra{\ss}e 37, 80333 Munich, Germany}
\affiliation{Department of Physics and Astronomy, University of Waterloo, Ontario, N2L 3G1, Canada}

\author{P.-N. Roy}
\affiliation{Department of Chemistry, University of Waterloo, Ontario, N2L 3G1, Canada}

\author{R. G. Melko}
\affiliation{Department of Physics and Astronomy, University of Waterloo, Ontario, N2L 3G1, Canada}
\affiliation{Perimeter Institute for Theoretical Physics, Waterloo, Ontario N2L 2Y5, Canada}

\author{A. Del Maestro}
\affiliation{Department of Physics, University of Vermont, Burlington, VT 05405, USA}
\affiliation{Vermont Complex Systems Center, University of Vermont, Burlington, VT 05405, USA}

\begin{abstract}
We introduce a quantum Monte Carlo algorithm to measure the R\'enyi entanglement entropies in systems of interacting bosons in the continuum. This approach is based on a path integral ground state method that can be applied to interacting itinerant bosons in any spatial dimension with direct relevance to experimental systems of quantum fluids. We demonstrate how it may be used to compute spatial mode entanglement, particle partitioned entanglement, and the entanglement of particles, providing insights into quantum correlations generated by fluctuations, indistinguishability and interactions. We present proof-of-principle calculations, and benchmark against an exactly soluble model of interacting bosons in one spatial dimension. As this algorithm retains the fundamental polynomial scaling of quantum Monte Carlo when applied to sign-problem-free models, future applications should allow for the study of entanglement entropy in large scale many-body systems of interacting bosons.
\end{abstract}
\maketitle

\section{Introduction}
Entanglement is a fundamental property of quantum mechanical systems, one which reflects the nonclassical information shared between distinct bipartitions of a quantum state.  It is well known that entanglement may be exploited for information processing \cite{Vedral:2013bk} via quantum algorithms that provide an exponential speedup over their classical counterparts \cite{Jozsa2003} as well as for secure communication \cite{Ekert:1991xy} and teleportation \cite{Bennett:1993ta}.  These and other applications have initiated a broad effort to find practical systems where it is feasible to create and manipulate persistent entangled states.  Additionally, the study of entanglement has had a significant impact on a variety of fields including condensed matter, atomic and molecular physics, quantum optics, quantum information, and high-energy theory. The description of entanglement in terms of the concepts of information theory \cite{Horodecki:2009gb} has proved particularly transformative in condensed-matter physics, providing a new paradigm with which to quantify quantum correlations~\cite{Amico2008}.  A striking application of these ideas is in the classification of exotic topological phases which cannot be fully described by local correlation functions alone \cite{Kitaev:2006dn, Levin:2006ij, Wolf:2008ht}.  Entanglement can also been used to identify the universality class of quantum critical points, and may be capable of quantifying the effective low-energy degrees of freedom that occur in the corresponding critical theories.

In order to access and study entanglement in interacting models of quantum many-body systems, large-scale simulations are a necessary tool.  For example, diagonalization techniques or the density matrix renormalization group allow for the measurement of entanglement quantities in a restricted class of systems through their essentially complete knowledge of the ground state wave function \cite{White92,Scholl05}.  Quantification of entanglement in quantum Monte Carlo (QMC) simulations had not been possible prior to 2010, when the introduction of ``replica trick'' methods \cite{Hastings2010,Melko2010} provided, for the first time, a scalable procedure for measuring the \ren entanglement entropy in the ground state of lattice Hamiltonians, without requiring knowledge of the full reduced density matrix.  The simplicity and broad applicability of the replica trick for QMC studies is illustrated by its rapid adoption to a wide range of ground state methods \cite{Tubman2012,Zhang2012,Inglis2013b,McMinis:2013dp,Selem2013,Pei2014,algorithm,Tubman2014}, while the finite temperature generalization \cite{Melko2010} has extended the types of systems one can examine \cite{Tommaso,Inglis:2013iv,Broecker,Chung2013,Chung2013a}, allowing observation of the competition between thermal mixing and quantum entanglement.

A common theme in all these works is that the entanglement is measured between two spatial subregions and investigated as the size of the bipartition is modified.  This has led to the widespread confirmation of an ``area law'' in the ground state of local bosonic Hamiltonians \cite{Sorkin,Shredder, Eisert2010}, where the entanglement entropy scales with the size of the boundary between spatial subregions.  More interestingly perhaps, this approach has facilitated the calculation of universal quantities that appear in subleading scaling terms, allowing for new methods to identify and characterize quantum phases and phase transitions.  The consequences of this approach are potentially far reaching.  For example, the ability of the \ren entropies to access the {\it central charge} $c$ of a $(1+1)$-dimensional quantum critical point's associated conformal field theory \cite{Callan1,EE1d1,EE1d2,EE1d3} was a powerful improvement over previous techniques that required calculation of subleading terms of the free energy and the elimination of non-universal velocities \cite{Cardyc,Affleckc}.  There is currently an active multidisciplinary effort to extend this paradigm to higher-dimensional quantum critical points, where a synergy between numerical lattice simulations \cite{Kallin_NLCE, Inglis2013b, Luitz}, field theory \cite{FradkinMoore,Max,Casini_FieldTheory}, and holography \cite{ryu_2,Robcorner} aims to identify similar quantities in the entanglement entropy that can serve to classify, characterize, and constrain interacting fixed points of general interest to condensed-matter physicists \cite{Zamo, Cardy_Cth,Casini12,Grover_C}.  

Given their potential, it is desirable to attempt to extend these methods to off-lattice itinerant systems, with continuous degrees of freedom.  While the investigation of entanglement in the spatial continuum is not new, (e.g. see Refs.~[\onlinecite{Simon2002,Vedral2003,Hines2003,Heaney2007,Heaney2007a,Kaszlikowski2007,Vedral2008,Heaney2009,Goold2009,Ding2009,Gagatsos2012,Calabrese2011a,Calabrese2011b,Burnett2001,Dunningham2002a}]), the class of models where it could actually be measured has been restricted to those without interactions, or consisting of a small number of particles. A general system with continuous degrees of freedom has an infinite Hilbert space, and thus there is no upper bound on the available entanglement \cite{Eisert:2002bx, Adesso:2007eq}. In fact, these infinite entanglement states are trace-norm dense in the Hilbert space \cite{Wehrl:1978ku}, but, physical states with finite energy (such as a quantum liquid, or a gas of trapped ions), have a bounded entanglement. 

For localized particles, it is most natural to partition the system into spatial subregions. However, when the particles are itinerant, additional subtleties arise, and one may also choose to partition into subsets of particles that are not localized to a region of space~\cite{Zanardi2002,Shi2003,Fang2003,Zozulya2008,Eckert2002,Haque2009,Haque:2007il,Zozulya2007a}. For systems of identical particles, this ``particle partitioned'' entanglement can arise from exchange statistics alone. Proposals to quantify and ultimately use this type of entanglement have been deterred by the fact that a subsystem of identical particles is not physically addressable through a measurement. Consequently, there has been much debate in the literature over what the most appropriate measures of entanglement of identical particles are~\cite{Dunningham2005,Balachandran2013,Benatti2012a,Wiseman2003,Wiseman2003fb}.

The need for new insights is pressing, as itinerant boson systems in the continuum are of particular experimental interest, and the capabilities for manipulating quantum fluids such as ultra-cold Bose gases and superfluid helium-4 are mature and highly developed. A canonical model for such systems consists of $N$ interacting itinerant particles in the spatial continuum that is described by the nonrelativistic Hamiltonian,
\begin{equation}
    H = \sum_{i=1}^N \left({ - \frac{\hbar^2}{2m_i} \nabla_i^2 + U_i}\right) + \sum_{i<j} V_{ij},
\label{eq:Ham}
\end{equation}
where $m_i$ is the mass of the $i^\text{th}$ particle subject to an external potential $U_i$ and two-body interaction $V_{ij}$. This Hamiltonian is general enough to describe a wide variety of systems, including trapped ultracold atomic gases at low density (where $U_i$ could be a harmonic potential and $V_{ij}$ a hard-core repulsion) or a high density quantum fluid such as helium-4 (with $U_i = 0$ and $V_{ij}$ an empirical dipole-dipole pair potential). Thus, a method capable of computing the entanglement entropy for bipartitions of the ground states of Eq.~(\ref{eq:Ham}) could find immediate application in experimentally accessible quantum many-body states of matter.  To this end, an alternative QMC formulation at $T=0$ based on the Feynman path integral description has been recently employed to compute the \ren entanglement entropy of a system of interacting itinerant bosons in one spatial dimension \cite{algorithm} under a ``particle'' bibipartitioning.  In this paper, we present the details of the algorithm presented in Ref.~[\onlinecite{algorithm}] and introduce extensions to allow for the measurement of entanglement for spatial bipartitions of itinerant bosons as well a method to compute the accessible entanglement that could be potentially transferred to a register for quantum information processing purposes.  

The paper is organized as follows. We first define the \ren entanglement entropy in terms of the reduced density matrices of a system and present a precise description of the various types of bipartitions that are possible for itinerant particles. After describing the implications of such definitions for some canonical states in a simple model of itinerant bosons on a lattice, we introduce our proposed QMC method and provide its algorithmic construction.  The numerical method is then benchmarked against an exactly soluble system of harmonically interacting bosons in a harmonic potential, where the entanglement entropy can computed analytically. After presenting results on the scaling properties of the algorithm with various model parameters, we discuss further algorithmic extensions as well as the classes of system where they can be immediately applied.

\section{R\'enyi entanglement entropies}

To define a measure of bipartite entanglement, one first chooses a bipartition that divides the system into two subsystems: $A$ and $B$. Given the density matrix of the system $\rho$, this bipartition defines the reduced density matrix of subsystem $A$, by ``tracing out'' all degrees of freedom in the other subsystem $B$,
\begin{equation}
\rho_A =  {\rm Tr}_B \rho. \notag
\end{equation}
Here we restrict our discussion to pure states of the full system $\vert \Psi \rangle$, where $\rho = \vert \Psi \rangle \langle \Psi \vert$. The bipartite entanglement entropy is a measure of the mixedness of $\rho_A$; in particular, we consider the R\'enyi entropies,
\begin{equation}
S_{\alpha}\left[ \rho_A \right] \equiv \frac{1}{1-\alpha} \log \left({ {\rm Tr} \rho_A^{\alpha} }\right), \label{eq:renyi}
\end{equation}
where $\alpha$ is the R\'enyi index. For $\alpha \rightarrow 1$ the R\'enyi entropy is equivalent to the von Neumann entropy: $S = -\mathrm{Tr}\, \rho_A \log \rho_A$. If $\rho$ can be written as a product state under this bipartition, $\rho_A$ will be  pure state with ${\rm Tr} \rho_A^{\alpha} = 1$ and all $S_{\alpha}[\rho_A]$ will vanish.

\begin{figure}
\begin{center}
\scalebox{1}{\includegraphics[width=\columnwidth]{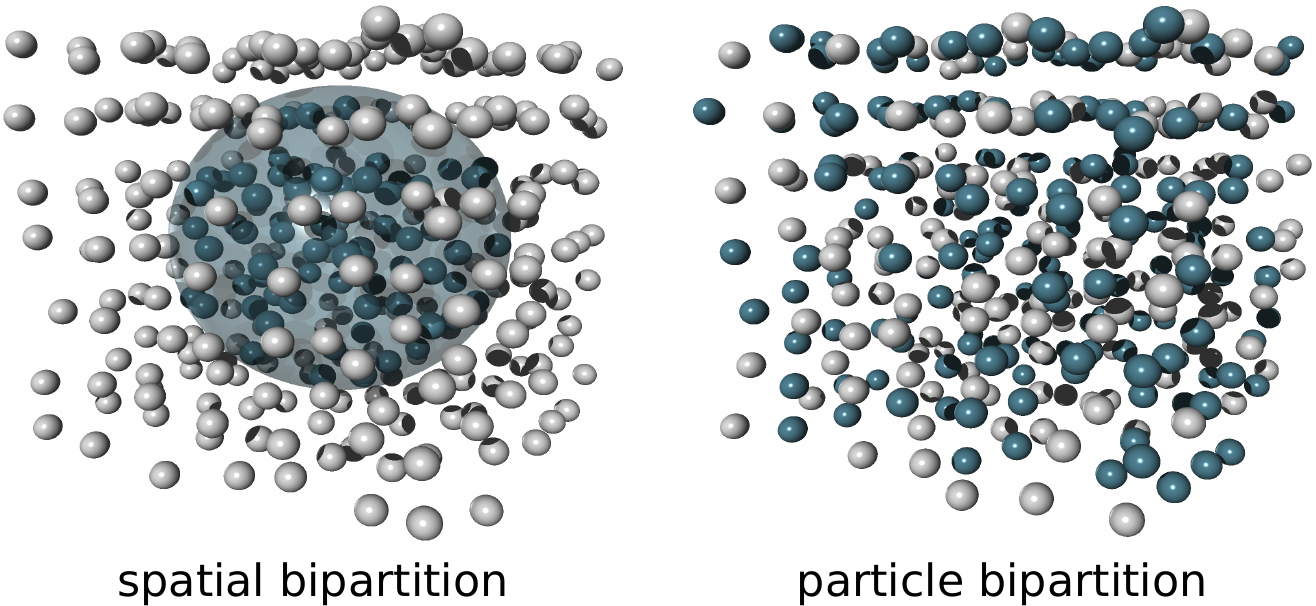}}
\end{center}
\caption{(Color online) A comparison of spatial and particle bipartitions in the continuum defined by particles shaded light (region $A$) or dark (region $B$).  Particle bipartitions are possible even in the case of their indistinguishability through a fictitious particle labeling scheme and subsequent average over all possible relabelings.}
\label{fig:bipart}
 \end{figure}

\subsection{Spatial mode entanglement}

For systems of localized particles, such as spin models, a natural choice of subsystem is a spatial subregion as shown in Fig.~\ref{fig:bipart}, and entanglement is most commonly studied under such a bipartition. For itinerant particle systems, a spatial subregion corresponds to a subspace of a Fock space of single particle spatial modes; thus,  spatial-partitioned entanglement is a type of {\it mode} partitioned entanglement. Here we discuss computing \ren entropy under a generic mode bipartition.

Given a single particle basis $\{\vert \psi_m \rangle\}$, we choose a subset of modes $\{m_A\}$ and bipartition the single particle Hilbert space accordingly such that $\{m\} \equiv \{m_A\} \cup \{m_B\}$. We can define a basis of the mode-occupation number eigenstates that is explicitly a tensor product under this bipartition,
\begin{equation}
\Bigl \vert \bigl\{\vec{n}_A , \vec{n}_B \bigr\} \Bigr \rangle \equiv \Bigl \vert \vec{n}_A \Bigr \rangle \otimes \Bigl \vert  \vec{n}_B  \Bigr \rangle, \notag
\end{equation}
where $n_m$ are the mode occupation numbers, $\vec{n} = \{n_m\}$ and the eigenstates in each subset of modes are defined in the usual second quantized form,
\begin{equation}
\Bigl \vert \vec{n} \Bigr \rangle \equiv \prod_{m} \frac{1}{\sqrt{n_{m}!}} \left( \psi_{m}^\dagger\right)^{n_{m}} \left \vert 0_A \right \rangle, \notag
\end{equation}
with field operators the $\psi_m^{\dagger}$. For spatial mode entanglement, the subset $A$ corresponds to a spatial subregion and $\{m_A\}$ are spatially localized modes. 

A general $N$-body state $\vert \Psi \rangle$ can be written in this mode-bipartitioned basis as
\begin{equation}
\left \vert \Psi \right \rangle = \sum_{ \vec{n}_A,  \vec{n}_B } c_{\vec{n}_A  \vec{n}_B}  \Bigl \vert \vec{n}_A \Bigr \rangle \otimes \Bigl \vert  \vec{n}_B  \Bigr \rangle; \notag
\end{equation}
particle number conservation is enforced if 
\begin{equation}
N = \sum_{m_A} n_{m_A}+\sum_{m_B} n_{m_B} \quad \forall \quad c_{\vec{n}_A  \vec{n}_B}\neq 0. \notag
\end{equation}
The reduced density matrix $\rho_A$ may then be defined as
\begin{align}
\rho_A &\equiv \sum_{\vec{n}_B} \Bigl \langle \vec{n}_B  \Bigr \vert \Psi \Bigl \rangle \Bigr \langle \Psi \Bigl \vert \vec{n}_B\Bigr \rangle \notag \\
&= \sum_{\vec{n}_A,\vec{n}_{A}'} \lambda_{\vec{n}_A \vec{n}_A' } \Bigl \vert \vec{n}_A \Bigr \rangle  \Bigl \langle  \vec{n}_A'  \Bigr \vert,   \notag
\end{align}
where the reduced density matrix elements are defined as
\begin{equation}
\lambda_{\vec{n}_A,  \vec{n}_{A}' } \equiv \sum_{  \vec{n}_B} c_{\vec{n}_A \vec{n}_B} c^*_{\vec{n}_B \vec{n}_{A}'}. \notag
\end{equation}

To quantify the mode entanglement, we consider the \ren entropies $S_\alpha (A) \equiv S_\alpha[\rho_A]$ as defined in Eq.~\eqref{eq:renyi}. A general unentangled product state under a particular mode bipartition takes the form
\begin{equation}
\left \vert \Psi \right \rangle = \sum_{ \vec{n}_A } c_{\vec{n}_A}  \Bigl \vert \vec{n}_A \Bigr \rangle \otimes \sum_{\vec{n}_B} c_{\vec{n}_B}  \Bigl \vert  \vec{n}_B  \Bigr \rangle; \notag
\end{equation}
clearly, all mode-occupation number eigenstates $\ket{\vec{n}}$ are unentangled under any bipartition of these modes. We note that this mode entanglement depends on both the single particle mode basis and the bipartition chosen, and may arise even in the absence of interactions between particles. For example, if the single particle ground state $\ket{\phi_0}$ has nonzero overlap with both $\{\vert \psi_{m_A} \rangle\}$ and $\{\vert \psi_{m_B} \rangle\}$, then $\ket{\phi_0}$ has nonzero mode entanglement due to particle fluctuations between modes; consequently, any noninteracting $N$-body ground state will also be entangled under this mode bipartitoning. However, if  $\{\vert \psi_m \rangle\}$  is chosen to be the single particle eigenbasis of the Hamiltonian, then mode entanglement will only arise due to interactions in systems of bosons.

\subsection{Particle partitioned entanglement}
For systems of itinerant particles, instead of partitioning the system into subsets of modes (including spatial subregions) we may choose to partition the system into subsystems of {\it particles}~\cite{Zanardi2002,Shi2003,Fang2003,Zozulya2008,Haque2009} as depicted in Fig.~\ref{fig:bipart}. A particle bipartition of a system of indistinguishable particles is entirely determined by the number of particles in the subsystem, $n$. The particle partitioned entanglement is a function of the $n$-body reduced density matrix $\rho_n$, which is most naturally defined in first quantized notation:
\begin{equation}
\rho_n \equiv \int d^d \boldsymbol{r}_{n}\dots d^d \boldsymbol{r}_{N-1}\left \langle \boldsymbol{r}_{n}\dots \boldsymbol{r}_{N-1} \right \vert \rho \left \vert \boldsymbol{r}_{n}\dots \boldsymbol{r}_{N-1} \right \rangle. \notag
\end{equation}
Note that we have chosen the normalization ${\rm Tr} \rho_n = 1$. The particle partitioned entanglement can be quantified through the R\'enyi entropies $S_{\alpha}(n) \equiv S_\alpha[\rho_n] $. The particle entanglement entropies only vanish when the many-body state is in a product state in first quantized notation, i.e., when all particles are condensed into one mode $\psi_m$:
\begin{equation}
S_\alpha \left( n \right) = 0 \Rightarrow \eket{\Psi} =\prod_{i=0}^{N-1} \eket{\psi_m}_i = \eket{n_m = N} . \notag
\end{equation}
Clearly many-fermion systems always have nonzero particle entanglement entropy, but for systems of bosons with a nondegenerate single particle ground state, the particle entanglement entropy will vanish in the noninteracting limit, when the ground state is a Bose-Einstein condensate. However, bosonic systems may have ``trivial" particle entanglement entropy as well, when the single particle ground state is degenerate, including the case where the system is taken to be a composite of two isolated noninteracting parts.  We emphasize, therefore, that particle entanglement entropy can arise both from interactions as well as a consequence of particle indistinguishability.

\subsection{Entanglement of particles}
\label{sec:EoP}

While both mode and particle partitioned entanglement entropies may give insight into the nature of a quantum state, neither is a direct measure of the physically accessible entanglement that may be experimentally accessed as a nonlocal resource for quantum information processing protocols, such as quantum teleportation \cite{Boschi:1998qt}. Accessing entanglement as a resource requires the ability to perform local physical operations on the subsystems. However, for those defined by mode and particle partitions, arbitrary local physical operations {\it cannot} be performed on the relevant subsystem.

In systems of identical particles, a subset of particles that defines $\rho_n$ is not accessible, even in principle, due to the indistinguishability of the particles. However, recent work by Killoran {\it et al.}~\cite{Killoran:2014gu} presents a protocol to transfer particle entanglement of identical particles into mode entanglement which is physically accessible. While Ref.~\onlinecite{Killoran:2014gu} relies on nonlocal operations that, in principle, can generate entanglement on their own, Killoran {\it et al.} provide conditions under which these nonlocal operations are sufficiently passive to merely transfer the particle entanglement of the initial state without generating additional entanglement~\cite{Killoran:2014gu}. For distinguishable particles, it may be possible to address physical operations on one species of particle, and therefore entanglement under a partition between species may be physically accessible \cite{Byrnes:2012mq}.

In contrast, a subset of modes is in general addressable by local physical operations. However, here we assume that the system of itinerant particles has an underlying conservation law that implies a particle number superselection rule; such a superselection rule forbids physical operations from creating superpositions of eigenstates of particle number with different numbers of particles~\cite{Aharonov1967} and thus restricts the local physical operations that are available. As discussed by  Wiseman and Vaccaro \cite{Wiseman2003}, entanglement generated purely by occupation number fluctuations between subsystems {\it cannot} be extracted by {\it local} physical operations that are constrained by such a superselection rule. To observe such occupation number mode entanglement (i.e., distinguishing the pure entangled state from a mixed state) requires a common reference phase to be shared by both subsystems~\cite{Aharonov1967}; such a shared reference phase requires a {\it nonlocal} resource which could introduce entanglement on its own~\cite{Dowling2006}. Consequently, mode-bipartitioned entanglement entropies generally {\it overestimate} the {\it physically accessible} entanglement of an itinerant particle system. However, there are quantum protocols which can take advantage of mode-occupation entanglement in the presence of superselection rules~\cite{Heaney2009}. 

To get a more direct measure of the entanglement that is accessible as a nonlocal physical resource, Wiseman and Vaccaro introduced the notion of the {\it entanglement of particles} based on an operational definition of entanglement \cite{Wiseman2003}. The entanglement of particles $E_p$ is defined as the amount of entanglement under a particular mode bipartition given the physical limitations of a superselection rule. For the mode bipartitioned reduced density matrix $\rho_A$, $E_p$ is determined by projecting onto a state of definite local particle number and taking the weighted average of an entanglement measure. Here we define $E_p$ for the R\'{e}nyi entropies as
\begin{equation}
E_p^\alpha \left( A \right) \equiv \sum_n P_n S_\alpha \left[ \rho_A^{(n)}\right], \label{eq:Ep}
\end{equation}
where $\rho_A^{(n)}$ is the projected reduced density matrix,
\begin{equation}
\rho_A^{(n)} \equiv  \frac{1}{P_n}\hat{P}_n \rho_A  \hat{P}_n, \notag
\end{equation}
$\hat{P}_n$ are projection operators onto eigenstate of particle number in $A$ with $n$ particles, and $P_n$ are the probabilities $P_n = \bra{\Psi} \hat{P}_n \ket{\Psi}$. 

Since $\rho_A^{(n)}$ has a definite particle number, $E_p$ is not sensitive to subsystem occupation number entanglement in $\ket{\Psi}$. Nonzero entanglement of particles requires that the projected state $\hat{P}_n \ket{\Psi}$ is not product state under the mode bipartition for at least one value of $n$; this is not, in general, true even when $\ket{\Psi}$ itself is not a product state, as the mixedness of $\rho_A$ may be solely due to particle fluctuations between subsystems. Consequently, for a given mode bipartition, the mode entanglement is an upper bound on the entanglement of particles:
\begin{equation}
E_p^\alpha \left(A \right) \leq S_\alpha \left(A \right). \notag
\end {equation}

Additionally, a nonzero particle partitioned entanglement is required to have nonzero entanglement of particles. For all particle entanglements to vanish, all particles mush be condensed into one single-particle mode $\ket{\psi_0}$ such that $\ket{\Psi} = \ket{n_0 = N}$. In an arbitrary mode basis, $\ket{\psi_0}$ will have nonzero overlap with both $A$ and $B$ modes such that
\begin{equation}
\ket{\psi_0} = \sum_{m_A} \psi_0 \left( m_A \right) \eket{\psi_{m_A}}+\sum_{m_B} \psi_0 \left( m_B \right) \eket{\psi_{m_B}}, \notag
\end{equation}
where $\psi_0(m_{A/B})$ are the overlaps with the $A$ and $B$ modes. We may define two modes that are completely localized to $A$ and $B$ accordingly,
\begin{align}
\eket{a_0} &\equiv \frac{1}{\sqrt{p_A}}\sum_{m_A} \psi_0 \left( m_A \right) \eket{\psi_{m_A}}, \notag \\
\eket{b_0} &\equiv \frac{1}{\sqrt{p_B}}\sum_{m_A} \psi_0 \left( m_B \right) \eket{\psi_{m_B}} , \notag
\end{align}
where $p_{A/B} \equiv \sum_{m_{A/B}} \vert \psi_0 (m_{A/B})\vert^2$.
The many-body condensate $\ket{\Psi}$ may be written as
\begin{equation}
\eket{\Psi}  = \frac{1}{\sqrt{N!}} \left(\sqrt{p_A} a_0^{\dagger} +\sqrt{p_B}b_0^{\dagger}\right)^N \eket{0}. \notag
\end{equation}
We can define a Fock space from the modes $a_0^\dagger$ and $b_0^\dagger$, which we represent as $\{ \ket{n_A,n_B} \}$. The condensate is then written in this basis as
\begin{equation}
\eket{\Psi} = \sum_{n=0}^N \sqrt{\binom{N}{n}} p_A^{n/2} p_B^{\left(N-n\right)/2} \ket{n,N-n} . \notag
\end{equation}
In this form it is clear that for the condensate, $\rho_A^{(n)}$ is a pure state for all $n$:
\begin{equation}
\eket{\Psi} = \eket{n_0 = N} \Rightarrow \rho_A^{(n)} = \eket{n,N-n}\ebra{n,N-n}. \notag
\end{equation}
Consequently, nonvanishing entanglement of particles requires a nonvanishing particle entanglement:
\begin{equation}
E_p^\alpha > 0 \Rightarrow S_\alpha(n) > 0. \notag
\end{equation}
We see then that while both mode and particle partitioned entanglement entropies may detect entanglement that is not physically accessible as a nonlocal resource, both must be nonzero for the entanglement of particles to be nonvanishing.

\section{Entanglement in systems of itinerant bosons}
\label{sec:BH}

To elucidate the behavior of the different entanglement measures described above, we consider several canonical phases that appear in lattice models of itinerant bosons (see also [\onlinecite{Ding2009,Dowling2006,Haque2009,Zozulya2008}] for related discussions). For concreteness we present a study of the 1D Bose-Hubbard model on a lattice of length $L$ with $N$ bosons interacting via the Hamiltonian
\begin{equation}
    H_{\rm{BH}} = \sum_j \left[ -t\left(b^\dagger_j b^{\phantom \dagger}_{j+1} + \rm{h.c.} \right)+ \frac{U}{2} n_j \left( n_j+1\right) - \mu_j n_j \right] \label{eq:Hbh}
\end{equation}
where $b_j^\dagger$ ($b^{\phantom\dagger}_j$) is the creation (annihilation) operator, $n_j$ is the number operator, $t$ is the hopping strength, $U$ is an onsite interaction, and $\mu_j$ is a site dependent chemical potential. Here we consider $t>0$ and unit filling $N=L$.

The ground state of noninteracting bosons ($U=0$) with uniform $\mu_j$ is a perfect Bose condensate where all particles condense into the single particle ground state mode
\begin{equation}
\phi_0^\dagger = \frac{1}{\sqrt{L}}\sum_j b_j^\dagger, \notag
\end{equation}
such that the $N$ particle state is 
\begin{equation}
\eket{\rm{BEC}} \equiv \frac{1}{\sqrt{N!}} \left( \phi_0^\dagger\right)^N \eket{\mathbf{0}}. \notag
\end{equation}
Given that $\eket{\rm{BEC}}$ is a product state in first quantized notation, all particle entanglements vanish: $S_\alpha^{\rm{BEC}} (n) =0$. However, due to the delocalized nature of $\phi_0$, $\eket{\rm{BEC}}$ is highly entangled under any spatial bipartition~\cite{Ding2009,Simon2002}. For a subregion $A$ of length $\ell$ the second \ren entropy is~\cite{Ding2009}
\begin{equation}
S_2^{\rm{BEC}} \left( \ell \right) = -\log \Bigl[\sum_{j=0}^L \Bigl(\frac{L!}{2^L \left(L-j\right)! j!} \ell^j \left(L-\ell\right)^{L-j} \Bigr)^2 \Bigr], \notag
\end{equation}
which scales as $(1/2)\log \ell$ for large $L$.

In the strongly repulsive limit, $U\rightarrow+\infty$, the ground state is a Mott insulator, where each site is singly occupied:
\begin{equation}
\eket{\mathrm{Mott}} \equiv \prod_j b_j^\dagger \eket{\mathbf{0}}. \notag
\end{equation}
The Mott insulator is manifestly a product state in the spatial mode basis, and thus all spatial entanglement entropies vanish: $S_\alpha^{\rm{Mott}} (\ell) =0$. However, the indistinguishability of the particles leads to a large particle entanglement:
\begin{equation}
S_2^{\rm{Mott}} \left( n \right) = \log \frac{L!}{\left(L-n\right)!n!}\,. \notag
\end{equation}
The scaling of $S_2^{\rm{Mott}}(n)$ ranges from $\log L$ for $n=1$ to $L \log 2$ for $n=L/2$ and $L\gg1$. 

For strongly attractively bosons, in the limit $U\rightarrow-\infty$ and all $\mu_j$ equal, the ground state is a ``Schr{\"o}dinger's cat''-like state which is an equal superposition of all states with $N$ particles occupying the same site:
\begin{equation}
\eket{\mathrm{Cat}} \equiv \sum_j \frac{1}{\sqrt{L} \sqrt{N!}} \left( b_j^\dagger\right)^N \eket{\mathbf{0}}. \notag
\end{equation}
This cat state has both nonzero particle and spatial entanglement entropies:
\begin{align}
S_2^{\rm{Cat}} \left( n \right) &= \log L, \notag \\
S_2^{\rm{Cat}} \left( \ell \right) &= -\log \Bigl[ 1-2\frac{\ell}{L}\left(1-\frac{\ell}{L}\right) \Bigr].  \notag
\end{align}

The cat state is unstable to local perturbations and if $\mu_j > \mu_i\; \forall_{i\neq j}$ for some site $j$, then in the $U\rightarrow-\infty$ limit the cat state will collapse to a state where particles are localized on the site $j$:
\begin{equation}
\eket{ N_j} \equiv \frac{1}{ \sqrt{N!}} \left( b_j^\dagger\right)^N \eket{\mathbf{0}}. \notag
\end{equation}
Such a state is trivially a product state in the spatial basis with only a single mode accessible to all particles; consequently, all particle and spatial entanglement entropies vanish.

We note that the entanglement of particles strictly vanishes for all the canonical states discussed above for any choice of spatial bipartition. As mentioned in Sec. \ref{sec:EoP}, $E_p \neq 0$ requires both particle and spatial entanglement entropies to be nonzero. Only the cat state satisfies this requirement, but the spatial entanglement of $\ket{\rm Cat}$ is solely due to fluctuations of all $N$ particles between sites and not interactions between the subsystems. This is precisely the sort of fluctuation entanglement that $E_p$ is not sensitive to and thus $E_p = 0$ for all states discussed above. Table \ref{tab:BH} shows the leading order scaling of various entanglement measures for large $L$ for these canonical states.
\begin{table}[t]
\begin{center}
    \renewcommand{\arraystretch}{1.2}
  \begin{tabular}{ c c c c c}
    \hline\hline
    State	& $S_2 \left( n = \frac{N}{2} \right)$	& $S_2 \left(\ell =  \frac{L}{2} \right)$	& $E_p \left( \ell \right)$ 	\\ \hline
    BEC 	&  $0$						&  $\frac{1}{2}\log L$			& $0$ 			\\  
    Mott 	&  $\left(\log 2\right) L$					& $0$ 						& $0$			\\ 
    Cat 	&  $\log L$					& $\log 2$				& $0$ 			\\ 
    $N_j$ 	&  $0$						& $0$ 						& $0$ 			\\ \hline \hline    \end{tabular}
\end{center}
\caption{\label{tab:BH} Leading order scaling of the particle and spatial 2nd R\'{e}nyi entropies under symmetric bipartitions ($n=N/2$ or $\ell=L/2$) at unit filling $N=L$ for large $L$ for the canonical states discussed in section \ref{sec:BH}. The entanglement of particles strictly vanishes for these states for all $N$ and $L$ for any bipartition. }
\end{table}

 \begin{figure}
\begin{center}
\scalebox{1}{\includegraphics[width=\columnwidth]{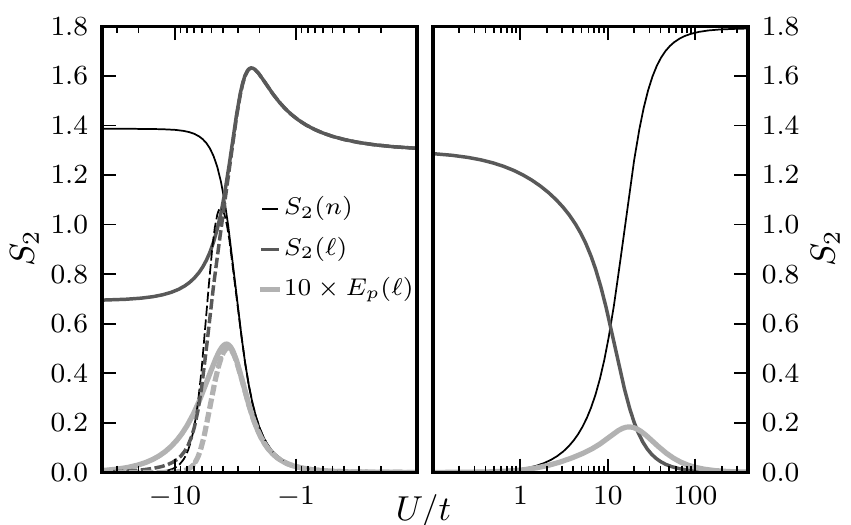}}
\end{center}
\caption{The spatial [$S_2(\ell)$] and particle [$S_2(n)$] second R\'{e}nyi entropies and the entanglement of particles [$E_p(\ell)$] for symmetric bipartitions ($\ell=n=L/2$) of the 1D Bose-Hubbard model Eq.~\eqref{eq:Hbh} for a $L=N=4$ system, as computed by exact diagonalization. The left and right panels correspond to negative and positive values of $U$, respectively. The dashed lines show the same quantities in the addition of a symmetry breaking chemical potential: $\mu_0 =0.02$ and $\mu_{j\neq0} = 0$. Note that the scale of $E_p$ has been magnified by an order of magnitude for visibility.} 
\label{fig:BHSvsU}
 \end{figure}
 To illustrate the behavior of the three entanglement measures---particle, spatial, and entanglement of particles---away from these canonical states, we have have computed the entropies for a small system via exact numerical diagonalization. Figure \ref{fig:BHSvsU} shows the spatial and particle second R\'{e}nyi entropies and the entanglement of particles for symmetric bipartitions ($\ell=n=L/2$) as a function of $U/t$ for $L=N=4$. For this system, $S_2^{\rm{BEC}} \left( \ell=2 \right)=\log(128/35)\approx1.3$ and $S_2^{\rm{Mott}} \left( n=2 \right)=\log6\approx1.8$. Note the small scale of $E_p$ relative to $S_2(n)$ and $S_2(\ell)$, which has been magnified by an order of magnitude for clarity in Fig.~\ref{fig:BHSvsU}. With a clear physical picture of the behavior of these \ren entanglement entropies for a specific itinerant boson system, we now introduce a general and scalable method for computing them for a wide class of physically relevant Hamiltonians with QMC.

\section{Method}

Path integral Monte Carlo (PIMC) is a powerful tool to study ground state and finite temperature properties of strongly interacting many-body systems~\cite{Ceperley1995}. For interacting bosons without a ``sign problem,'' the polynomial scaling of computational resources required for PIMC allows for the study of large-scale systems in any dimension described by the Hamiltonian of Eq.~(\ref{eq:Ham}).  This includes experimentally relevant systems such as liquid helium-4 and cold atomic gases. At $T=0$, the path-integral ground state (PIGS) algorithm \cite{Sarsa2000,Cuervo2005a} provide access to ground state properties of a many-body system by statistically sampling the imaginary time propagator $e^{-\beta H}$. Given a trial wave function $\Psi_{\mathrm{T}}$, in the large imaginary time limit $\beta\rightarrow\infty$, $e^{-\beta H} \ket{\Psi_{\mathrm{T}}}$ converges to the ground state, as long as $\ket{ \Psi_\mathrm{T} }$ has any finite overlap with it.  Therefore, we may compute ground state properties by statistically sampling the expectation value of an observable $\mathcal{\hat{O}}$,
\begin{equation}
\eexp{ \hat{\mathcal{O}} } = \lim_{\beta \rightarrow \infty}\frac{ \ebra{\Psi_{\mathrm{T}}}e^{-\beta H} \hat{\mathcal{O}} e^{-\beta H} \eket{\Psi_{\mathrm{T}}}}{\ebra{\Psi_{\mathrm{T}}}e^{-2 \beta H} \eket{\Psi_{\mathrm{T}}}}. \notag
\end{equation}
A considerable benefit of PIGS over other zero temperature methods is is that the choice of $\Psi_{\mathrm{T}}$ does not introduce any systematic bias in the measurement of estimators \cite{Boninsegni2012} provided that $\beta$ is large enough. 

\begin{figure}
\begin{center}
\scalebox{1}{\includegraphics[width=\columnwidth]{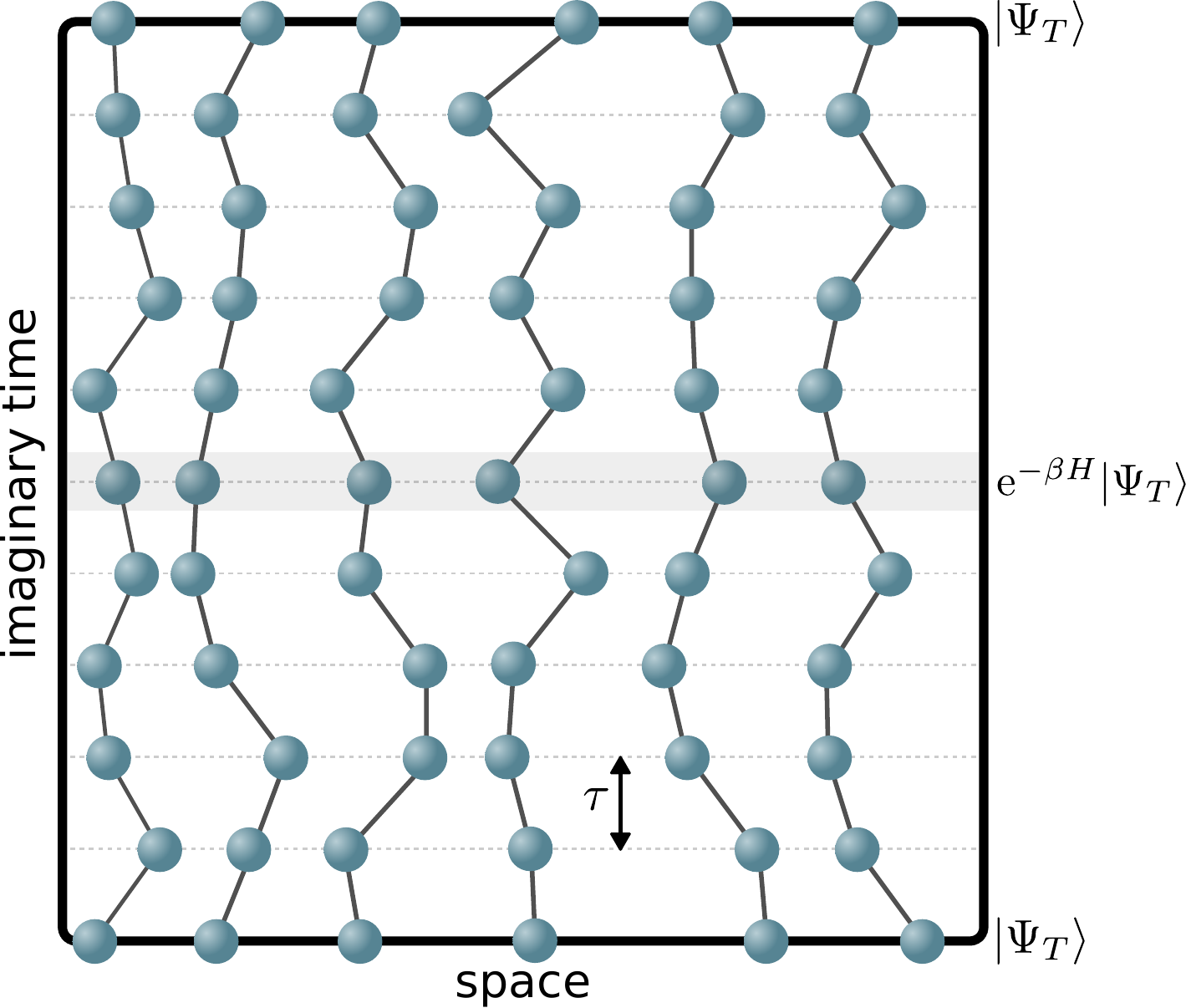}}
\end{center}
\caption{(Color online) A configuration with $N=6$ bosons at zero temperature in
one spatial dimension. Diagonal ground state estimators can be measured at the central time slice as indicated by the shaded bar.} 
\label{fig:worldlines}
 \end{figure}

In practice, ground state estimators for a $D$-dimensional interacting continuum system are computed with PIGS by working in an extended $(D+1)$-dimensional configuration space where the degrees of freedom involve imaginary time world lines of the particles.  The absence of a lattice requires that the imaginary time direction be broken into an integer number of \textit{time steps} of size $\tau$ and we approximate the full propagator $\mathrm{e}^{-\beta H}$ as a product of an approximate short time propagator $\rho_\tau \simeq \mathrm{e}^{-\tau H}$.  The error made in using the short-time propagator is determined by the size of $\tau$ and the specific decomposition employed to deal with the noncommuting parts of $H$.  As described in Sec.~\ref{subsec:shortTime}, we use a form for $\rho_\tau$ that allows us to use a sufficiently small $\tau$ to ensure that any systematic errors are smaller than statistical uncertainty. The imaginary time world lines are composed of discrete particle positions, referred to as ``beads'', connected by links representing insertions of the short-time propagator. The indistinguishability and bosonic symmetry of the particles is enforced through the choice of the trial wave function and a proper symmetrization of any estimator measured at the central time slice. A sample configuration of world lines is shown in Fig.~\ref{fig:worldlines}.

\subsection{Replicated configuration space}
\label{sec:replicated}

Despite not being a conventional observable, recent work by Hastings {\it et al.} demonstrated that R\'{e}nyi entanglement entropies can be computed in Monte Carlo simulations by defining a so-called ``SWAP'' operator in a {\it replicated} Hilbert space~\cite{Hastings2010,Melko2010}. There is a large volume of subsequent literature that applies these and related methods to compute spatial entanglement entropy in lattice systems with QMC (e.g., see Ref.~\onlinecite{Zhang2012,Inglis2013b,Selem2013,Tommaso,Inglis:2013iv,Pei2014,Broecker,Chung2013,Chung2013a}). However, its measurement in continuous-space systems has been limited to a variational Monte Carlo method for fermionic systems~\cite{Tubman2012,McMinis:2013dp,Tubman2014} and a recent study by the authors of $1D$ short-range interacting bosons~\cite{algorithm}. Here we report on  details and various extensions of the latter method, which is based on PIGS and allows for the computation of particle and spatial-partitioned R\'{e}nyi entanglement entropy in the ground state of $D$-dimensional bosonic quantum fluids.

Motivated by the algorithm of Hastings {\it et al.}, we consider a replicated Hilbert space of a continuous-space system of $N$ bosons in first quantized notation. A basis state of the original system can be written as $\ket{\R}$, where $\R = \{\vec{r}_0,...,\vec{r}_{N-1}\}$ is a vector of length $D\times N$ describing the position of all particles.  This Hilbert space is then replicated, producing $\{\ket{\tilde{\R}}\}$ and allowing for the formation of a tensor product Hilbert space $\{\ket{\R}\otimes\ket{\tilde{\R}}\} \equiv \{\ket{\R\otimes\tilde{\R}}\}$.  $\R$ and $\tilde{\R}$ are noninteracting, physically equivalent systems and the definition of an operator that connects observables between them will allow for the estimation of the second \ren entropy $S_2$.  A straightforward extension to measuring $S_\alpha$ would require replicating the system $\alpha$ times:
$\tilde{\R} \to \R_j$ for $j=1,\ldots,\alpha$.

\begin{figure}
\begin{center}
\scalebox{1}{\includegraphics[width=\columnwidth]{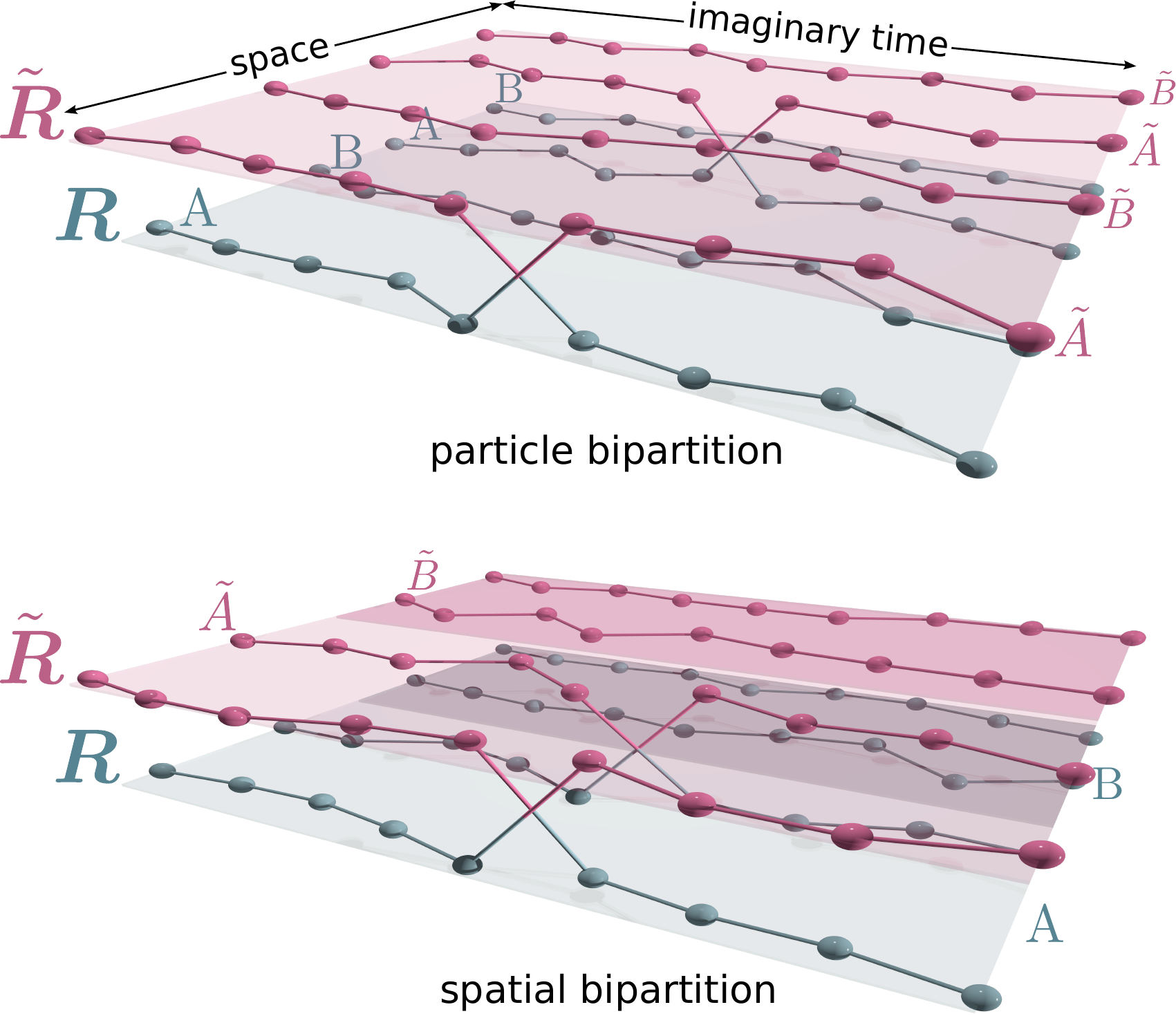}}
\end{center}
\caption{(Color online) Replicated $(1+1)$-dimensional Monte Carlo configurations of $N=4$ bosons showing world lines broken at time slice $\beta$ for a particle bipartition with $n=2$ (top) and a spatial bipartition with $\ell=L/2$ (bottom).  For both types of bipartitions, particle world lines in subsystem $A$ are discontinuous, while those in subsystem $B$ are continuous.  Two-tone lines connecting $\R$ and $\tilde{\R}$ correspond to insertions of the short-time propagator $\rho_\tau$ that are used to measure the \ren entropies.}
\label{fig:replica}
 \end{figure}

To compute a bipartite R\'{e}nyi entropy, we must first define a subsystem by a particular choice of bipartition. As described in Ref.~[\onlinecite{algorithm}], to measure the particle entanglement, we choose a subset of particles $A$, such that $\R = \{\R_A,\R_B\}$. Bosonic symmetry implies that any physical properties of this bipartition will only depend on the number $n$, of particles in $A$.  For spatial-mode partitioning, we must define a spatial subregion $A$ and decompose each configuration into $\{\R_A,\R_B\}$, where we implicitly have assumed that
\begin{equation}
    \vec{r} \in \R_A \Rightarrow \vec{r}_A \in A. \notag
\end{equation} 
To compute the R\'{e}nyi entropy of the ground state of a system with a Monte Carlo method, we must be able to sample the ground state in the replicated Hilbert space and define an appropriate generalization of the SWAP estimator given a choice of bipartition. This can be achieved by generating configurations that have a mix of closed and broken world lines. The breaks are constrained to occur only at the central imaginary time slice, and allow for the insertion of an off-diagonal operator with nonvanishing weight. For particle partitioning, there are $n$ broken world lines in the ensemble, whereas for spatial partitioning all world lines within $A$ are broken and consequently the number of such broken world lines fluctuates.  Examples of the replicated configuration space corresponding to both types of bipartition are shown in Fig.~\ref{fig:replica}.

We statistically sample an ensemble of such configurations with a weight given by
\begin{equation}
W \left( \R_{\beta}, \R_{\beta+\tau} \right) = W_\beta \left( \R_{\beta} \right) \rho_{\tau}^B \left( \R_{\beta}, \R_{\beta+\tau}\right) W_\beta \left(\R_{\beta+\tau}\right), \label{eq:weights}
\end{equation}
where $W_\beta \left( \R \right)$ is the weight of the closed path on either side of center time step,
\begin{equation}
W_\beta \left( \R \right)  = \bra{\R} e^{-\beta H} \ket{\Psi_T}, \notag
\end{equation}
$\R_{\beta}$ and $\R_{\beta+\tau}$ are the configurations on either side of the center time step, and we have defined the reduced propagator for the $B$ subsystem as
\begin{align}
 \rho_\tau^B( \R,\R') &\equiv \frac{n!\left(N-n\right)!}{N!}\sum_{\R_{n_B}} \rho_\tau \left( \R_{n_B}, \R_{B}' \right), \label{eq:rhoB}
\end{align}
where $n_B=N-n$ is the number of particles in subsystem $B$ in $\R'$ and $\R_{n_B}$ is one possible subset of $n_B$ particles of $\R$ such that $\R_{n_B} \in \R$. Additionally, we define $\rho_{\tau} \left( \R, \R'\right)$ to be the matrix elements of the (implicitly) Bose symmetrized propagator: 
\begin{equation}
\rho_{\tau} \left( \R, \R'\right) = \bra{\R} \rho_\tau \ket{\R'}.  \label{eq:rhoRR'}
\end{equation}
The weights given in Eq.~\eqref{eq:weights} correspond to the weights for paths with $N-n$ closed world lines at the center time slice, such as those shown in \figref{fig:replica} without the links connecting the two replicas. Note that $R^{\beta+\tau}_B$ is uniquely determined by $R^{\beta+\tau}$ (either through the spatial locations of the particles or a fixed set of labels, depending on the type of partition) but the $R^{\beta}_B$ must be sampled over for any given $R^{\beta}$. Such an ensemble may be Monte Carlo sampled by standard PIMC methods using a variety of updates to ensure detailed balance~\cite{Ceperley1995}. In the replicated configuration space, we have independent weights for each replica, so the total weight is simply $W(\R_{\beta},\R_{\beta+\tau})W(\tilde{\R}_{\beta},\tilde{\R}_{\beta+\tau})$.

\subsection{Monte Carlo Updates}
\label{sec:updates}

To ergodically sample configuration spaces consisting of broken and closed world lines described above, we use a set of updates that depend on the type of partitioning that is of interest. In all cases, the closed world lines away from the center or ends of the paths may be sampled with a variety of conventional PIMC updates that are well described in the literature \cite{Ceperley1995}. For all partition choices, we update the open ends of the world lines by generating a new free particle path of length $M/2$ starting from the configuration $M/2$ time steps from the end. The acceptance rate is then determined by the ratio of the diagonal weights of the old and new paths.

\subsubsection{Particle partitioning}
\begin{figure}
\begin{center}
\scalebox{1}{\includegraphics[width=\columnwidth]{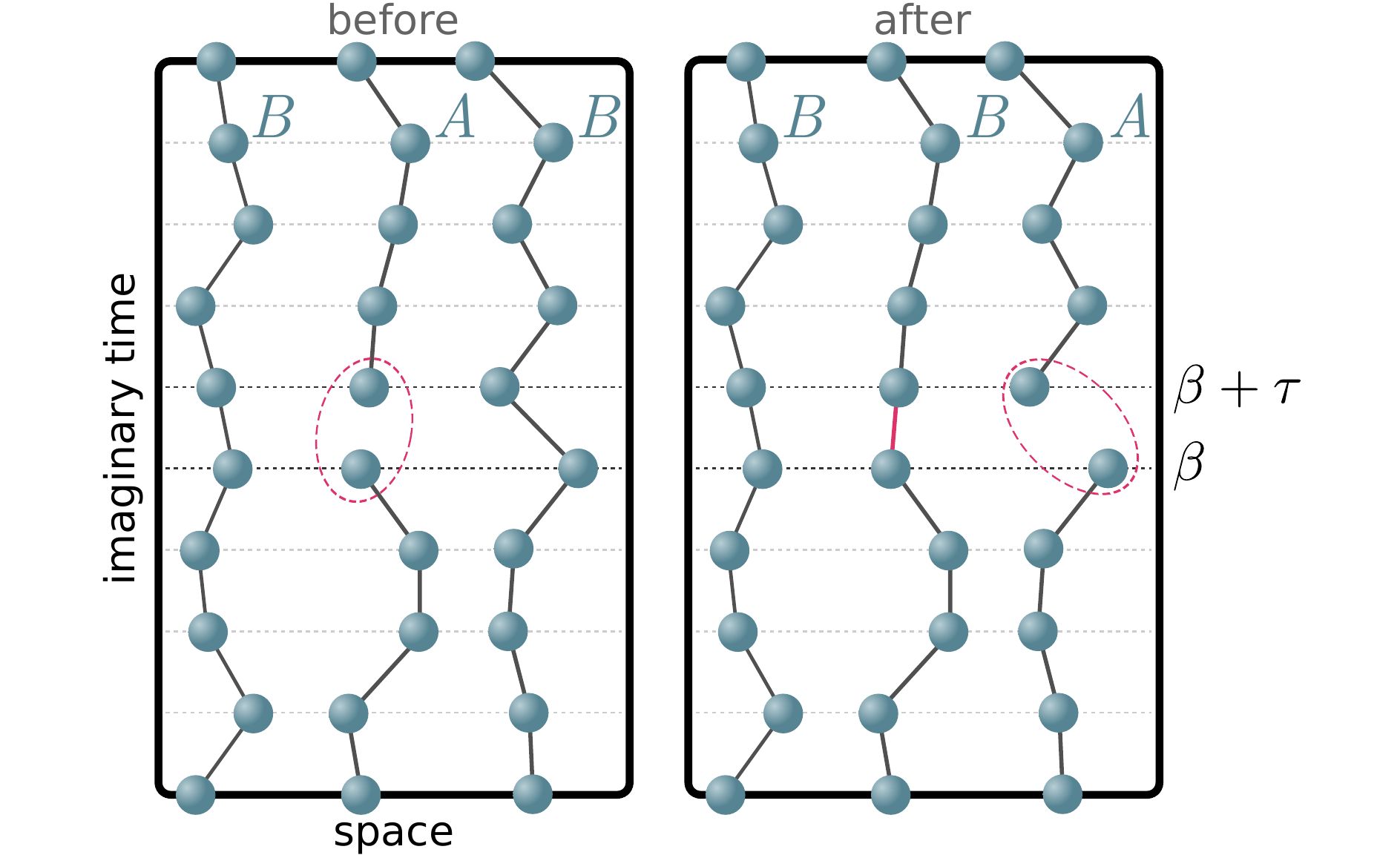}}
\end{center}
\caption{(Color online) The break-swap update used in the measurement of the \ren entropy for a particle bipartition.  A $(1+1)$-dimensional configuration of $N=3$ bosons is updated by proposing a move that \textit{swaps} the location of a missing link joining time slices $\beta$ and $\beta+\tau$ between a broken and a connected world line.  The acceptance probability is given in Eq.~(\ref{eq:breakswap}) of the text.}
\label{fig:breakswap}
 \end{figure}

To compute particle entanglement entropies, we sample a configuration space with a fixed number, $n$, of broken world lines. The disconnected beads at the center of the path are not restricted in space, and the number of such beads remains fixed at $n$ throughout the simulation. The broken world lines at the center of the path can be updated in the same manner as the ends of the paths. If the same world lines remain broken during the simulation, then the estimator will not, in general, be symmetric over all particles, as the broken world lines introduce an artificial label which renders them distinguishable.  To ensure that the estimator is symmetric over particle permutations, we have implemented a ``break-swap'' update that reconnects a broken worldline and breaks a connected worldline.  This update is summarized in Fig.~\ref{fig:breakswap} and the procedure to implement it is as follows.
%

\begin{enumerate}
\item Randomly choose a bead on each side of the broken path, ($\rv^{o}_{\beta}$ and $\rv^o_{\beta+\tau}$) and a closed worldline with a bead at position $\rv^c_{\beta}$.
\item Propose the formation of a link between beads $\rv^o_{\beta}$ and $\rv^o_{\beta+\tau}$ and the removal of the link between $\rv^c_{\beta}$ and $\rv^c_{\beta+\tau}$.
\item Accept the update with probability 
\begin{equation}
P_{\mathrm{acc}}^{b-s} = \text{min} \left[ \frac{N-n}{n^2} \frac{\rho_\tau \left(\rv^o_{\beta},\rv^o_{\beta+\tau}\right)}{\rho_\tau \left(\rv^c_{\beta},\rv^c_{\beta+\tau}\right)},1\right].
\label{eq:breakswap}
\end{equation} 
\item If the update is accepted, form a link between beads $\rv^o_{\beta}$ and $\rv^o_{\beta+\tau}$ while breaking the link between $\rv^c_{\beta}$ and $\rv^c_{\beta+\tau}$.
\end{enumerate}

For efficiency, one can build a nearest neighbor table at the center time slice based on the free particle propagator and only attempt to link beads within a certain length scale, as the free particle propagator decays exponentially in distance. In practice such an update may be unnecessary if the beads at the center of the path do not break permutation symmetry, which will depend on the nature of the physical ground state.

\subsubsection{Spatial partitioning}

\begin{figure}
\begin{center}
\scalebox{1}{\includegraphics[width=\columnwidth]{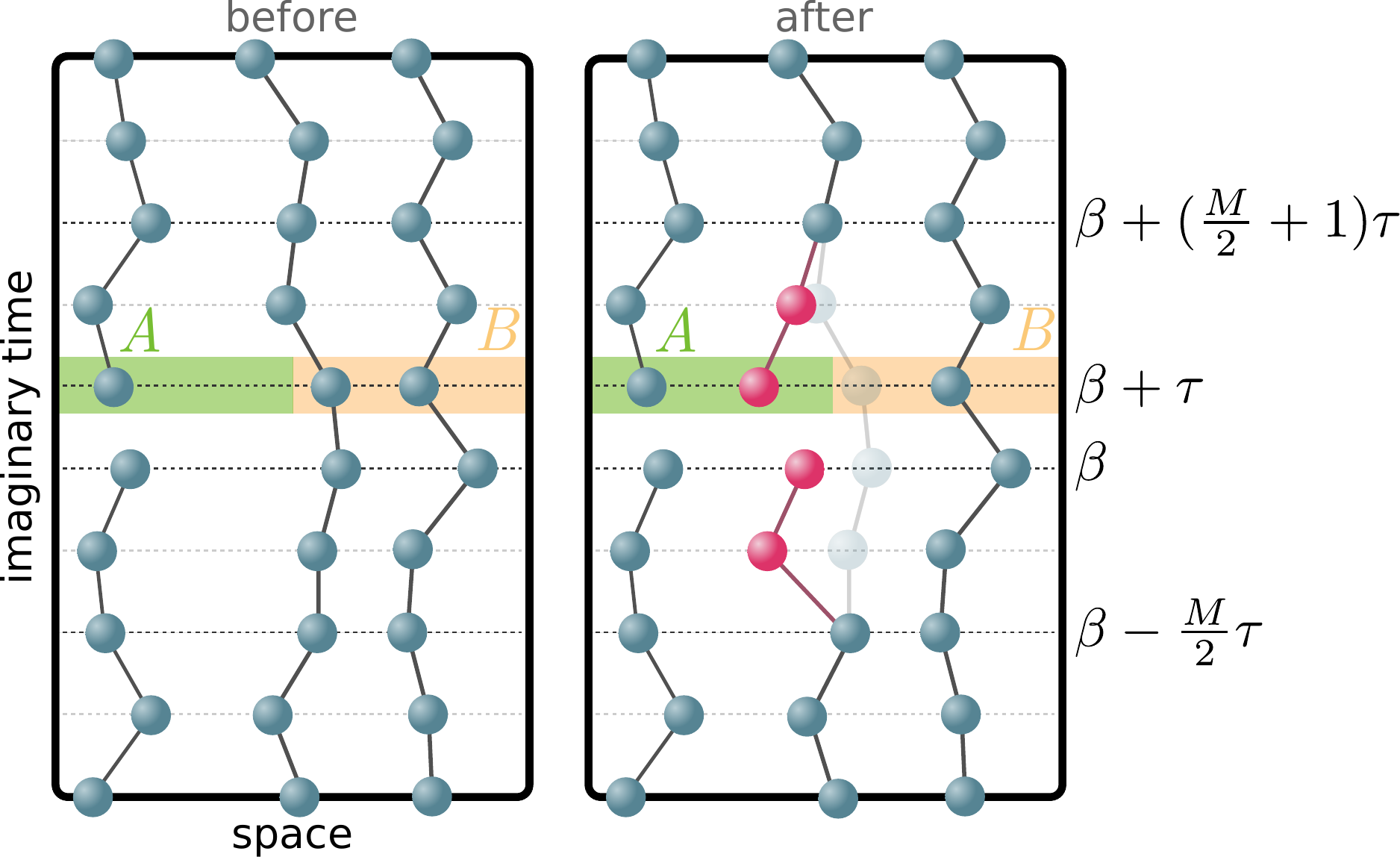}}
\end{center}
\caption{(Color online) The spatial-reconnect update used in the measurement of the \ren entropy for a spatial bipartition.  A $(1+1)$-dimensional configuration of bosons is updated by proposing a move that generates a new free particle trajectory between time slices $\beta-M/2\tau$ and $\beta+(M/2+1)\tau$.  If the bead at $\beta$ is moved into region $A$ after the update, a link across the central time slice is removed.  If the modified path moves a bead at time slice $\beta$ from region $A$ into region $B$, the worldline is reconnected. The acceptance probabilities for the various possibilities are given in Eqs.~(\ref{eq:AA})--(\ref{eq:BB}) in the text.}
\label{fig:spatialrecon}
 \end{figure}

For a spatial bipartition, broken beads on one side of the center imaginary time link at $\beta + \tau$ are constrained to reside in the spatial subregion $A$ and the number of broken world lines will fluctuate as particles move in and out of the region. Consequently, an update which changes the number of broken beads as they move between subregions is required, and a schematic of the ``spatial-reconnect'' move is shown in Fig.~\ref{fig:spatialrecon}.  
%
%
The update is implemented as follows.
 
\begin{enumerate}
\item  Choose a bead at imaginary time $\beta-M/2\tau$, which is at position $\rv_{\beta-M/2\tau}$. 
\item Traverse the worldline to the center link $\beta$ which defines the bead at $\rv_{\beta}$. 
\item If $\rv_{\beta}$ is on a broken worldline, choose a disconnected bead $\rv_{\beta+\tau}$ at random; otherwise, $\rv_{\beta+\tau}$ is defined by the center link. 
\item Move $M/2$ additional steps along the chosen worldline to reach $\rv_{\beta+(M/2+1)\tau}$. This defines the world lines that will be potentially updated.
\item Generate a new free particle path between $\rv_{\beta-M/2\tau}$ and $\rv_{\beta+(M/2+1)\tau}$, and label this path of $M$ new bead positions as $\rv'$. This can be done in a rejection-free manner by standard PIMC methods~\cite{Ceperley1995}.
\item The probability of accepting the update depends on which of the four possible sceneries occur, where $e^{-\delta V}$ is the ratio of the initial to final diagonal weights:
\begin{enumerate}[(a)]
    \item $(\rv_{\beta+\tau} \in A) \land (\rv'_{\beta+\tau} \in A)$,
\begin{equation}
P_{\mathrm{acc}}^{AA} =  \text{min} \left[ \frac{\rho_\tau \left(\rv_{\beta},\rv_{\beta+\tau}\right)}{\rho_\tau \left(\rv_{\beta}',\rv_{\beta+\tau}'\right)}e^{-\delta V},1\right];
\label{eq:AA}
\end{equation}

\item $(\rv_{\beta+\tau} \in A) \land (\rv'_{\beta+\tau} \in B)$,
\begin{equation}
P_{\mathrm{acc}}^{AB} =  \text{min} \left[ \rho_\tau \left(\rv_{\beta},\rv_{\beta+\tau}\right) e^{-\delta V},1\right];
\label{eq:AB}
\end{equation}

\item $(\rv_{\beta+\tau} \in B) \land (\rv'_{\beta+\tau} \in A)$,
\begin{equation}
P_{\mathrm{acc}}^{BA} =  \text{min} \left[ \frac{1}{\rho_\tau \left(\rv'_{\beta},\rv'_{\beta+\tau}\right)} e^{-\delta V},1\right];
\label{eq:BA}
\end{equation}

\item $(\rv_{\beta+\tau} \in B) \land (\rv'_{\beta+\tau} \in B)$,
\begin{equation}
P_{\mathrm{acc}}^{BB} =  \text{min} \left( e^{-\delta V},1\right).
\label{eq:BB}
\end{equation}
\end{enumerate}

\item Accept the update with probability $P_{\mathrm{acc}}$ given above. 
    
\item If the move is accepted and $(\rv_{\beta+\tau} \in B) \land (\rv'_{\beta+\tau} \in A)$,  remove the link between $\rv'_{\beta}$ and $\rv'_{\beta+\tau}$. If the move is accepted and 
$(\rv_{\beta+\tau} \in A) \land (\rv'_{\beta+\tau} \in B)$, preserve the link between $\rv'_{\beta}$ and $\rv'_{\beta+\tau}$.
\end{enumerate}

\subsection{Measuring R\'{e}nyi entropies in PIGS}
\label{sec:estimator}

We define a permutation operator $\Pi_{\alpha}^A$ that maps $\R_A$ from one replica to another, modulo $\alpha$, and acts as the identity on all $\R_B$ \footnote{For $S_2$, $\Pi_{2}^A$ is called the ``SWAP'' operator in the literature for spatial entanglement.  However, we do not use this notation here to avoid confusion with a {\it swap} update used in a continuous space worm algorithm.}.  In the case of the second R\'enyi entropy, 
$\Pi_{2}^A$ then simply interchanges the subset $A$ and $\tilde{A}$ between the two subsystems,
\begin{multline}
    \Pi_{2}^A \left[  \{\R_A,\R_B\} \otimes \{\tilde{\R}_{\tilde{A}},\tilde{\R}_{\tilde{B}}\} \right]  = \\ \{\tilde{\R}_{\tilde{A}},\R_B\} \otimes  \{\R_A,\tilde{\R}_{\tilde{B}}\}, \notag
\end{multline}
which when written in operator notation is
\begin{equation*}
    \Pi_2^A \Bigl| \R \otimes \tilde{\R} \Bigr \rangle  = \Bigl| \Pi_2^A\left[\R \otimes \tilde{\R}\right] \Bigr\rangle.
\end{equation*}

The expectation value of this permutation operator of state $\ket{\Psi}$ in the doubled Hilbert space is related to the second R\'{e}nyi entropy of $\ket{\Psi}$, $S_2$~\cite{Hastings2010, Cardy}:
\begin{equation}
\ebra{\Psi \otimes \tilde{\Psi}} \Pi_{2}^A \eket{ \Psi \otimes \tilde{\Psi} } = e^{-S_2}. \notag
\end{equation}
The estimator for the $\Pi_2^A $ operator corresponds to sampling the statistical weight linking the world lines of the $A$ particles with $\tilde{A}$ particles across the central time slice, as illustrated in \figref{fig:replica}. In this ensemble, the estimator for the permutation operator is
\begin{equation}
\left \langle \Pi_2^A \right \rangle = 
\frac{1}{Z_2^A} \left \langle\rho^A_{\tau} \Biggl( \R^{\beta}\otimes\tilde{\R}^{\beta};\Pi_{2}^A \left[\R^{\beta+\tau}\otimes\tilde{\R}^{\beta+\tau}\right]\Biggr)\right \rangle, 
\label{eq:PERM}
\end{equation}
where we have defined a reduced propagator
\begin{multline}
    \rho_\tau^A\left(\R \otimes \R'; \R'' \otimes \R'''\right) \equiv  \\
\frac{\rho_\tau(\R;\R'') \rho_\tau(\R';\R''')}{\rho_\tau(\R_{B};\R''_{B}) \rho_\tau(\R'_{B};\R'''_{B})},
\label{eq:rhoA}
\end{multline}
and $Z_2^A$ is the closed path normalization:
\begin{equation}
\left \langle Z_2^A \right \rangle = 
\left \langle\rho^A_{\tau} \left( \R^{\beta}\otimes\tilde{\R}^{\beta};\R^{\beta+\tau}\otimes\tilde{\R}^{\beta+\tau}\right)\right \rangle . \notag
\end{equation}
The expectation value in Eq.~\eqref{eq:PERM} corresponds to the statistical weight of the ``permuted'' path and $Z_2^A$ is a normalization factor arising from the weight of the paths under the identity permutation. This form of the estimator is independent of the choice of the short-time propagator, which in general will involve diagonal weights at each bead as well as off-diagonal weights for the links. 

Notice that one has the freedom to choose to account for diagonal weights in either the configuration weights given in Eq.~\eqref{eq:weights} or the estimator Eq.~\eqref{eq:PERM}. For simplicity, in the results presented below we have chosen to include the diagonal weights of $\rho_\tau^A$ in the configuration weights such that the estimator Eq.~\eqref{eq:PERM} must be reweighted accordingly.

The estimator in Eq.~(\ref{eq:PERM}) is composed of the product of Gaussian short-time propagators for both $A$ and $\tilde{A}$ and thus both its numerator and denominator will be exponentially suppressed in the size of the chosen bipartition.  This behavior is expected from an understanding of the behavior of the SWAP operator for spatial bipartitions in local lattice models due to the presence of the area law~\cite{Hastings2010}.  We expect then, for bipartitions that are large in either their spatial regions or number of particles $n$, that a generalized ratio sampling, involving computing $S(A)$ from smaller bipartitions, will be required~\cite{Hastings2010}, thus curing the exponential decay of the bare estimator at an additional {\it quadratic} cost in computational time. 

\subsubsection{Entanglement of particles}

The entanglement of particles, $E_p$, is a weighted average of spatial \ren entropies, as defined by Eq.~\eqref{eq:Ep}. We can compute $E_p$ simultaneously with $S_\alpha (A)$ from an ensemble described in Sec.~\ref{sec:replicated} under a spatial partitioning. This is accomplished in practice, by binning the numerator and denominator of Eq.~\eqref{eq:PERM} according to the number of particles in subregion $A$. We define an estimator for this {\it projected} permutation operator,
\begin{multline}
\left \langle \Pi_2^A \left( n \right) \right \rangle = \frac{1}{Z_2^A \left( n \right) } \\
\times \left \langle \rho^A_{\tau} \Biggl( \R^{\beta}\otimes\tilde{\R}^{\beta};\Pi_{2}^A \left[\R^{\beta+\tau}\otimes\tilde{\R}^{\beta+\tau}\right]\Biggr)  \delta_{n,n_A }  \delta_{n,\tilde{n}_{\tilde{A}}} \right \rangle, \notag
\end{multline}
where $n_A$ and $\tilde{n}_{\tilde{A}}$ are the number of particles in subregion $A$ and $\tilde{A}$ at time slice $\beta+\tau$, respectively, and the normalization factor is 
\begin{align}
    &\left \langle Z_2^A \left( n \right) \right \rangle = \left \langle\rho^A_{\tau} \left( \R^{\beta}\otimes\tilde{\R}^{\beta};\R^{\beta+\tau}\otimes\tilde{\R}^{\beta+\tau}\right) \delta_{n,n_A}  \delta_{n,\tilde{n}_{\tilde{A}}} \right \rangle. \notag
\end{align}
The entanglement of particles may be then computed from this projected permutation estimator,
\begin{equation}
E_p = -\sum_{n=0}^N P_n \log \Bigl( \left \langle \Pi_2^A \left( n \right) \right \rangle \Bigr), \notag
\end{equation}
where $P_n$ the probability of having $n$ particles in subregion $A$. In this ensemble, $P_n$ may be computed from
\begin{equation}
P_n = \sqrt{\frac{Z_2^A \left( n \right)}{Z_2^A}.}  \notag
\end{equation}

\subsection{Explicit form of the $\langle \Pi_2^A \rangle$ estimator for an $\mathrm{O}(\tau^4)$ propagator}
\label{subsec:shortTime}

For clarity, we present the explicit form of the $\Pi_2^A$ estimator and ensemble weights for the commonly used fourth order propagator described in Ref.~[\onlinecite{Jang2001}] for a Hamiltonian decomposed as $H = \T + \V$. To implement this approximation, we decompose the short-time propagator, $\mathrm{e}^{-\tau H}$, into two off-diagonal time steps and add an additional ancillary bead between the physical beads,
\begin{equation}
e^{-2\dt H} \simeq e^{-c_e \dt \V} e^{-\dt \T} e^{-c_o \dt \tilde{\V} } e^{-\dt \T} e^{-c_e \dt \V},
\label{eq:jjv}
\end{equation}
where $\tilde{\V}$ is a diagonal weight determined by the total potential energy $\mathcal{V} = U + V$ and a higher order correction term,
\begin{equation}
\tilde{\V} \equiv \V + c_c \dt^2 [\V,[\T,\V]], \notag
\end{equation}
and $e^{-\dt \T}$ is the free kinetic propagator for all particles. To reduce the overall number of costly numerical evaluations of $\tilde{\V}$, we have chosen $c_e = 2/3$, $c_o = 4/3$ and $c_c =1/12$.  The short time action is computed over two links involving three beads. Notice that each even bead has a factor of $e^{-c_e \dt \V}$ for each link: In the middle of a path this gives a factor of $2$ in the potential action, whereas on the end of a path, there is only one factor. For finite temperature PIMC simulations with periodic imaginary time boundary conditions, this action requires the path to have an even number of beads, $N_\tau=2p$, for integer $p$, $2\beta = N_\tau \tau$. For PIGS with open imaginary time boundary conditions, we require $N_\tau = 2p+1$ as the time slices at the two ends are not identified.

To make an off-diagonal estimator symmetric in imaginary time, we choose to have a central  double time slice of length $2\tau$ corresponding to one complete application of $e^{-2\dt H}$; consequently, this requires the path length $N_\tau =4p+1$. A general off-diagonal operator can be estimated from an ensemble of particle world lines that are broken adjacent to the center time slice. Here we label the central time slice $\R^\beta$ and the adjacent time slices $\R^{\beta-\tau}$ and $\R^{\beta+\tau}$. We decompose $\R^{\beta+\tau}$ into particles in the two subsystems: $\R^{\beta+\tau} = \{\R_A^{\beta+\tau},\R_B^{\beta+\tau}\}$. We can generate an ensemble where the world lines of $\R_A^{\beta+\tau}$ are broken but those of $\R_B^{\beta+\tau}$ are connected to $\R^{\beta}$ with a free propagator as shown in Fig.~\ref{fig:replica}. The corresponding weights for such paths are
\begin{widetext}
\begin{align}
&W\left( \R^{\beta-\tau}, \R^{\beta}, \R^{\beta+\tau} \right) = \notag \\
&\qquad \qquad W_\beta \left( \R^{\beta-\tau} \right) e^{-c_e \dt \V\left( \R^{\beta-\tau}\right)} \rho_0 \left(\R^{\beta-\tau},\R^{\beta} \right)  e^{-c_o \dt \tilde{\V}\left(\R^{\beta}\right) } \rho_0^B \left( \R^{\beta}, \R^{\beta+\tau}\right) e^{-c_e \dt \V\left( \R^{\beta+\tau}\right)} W_\beta \left( \R^{\beta+\tau} \right), \label{eq:WGSF}
\end{align}
\end{widetext}
where we have defined the matrix elements of the free propagator $\rho_0( \R,\R')$ and the reduced propagator $\rho_0^B( \R,\R_B')$ in analogy with Eqs.~\eqref{eq:rhoB} and \eqref{eq:rhoRR'}. For a particle bipartition, $n$ will correspond to the fixed subsystem size used to compute $S_\alpha(n)$; for a spatial bipartition, $n$ will fluctuate as particles move in and out of $\R^{\beta+\tau}_B$. In practice, one term in the sum $\rho_0^B$, corresponding to a particular choice of $\R^\beta_{n_B}$ will be sampled at a time as a particular choice of links between $\R^{\beta+\tau}_B$ and $\R^{\beta}$ will represent a given configuration; other link choices are then appropriately sampled via the updates described in Figs.~\ref{fig:breakswap} and \ref{fig:spatialrecon}. There is no need to explicitly sample over the permutations of $R_A^\beta$ as all such configurations come with equal weight, thus generating an additional factor of $n!$ in the weight given by \eqref{eq:WGSF}. Notice that in Eq.~\eqref{eq:WGSF} we have chosen to include the full {\it closed} path diagonal weight for $\R^{\beta+\tau}$; instead, one could choose to only include the diagonal weights for the connected world lines $\R^{\beta+\tau}_B$ and adjust the estimator below accordingly. For the doubled path configuration space, the total weight is simply the product of the weights of both paths:
\begin{align}
&W\left( \R^{\beta-\tau}, \R^{\beta}, \R^{\beta+\tau};\tilde{\R}^{\beta-\tau}, \tilde{\R}^{\beta}, \tilde{\R}^{\beta+\tau} \right) = \notag \\
&\qquad W\left( \R^{\beta-\tau}, \R^{\beta}, \R^{\beta+\tau} \right) W\left( \tilde{\R}^{\beta-\tau}, \tilde{\R}^{\beta}, \tilde{\R}^{\beta+\tau} \right). \notag
\end{align}

The estimator for $\Pi_2^A$ then takes the form
\begin{widetext}
\begin{align}
\left \langle \Pi_2^A \right \rangle =  \frac{1}{Z_2^A} \Biggl \langle \rho_0^A \left( \R^{\beta}\otimes \tilde{\R}^{\beta}; \Pi_{2}^A \left[\R^{\beta+\tau}\otimes\tilde{\R}^{\beta+\tau}\right] \right) \exp \left\{ {-c_e \dt \left[ \V\left(\Pi_{2}^A \left[\R^{\beta+\tau}\otimes\tilde{\R}^{\beta+\tau}\right] \right)\! -\! \V \left( \R^{\beta+\tau} \right)\!-\! \V \left( \tilde{\R}^{\beta+\tau} \right) \right]} \right\} \Biggr \rangle, \notag \end{align}
\end{widetext}
where 
\begin{equation}
    \V\left(\R_1 \otimes \R_2\right) \equiv \V(\R_1) +  \V(\R_2), \notag
\end{equation}
with the normalization factor
\begin{equation}
Z_2^A = \Bigl \langle  \rho_0^A \left( \R^{\beta}\otimes\R^{\beta+\tau}; \tilde{\R}^{\beta}\otimes \tilde{\R}^{\beta+\tau} \right)  \Bigr \rangle \notag 
\end{equation}
and the reduced free propagator $\rho_0^A$ is defined in analogy with Eq.~\eqref{eq:rhoA}.

\section{Harmonically interacting bosons in a harmonic potential}

Although the PIGS method for computing R\'{e}nyi entropies that we have presented above is general to all systems described by Eq.~(\ref{eq:Ham}) in any spatial dimension $D$, we have chosen to benchmark it for an interacting many-body system where the \ren entropies are analytically soluble. We consider a system of $N$ bosons of mass $m$ in one spatial dimension, interacting via a harmonic two-body potential and subject to an external harmonic potential. The Hamiltonian is given by
\begin{align}
    H &=  -\frac{\hbar^2}{2 m} \sum_i\frac{d^2}{dx_i ^2} \nonumber \\
    \qquad & +\frac{1}{2}m \omega_0^2  \sum_i x_i^2 + \frac{1}{2}m \omega_{\mathrm{int}}^2  \sum_{i<j} \left( x_i - x_j\right)^2,
\label{HcoupledSHO}
\end{align}
where $x_i$ is the spatial position of boson $i$.  In Eq.~\eqref{HcoupledSHO}, $\omega_0$ is the oscillator frequency of the external potential, and $\omega_{\mathrm{int}}$ characterizes the strength of the interaction. We define the length scale of the noninteracting oscillator to be $\sigma_0 = \sqrt{\hbar/m \omega_0}$.

For $N=2$, the ground state of Eq.~\eqref{HcoupledSHO} takes the form of two {\it decoupled} oscillators~\cite{Peschel2012,Han1999,Benavides-Riveros2014} with frequencies $\omega_{0}$ and $\omega_{1}$, where $\omega_1 = \eta \omega_0$ with $\eta \geq 1$ given by
\begin{equation}
\eta = \sqrt{1+N\frac{\omega_{\mathrm{int}}^2}{\omega_0^2}}. \notag
\end{equation}
The ground state wave function $\Psi_0 (x_0,x_1)$ is the product of two Gaussians:
\begin{align}
    \Psi_0 (x_0,x_1) = \frac{\eta^{1/4}}{\sqrt{\pi}}  \exp \Biggl\{-\frac{1}{2\sigma_0^2}\Biggl[\frac{1}{2}\left(1+\eta\right)\left(x_0^2+x_1^2\right)& \notag \\
+\left(1-\eta\right)x_0 x_1\Biggr]&\Biggr\}. \label{eq:psi0}
\end{align}

\subsection{Particle partitioning}
We begin with a discussion of the analytical solution of the single particle ($n=1$) \ren entanglement entropy for $N=2$ and demonstrated agreement with QMC calculations. We then compare QMC calculations for $n=1$ and $N\geq2$ to the analytical solution presented in Ref.~[\onlinecite{Benavides-Riveros2014}]. Finally, we extend our analysis to the measurement of the two-particle ($n=2$) entropy for the specific case of $N=3$. 

\subsubsection{Single particle entanglement for $N=2$}
\begin{figure}
\begin{center}
\scalebox{1}{\includegraphics[width=\columnwidth]{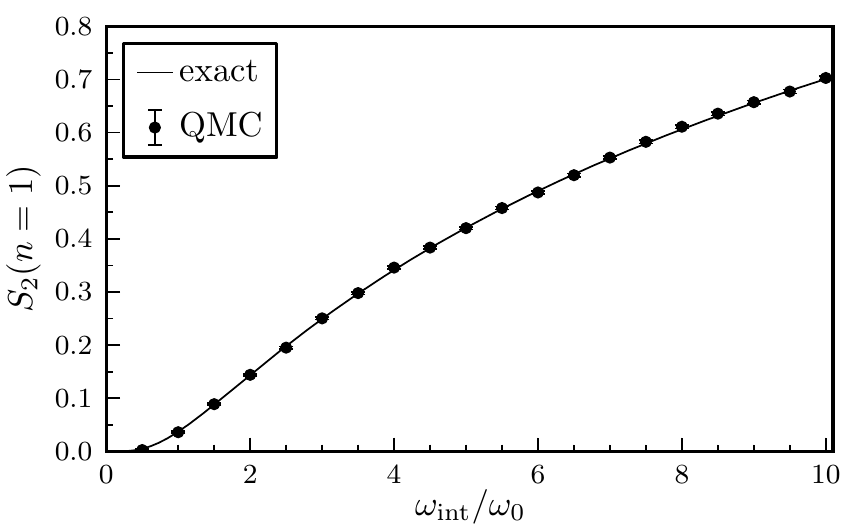}}
\end{center}
\caption{The single particle second R\'{e}nyi entropy $S_2(n=1)$ vs interaction strength $\omint/\omega_0$ computed by QMC (points) and the exact result (solid line) for two harmonically interacting bosons in a harmonic potential, described by Eq.~\eqref{HcoupledSHO}.} 
\label{fig:S1SHO}
 \end{figure}

Given the exact form of the ground state in Eq.~\eqref{eq:psi0}, the one-particle reduced density matrix $\rho_1$ is easily determined~\cite{Peschel2012,Han1999}:
\begin{align}
&\rho_1\left(x,x'\right) = \int_{-\infty}^\infty dx'' \Psi_0^* (x,x'') \Psi_0 (x',x'') \nonumber \\
&\qquad = \frac{\sqrt{2}}{\sqrt{\pi} \sigma_0} \sqrt{\frac{\eta}{1+\eta}} \exp\left\{-\frac{1}{4\sigma_0^2}\Bigl[\left(1+\eta\right)\left(x^2+x'^2\right) \right. \nonumber \\
&\left. \qquad \qquad \qquad - \frac{1}{2}\frac{\left(1-\eta\right)^2}{1+\eta} \left(x+x'\right)^2\Bigr]\right\}. \notag
\end{align}
The single particle second R\'{e}nyi entropy is then:
\begin{align}
S_2 \left(n=1\right) &=-\log \left[  \int_{-\infty}^\infty dx \int_{-\infty}^\infty dx' \rho_1^2\left(x,x'\right)\right] \nonumber \\
                     &=  \log \left[ \frac{1}{2} \left( \eta^{1/2}+  \eta^{-1/2}\right) \right].
\label{eq:S2particle}
\end{align}
Figure \ref{fig:S1SHO} shows the single particle 2nd R\'{e}nyi entropy as computed by the permutation estimator, compared to the exact result, for $N=2$ harmonically coupled bosons over a range of interaction strengths. Note that $S_2 (n=1)$ vanishes in the noninteracting limit ($\omint =0$). Table \ref{tab:S2n1} shows the numerical values of $S_2(n=1)$ as computed by QMC and the exact value, for several choices of $\omega_{\rm{int}}$; we find systematic errors smaller than $10^{-3}$.

\begin{table}
  \centering
\begin{tabular}{ l l l l l}
\hline \hline
	$\omega_{\rm{int}}/\omega_0$ 	&  \multicolumn{1}{ c  }{$1.0$}	& \multicolumn{1}{ c }{$2.0$}	&  \multicolumn{1}{ c  }{$4.0$}	& \multicolumn{1}{ c  }{$8.0$} \\ \hline 
 	exact 			& $0.0373$ 	&   $0.1438$	& $0.3415$											& $0.6062$	\\ 
	QMC  			& $0.0374(2)$	&  $0.1437(2)$	& $0.3417(2)$										& $0.6066(3)$ \\ \hline \hline
  \end{tabular}
  \caption{The single particle 2nd R\'{e}nyi entropy $S_2(n=1)$ for several values of the interaction strength $\omint/\omega_0$ computed by QMC vs. exact result for two harmonically interacting bosons in a harmonic potential, described by Eq.~\eqref{HcoupledSHO}.}
  \label{tab:S2n1}
\end{table}

\subsubsection{Single particle entanglement for general $N$}

\begin{figure}
\begin{center}
\scalebox{1}{\includegraphics[width=\columnwidth]{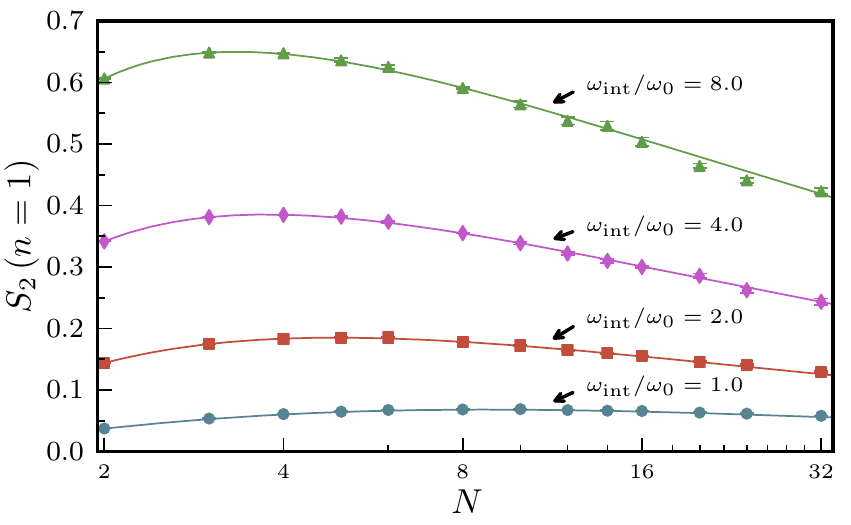}}
\end{center}
\caption{(Color online) Single particle second \ren entropies $S_2(n=1)$ for harmonically interacting bosons in a harmonic potential, described by Eq.~\eqref{HcoupledSHO}, as a function of system size $N$ as computed by QMC (points), for several values of interaction strength $\omint/\omega_0$. The solid lines correspond to the exact result given by Eq.~\eqref{eq:S2N}.} 
\label{fig:S1vsN}
\end{figure}

The single particle density matrix for general $N$ was recently computed in Ref.~[\onlinecite{Benavides-Riveros2014}] using a Wigner quasi-distribution approach. The authors find that the eigenvalues of $\rho_1$ which correspond to the occupation numbers $\{n_k\}$ of the ``natural orbitals" and are given by
\begin{equation}
n_k = \frac{2 \lambda_N}{1+\lambda_N} \bigl( \frac{1-\lambda_N}{1+\lambda_N} \bigr)^k, \notag
\end{equation}
where $\lambda_N \leq 1$ is defined as
\begin{equation}
\lambda_N \equiv \Biggl[ \left( \frac{N-1}{N} \frac{1}{\sqrt{\eta}}+\frac{\sqrt{\eta}}{N} \right) \left( \frac{1}{N} \frac{1}{\sqrt{\eta}}+\frac{N-1}{N} \sqrt{\eta} \right) \Biggr]^{-\frac{1}{2}}. \notag
\end{equation}
The second R\'{e}nyi entropy is therefore given by
\begin{align}
S_2 &\left( n=1 \right) = -\log \biggl[ \sum_k n_k^2 \biggr] = -\log \lambda_N \notag \\
& = \frac{1}{2} \log \Biggl[ \left( \frac{N-1}{N} \frac{1}{\sqrt{\eta}}+\frac{\sqrt{\eta}}{N} \right) \left( \frac{1}{N} \frac{1}{\sqrt{\eta}}+\frac{N-1}{N} \sqrt{\eta} \right) \Biggr]. \label{eq:S2N}
\end{align}
The single particle second \ren entropy is shown in \figref{fig:S1vsN} as a function of system size for $N=2-32$ for several values of the interaction strength $\omint$ as computed by QMC; the lines correspond to the exact values in Eq.~\eqref{eq:S2N}.
%

\subsubsection{Two particle entanglement for $N=3$}
\begin{figure}
\begin{center}
\scalebox{1}{\includegraphics[width=\columnwidth]{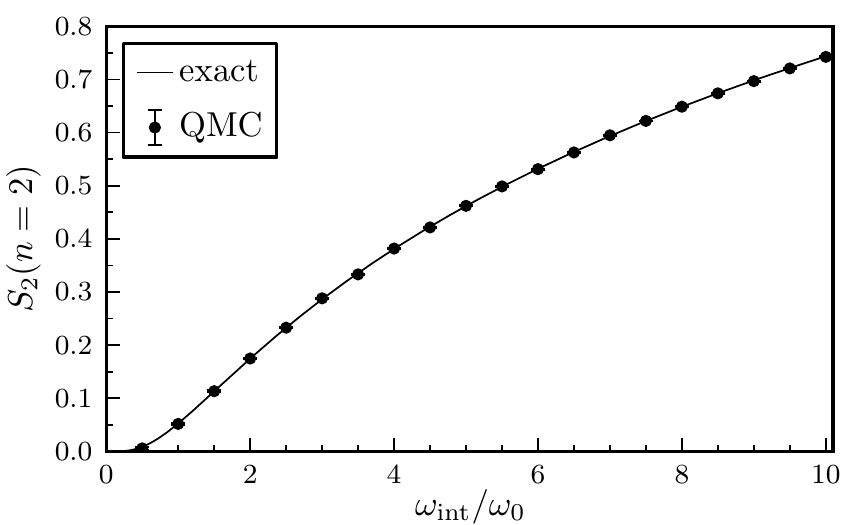}}
\end{center}
\caption{Two particle second \ren entropy $S_2(n=2)$ for an $N=3$ system of harmonically interacting bosons in a harmonic potential, described by Eq.~\eqref{HcoupledSHO}, as computed by QMC. The solid line corresponds to the exact value of $S_2(n=1)$ given by Eq.~\eqref{eq:S2N}. The demonstrated equivalence, due to the identity $S(n)=S(N-n)$, is a proof of principle of the QMC method for $n>1$.} 
\label{fig:S1vsS2}
\end{figure}

For $N=3$, we can benchmark calculations of the two particle entanglement, $S_2(n=2)$, in a simple system where it must be equal to the single particle entanglement $S_2(n=1)$, due to the identity $S(n)=S(N-n)$.  The results are shown in Fig.~\ref{fig:S1vsS2} as a function of the interaction strength $\omint/\omega_0$, displaying this agreement. The extension to $n>2$ is straightforward, albeit more computationally difficult without the aid of a generalized ratio trick discussed in Sec.~\ref{sec:algadv} below.
%

\subsection{Spatial partitioning}

Next, we consider the spatial mode entanglement and the entanglement of particles for two choices of the spatial subregion $A$ which are parametrized by the dimensionless parameter $a$:
\begin{align}
A_1:\quad& x \in \left(-a\sigma_0,a\sigma_0 \right), \notag \\
A_2:\quad& x \in\left(-\infty,a\sigma_0 \right), \notag
\end{align}
where $B_j: x \in \mathbb{R} \setminus A_j$.

\subsubsection{Spatial mode entanglement entropy}

First we present the exact solution for the spatial mode-bipartitioned R\'{e}nyi entropy of $N$ noninteracting bosons in a harmonic potential, with the Hamiltonian given by Eq.~\eqref{HcoupledSHO} with $\omega_\mathrm{int}=0$. The $N$-body ground state has all particles condensed into the single particle ground state:
\begin{equation}
\psi_0 \left(x\right) = \frac{1}{\pi^{1/4}\sqrt{ \sigma_0}} e^{-x^2/2 \sigma_0^2}. \notag
\end{equation}
We may now use a spatial Fock space basis $\{\ket{n_A,n_B}\}$ following the discussion in Sec.~\ref{sec:EoP} and write the $N$ particle space in this basis:\begin{equation}
\left \vert \Psi_N \right \rangle = \sum_{n_A=0}^N \sqrt{\binom{N}{n_A}} p_A^{n_A/2} p_B^{\left(N-n_A\right)/2} \left \vert n_A, N-n_A \right \rangle. \notag
\end{equation}
This Fock space is the Schmidt basis which diagonalizes the reduced density matrix:
\begin{equation}
\rho_A = \sum_{n_A=0}^N  \binom{N}{n_A}  p_A^{n_A} p_B^{N-n_A} \left \vert n_A, N-n_A \right \rangle \left \langle n_A, N-n_A \right \vert. \label{eq:rhoASHO}
\end{equation}

\begin{figure}
\begin{center}
\scalebox{1}{\includegraphics[width=\columnwidth]{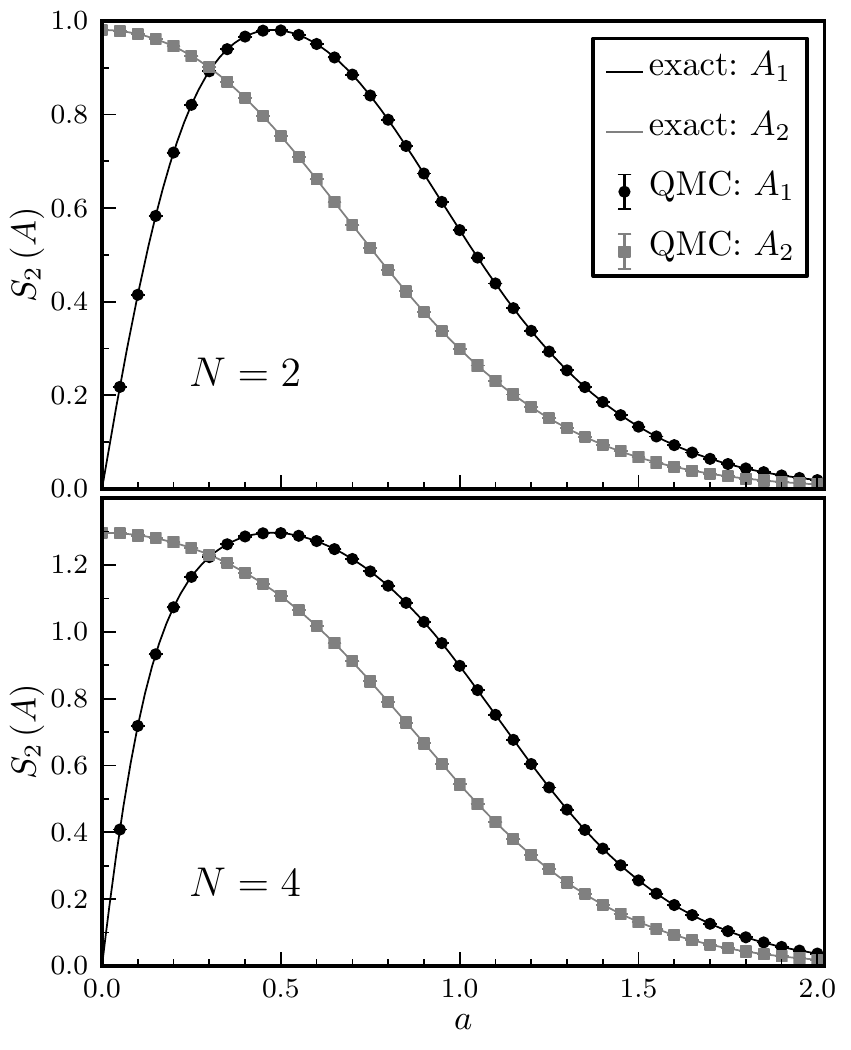}}
\end{center}
\caption{Comparison of the spatially bipartitioned second R\'{e}nyi entropy $S_2 (A)$ computed by QMC (points) and the exact result (line) for $N=2$ (top) and $N=4$ (bottom) noninteracting bosons in a harmonic potential vs bipartition size $a$ for the symmetric ($A_1$) and asymmetric ($A_2$) subregions.} 
\label{fig:shoN2N4spEE}
 \end{figure}

Given the form of the reduced density matrix $\rho_A$ in Eq.~\eqref{eq:rhoASHO}, the second R\'{e}nyi entropy may be written in terms of $p_A$ and $p_B$:
\begin{equation}
    S_2 \left( N \right) = -\log \left[ \sum_{n_A=0}^N  \binom{N}{n_A}^2  p_A^{2n_A} p_B^{2\left(N-n_A\right)} \right]. \notag
\end{equation}
The probabilities $p_A$ and $p_B$ are defined by the single particle ground state,
\begin{equation}
p_A = \int_{x \in A} dx \left \vert \psi_0 \left(x\right) \right \vert^2, \quad p_B = 1-p_A, \notag
\end{equation}
and these are readily computed for both bipartition choices,
\begin{align}
    p_{A_1} &= \int_{-a\sigma_0}^{a\sigma_0} dx \frac{1}{\sqrt{\pi} \sigma_0} e^{-x^2/ \sigma_0^2} = \mathrm{Erf}\left(a\right), \notag \\
    p_{A_2} &= \int_{-\infty}^{a\sigma_0} dx \frac{1}{\sqrt{\pi} \sigma_0} e^{-x^2/ \sigma_0^2} = \frac{1}{2} \left( 1+\mathrm{Erf}\left(a\right) \right), \notag
\end{align}
where $\mathrm{Erf}(x) = 2\int_0^x dt \mathrm{e}^{-t^2}/\sqrt{\pi}$  is the error function.
Figure~\ref{fig:shoN2N4spEE} shows the second R\'{e}nyi entropy under spatial bipartitions $A_1$ and $A_2$ as a function of bipartition size for $N=2$ and $N=4$ noninteracting bosons described by Eq.~(\ref{HcoupledSHO}) with $\omega_{int}=0$.

%
%

\subsubsection{Entanglement of particles}

As the entanglement of particles vanishes in the noninteracting limit ($\omint =0$), we must consider $\omint > 0$ such that $E_p > 0$. Given the definition of $E_p$ from Eq.~\eqref{eq:Ep}, we see that only local particle number sectors where the projected reduced density matrix $\rho_A^{(n)}$ is not a pure state will contribute to $E_p$. For $n=0$ and $n=N$, one of the subregions will be in the vacuum state,
\begin{align}
\rho_A^{(0)} = \eket{0_A} \ebra{0_A},\quad \rho_B^{(N)} = \eket{0_B}\ebra{0_B}; \notag
\end{align} 
consequently, $\rho_A^{(n)}$ is pure in each case so such sectors do not contribute to $E_p$. For $N=2$, the entanglement of particle simplifies to
\begin{align}
E_p = -P_1 \rm{log}\Bigr[ \rm{Tr}\bigl[ \left( \rho_A^{(1)} \right)^2 \bigr] \Bigr], \notag
\end{align}
where $P_1$ is given by
\begin{align}
P_1 &\equiv 2\int_{\in A} dx_0 \int_{\in B} dx_1 \left \vert \Psi_0\left(x_0,x_1\right) \right \vert^2, \label{eq:P1SHO}
\end{align}
and the one-particle spatial reduced density matrix is
\begin{align}
\rho_A^{(1)}  &\equiv \int_{\in A} dx \rho_A^{(1)} \left( x,x' \right)  \eket{x}_A  \ebra{x'}_A, \notag \\
\rho_A^{(1)} \left( x,x' \right) &\equiv \frac{2}{P_1} \int_{\in B} dx'' \Psi_0^*\left(x,x''\right)\Psi_0\left(x',x''\right) . \notag
\end{align}
We then must compute the trace:
\begin{align}
\mathrm{Tr} \Bigl[ \left( \rho_A^{(1)} \right)^2 \Bigr] = \int_{\in A} dx \int_{\in A}dx' \rho_1^A \left( x,x' \right)^2. \label{eq:trrho2SHO}
\end{align}
%
\begin{figure}
\begin{center}
\scalebox{1}{\includegraphics[width=\columnwidth]{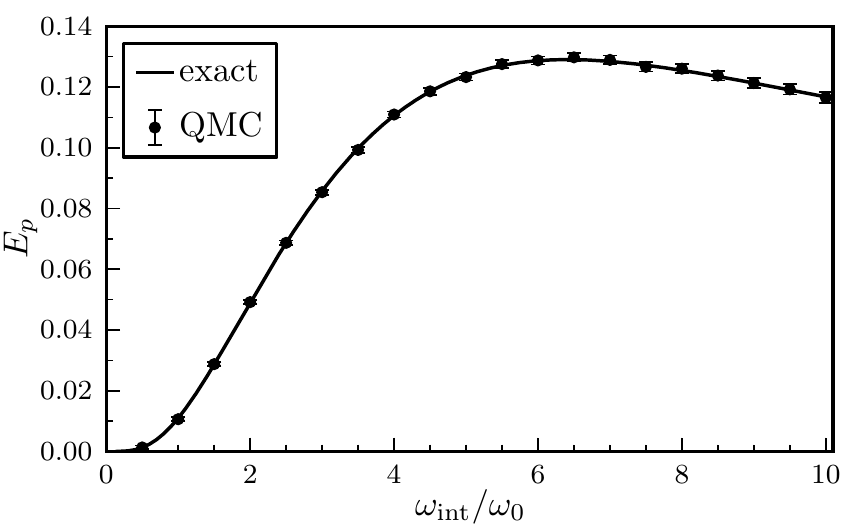}}
\end{center}
\caption{Comparison of the entanglement of particle $E_p$ vs interaction strength $\omint/\omega_0$ computed by QMC (points) and numerical integration of exact ground state (line) for two harmonically interacting bosons in a harmonic potential, with a spatial bipartition of type $A_1$ with $a=0.6$.} 
\label{fig:EpSHO}
 \end{figure}
%
Equations~\eqref{eq:P1SHO} and \eqref{eq:trrho2SHO} require two and three finite integrals over Gaussian functions. As the integration of error functions cannot be done analytically, we use numerical integration to compute $E_p$ with arbitrary precision.  \figref{fig:EpSHO} presents a comparison of $E_p$ computed by numerical integration of the exact ground state with that computed by QMC.

\subsection{Scaling of \ren entropy in PIGS simulations}

The coupled boson pair system studied above provides an excellent arena to benchmark our PIGS method for computing \ren entropies. In this section, we present the details of how the $\Pi_2$ estimator for the second \ren entropy scales with the standard PIGS parameters, the length of imaginary time paths, $\beta$, and the size of the imaginary time step, $\tau$, that control the systematic error of simulations. We focus on $S_2(n=1)$ for the one dimensional system described by Eq.~(\ref{HcoupledSHO}) for $N=2$ with fixed interaction strength $\omint=4\omega_0$.   In our PIGS simulations, we use the $\mathrm{O}(\tau^4)$ decomposition of the short-time propagator $\rho_\tau = \mathrm{e}^{-\tau H}$~\cite{Jang2001} described in Eq.~(\ref{eq:jjv}) and employ identity trial wave functions $\ket{\Psi_T} = 1$ at the terminus of all world lines. In principle, one can use a variationally optimized wave function to get convergence to the ground state with a smaller $\beta$, but in practice we found this unnecessary for this model.

\begin{figure}
\begin{center}
\scalebox{1}{\includegraphics[width=\columnwidth]{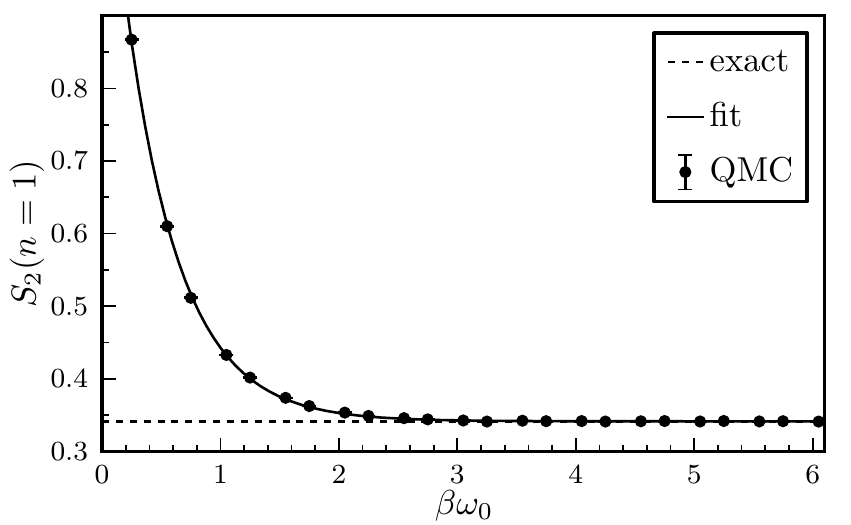}}
\end{center}
\caption{Scaling of the single particle second R\'{e}nyi entropy $S_2(n=1)$ with imaginary time length $\beta$ for the $N=2$ coupled boson system with $\omint=4\omega_0$ and $\tau = 0.05/ \omega_0$. The dashed line represents the exact ground state value and the solid line represents the best exponential to the finite imaginary time error.} 
\label{fig:betascale}
 \end{figure}

\figref{fig:betascale} shows the exponential convergence of $S_2(n=1)$ with imaginary time length $\beta$ to the exact value, $S_{\text{exact}}$ of Eq.~(\ref{eq:S2particle}).
We fit the error to an exponential,
\begin{equation}
S \left( \tau \right) = S_{\rm{exact}} + c_\beta e^{-\Delta \beta}, \notag
\end{equation}
where $c_\beta$ is a constant and find $\Delta/\omega_0 =2.183 \pm 0.005$ and $c_\beta \simeq 0.90$. This rapid exponential decay allows us to work with $\beta \omega_0 = 4.0$.

\begin{figure}
\begin{center}
\scalebox{1}{\includegraphics[width=\columnwidth]{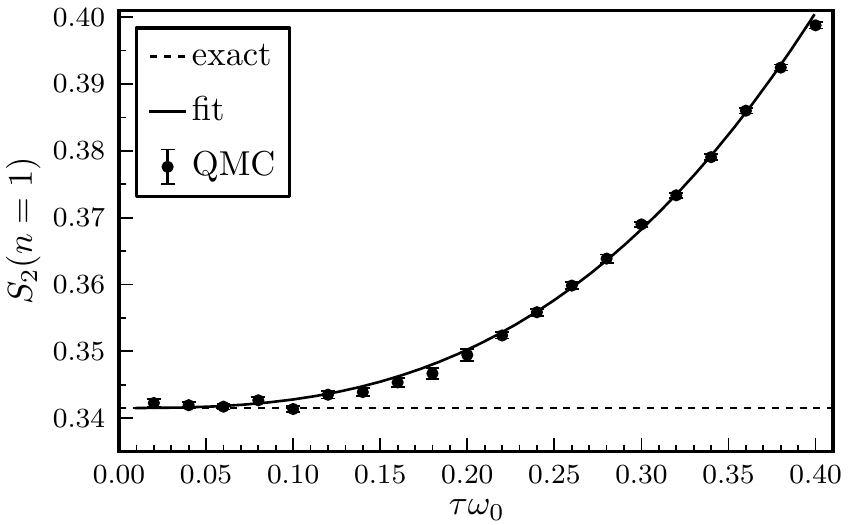}}
\end{center}
\caption{Scaling of the single particle second R\'{e}nyi entropy $S_2(n=1)$ with discrete imaginary time step $\tau$ for the $N=2$ coupled boson system $\omint=4\omega_0$ and $\beta=4.0/\omega_0$. The dashed line represents the exact ground state value and the solid line represents the best fit power-law fit to the finite time-step error.} 
\label{fig:tauscale}
\end{figure}

Fixing $\beta$, we now investigate the scaling of $S_2(n=1)$ with imaginary time step $\tau$, with the results shown in Fig.~\ref{fig:tauscale}. 
Again comparing with the exact result of Eq.~(\ref{eq:S2particle}) we fit the finite time-step error to a power law,
\begin{equation}
S \left( \delta \right) = S_{\rm{exact}} + c_\tau \tau^\delta \notag
\end{equation}
where $c_\tau$ is a constant and find $\delta =  2.70\pm0.03$ and $c_\tau=0.743\pm0.001$. This analysis demonstrates that these systematic errors can be chosen to be smaller than any statistical error, while retaining the fundamental power-law scaling of computational resources required for the QMC algorithm. From this analysis of the scaling with PIGS parameters, we choose $\tau = 0.2/\omint$ and $\beta=4.0/\omega_0$, which reduces the systematic errors in our PIGS calculations of $S_2$ to less than $10^{-3}$.

\section{Future algorithmic development} 
\label{sec:algadv}

We have presented a fundamental algorithm for computing \ren entropies using the PIGS QMC method. Given the polynomial resources required for such stochastic computations, this approach offers the potential of studying entanglement entropies in large-scale interacting systems of bosons in the spatial continuum. The  biggest impediment to immediate application of the method for $N\gg1$ is the fact that the expectation value of the permutation estimator decays exponentially with the bipartition size. This exponential decay of the bare $\Pi_2$ estimator is due to the Gaussian free propagator associated with each particle in the bipartition. This is indicative of the linear scaling of the particle entanglement with bipartition size and the ``area law'' scaling of the spatial entanglement entropy. While this might first appear to be a fundamental limitation of Monte Carlo methods to compute R\'{e}nyi entropies, there has been much algorithmic development in lattice formulations to address this issue. The solutions that have already been successfully implemented in lattice Monte Carlo methods use some variant of a ``ratio sampling'' method~\cite{Hastings2010,Inglis2013b,Kallin2011}, reducing an exponentially decaying expectation value to a product of finite values. 

The analogous approach for computing particle entanglement via our PIGS method is as follows. First, observe that the ratios
\begin{equation}
    \mathcal{R}^{dn}_\alpha(n) \equiv \frac{\Pi_\alpha^{n+dn}}{\Pi_\alpha^{n}} \notag
\end{equation}
can be directly computed by sampling a configuration space where $n$ world lines are permuted between the $\alpha$ replicas (i.e., $n$ world lines from each replica are connected to another replica at the center time slice by a ``link'') and $dn$ world lines are broken. The estimator for $\mathcal{R}^{dn}_\alpha(n)$ just involves permuting the remaining $dn$ broken world lines. Following Ref.~[\onlinecite{Hastings2010}], we then note that the expectation value of the overall permutation operator $\Pi^n_\alpha$ is related to a product of $\mathcal{R}(n_i)$:
\begin{equation}
    \left \langle \Pi_\alpha^n \right \rangle = \left \langle \mathcal{R}^{dn}_{\alpha}(0) \right \rangle \dots \left \langle \mathcal{R}^{dn}_\alpha(n-dn) \right \rangle. \notag
\end{equation}
Consequently, a single calculation of $\Pi^n_\alpha$, which would be an exponentially small quantity, can be replaced by $n/dn$ calculations, each of which has a sufficiently large value, where $dn \geq 1$ can be chosen for maximum efficiency. This adds an additional linear scaling in $n$, and if one is interested in studying extensive values of $n$ (e.g., $1\leq n \leq N/2$), this adds a factor of $N^2$ to the overall scaling, where the second factor of $N$ is required to keep the statistical error fixed when multiplying many ratios together. 

For a spatial bipartition, an analogous approach may be used where the ratios are taken between different spatial regions and the number of closed permuted world lines will fluctuate.  While such calculations are more computationally expensive than computing traditional observables, the polynomial scaling, as well as the demonstrated success of related methods for lattice systems, suggests that this approach could be fruitfully applied to a range of models in the $D$-dimensional spatial continuum.

A major advantage of using a method based on PIMC is that it allows us to import many PIMC techniques for efficiently sampling the configuration space required to compute the $\Pi_{\alpha}$ estimator. In particular, the worm algorithm~\cite{Boninsegni:2006ed,*Boninsegni:2006gc} is a powerful method for sampling the configuration space of bosonic world lines, involving both closed and open imaginary time paths. In the worm algorithm, broken world lines, or {\it worms}, are allowed to wander in space and imaginary time. Such worms could be used to efficiently update the broken world line configuration space by allowing the worm head and tail to attach and detach from the center time slice during an update. These moves would be ergodic on their own and alleviate the necessity for the updates described in Sec.~\ref{sec:updates}.  In the language of the worm algorithm, closed path configurations belong to the so-called $Z$ sector while configurations with open world lines belong to the so-called $G$ sector (they contribute to the one-particle Matsubara Green function and are therefore labeled $G$ as they relate to an off-diagonal density matrix).  At present, we perform $Z$ sector simulations to compute properties such as the energy in order to optimize the projection time $\beta$ and time step $\tau$ parameters. These parameters are then used in open path ($G$-sector) simulations for the computation of the $\Pi_{\alpha}$ estimator.  The use of a worm type algorithm would allow us to compute all properties in a {\it single} simulation.  Note that since both closed and open word lines are sampled in a worm algorithm simulation, the ensemble partition function, $Z_W$, corresponds to a generalization of the form $Z_W =Z+Z'$, where $Z$ is the regular closed path partition function and $Z'$ is the $G$-sector partition function \cite{Boninsegni:2006ed,*Boninsegni:2006gc}.  For the $G$-sector configurations, the number of continuous variables is not constant and therefore suggests the use of diagrammatic Monte Carlo techniques \cite{PhysRevLett.81.2514} for updates, as the number of variables is also a degree of freedom.  The possibility of using diagrammatic techniques for the design of more efficient simulations to compute the $\Pi_\alpha$ estimator along with other observables is a very promising area for future investigation.

\section{Discussion} 
In this paper, we have presented a general, scalable Monte Carlo simulation method for computing R\'{e}nyi entanglement entropies in continuum systems of itinerant bosons based on the {\it replica trick}.  We have implemented the algorithm in ground state PIGS~\cite{Sarsa2000,Cuervo2005a} and benchmarked its accuracy in a simple system of $N$ harmonically trapped and interacting bosons in one spatial dimension.  Detailed convergence tests of the algorithm demonstrate the fundamental power-law scaling of computational resources required for simulations in both total system size $N$ and the size of the particle bipartition $n$.  This work opens the door to several immediate extensions of the replica trick R\'enyi entropy algorithm to other nontrivial models of interacting bosons, both in one and higher dimensions. A straightforward adaptation of the algorithm to finite temperature path-integral methods based on the partition function \cite{Ceperley1995,Boninsegni:2006ed,*Boninsegni:2006gc} will also provide access to R\'enyi entropies at $T > 0$ and associated quantities, like the mutual information \cite{Wolf:2008ht,Singh:2011jx}.

The immanent adoption of replica trick methods to PIMC simulations based on the presented algorithm is poised to make significant headway in a variety of problems of physical interest in the continuum.  This could have a significant impact on our understanding of such interacting many-body systems in analogy to what has been learned for lattice models since 2010 \cite{Hastings2010,Melko2010}.  One of the primary advantages of this technique is that after being implemented in a QMC code base, it can be easily applied to any quantum many-body system described by the Hamiltonian of Eq.~(\ref{eq:Ham}), regardless of the spatial dimension or the form of the external and interaction potentials $U$ and $V$, with extremely minimal programmatic modifications.  This generality opens up the ability to quantitatively measure the entanglement properties of experimentally relevant systems of identical bosons, including quantum fluids of helium-4 and ultracold atomic gases.  For the latter, there is currently a coordinated experimental and theoretical effort under way to create and manipulate entangled multiparticle states \cite{Sorensen2001,Esteve:2008ij,Riedel:2010gea,Benatti:2011hf, He:2011jf, Hyllus:2012dl,Lucke:2014hz,Dalton:2014dv} for quantum metrology and information processing purposes.  Many other applications and extensions of this work become immediately apparent, including the study of phase transitions in itinerant boson systems; the correlation between superfluidity, condensate fraction, and entanglement in superfluid droplets \cite{Li2010,raston_coh2_superfluid,zeng2014microscopic}; and much more.

Finally, as has been previously demonstrated with similar techniques, we expect that the ability to measure R\'enyi entropies in large-scale computer simulations of interacting quantum systems in the continuum will synergistically feed back into related areas of quantum information science and beyond.  For example, PIGS simulations of interacting bosonic systems of relevance for condensed-matter physics will allow for the evaluation of the appropriateness of {\it tensor network} ans\"atze for the spatial continuum, which require significant restrictions in the scaling of entanglement entropy to be valid \cite{Verstraete2010}.  Also, PIGS measurements of entanglement in quantum phases like Bose-Einstein condensates may become essential when evaluating the resource capabilities of such states for quantum information processing. This may prove particularly important, as recent work indicates that identical-particle entanglement may be useful as a resource for standard quantum information tasks \cite{Killoran:2014gu}.

\section{Acknowledgments}
We acknowledge the use of the computing facilities of Compute Canada (RQCHP's Mammouth cluster) and the Vermont Advanced Computing Core supported by NASA (Grant No. NNX-08AO96G).  This research was supported by NSERC of Canada, the Canada Research Chair Program, the Perimeter Institute for Theoretical Physics, the John Templeton Foundation, and the University of Vermont.  Research at PI is supported by the Government of Canada through Industry Canada and by the Province of Ontario through the Ministry of Economic Development \& Innovation.

\bibliographystyle{apsrev4-1}
\bibliography{refs}

\begin{thebibliography}{109}%
\makeatletter
\providecommand \@ifxundefined [1]{%
 \@ifx{#1\undefined}
}%
\providecommand \@ifnum [1]{%
 \ifnum #1\expandafter \@firstoftwo
 \else \expandafter \@secondoftwo
 \fi
}%
\providecommand \@ifx [1]{%
 \ifx #1\expandafter \@firstoftwo
 \else \expandafter \@secondoftwo
 \fi
}%
\providecommand \natexlab [1]{#1}%
\providecommand \enquote  [1]{``#1''}%
\providecommand \bibnamefont  [1]{#1}%
\providecommand \bibfnamefont [1]{#1}%
\providecommand \citenamefont [1]{#1}%
\providecommand \href@noop [0]{\@secondoftwo}%
\providecommand \href [0]{\begingroup \@sanitize@url \@href}%
\providecommand \@href[1]{\@@startlink{#1}\@@href}%
\providecommand \@@href[1]{\endgroup#1\@@endlink}%
\providecommand \@sanitize@url [0]{\catcode `\\12\catcode `\$12\catcode
  `\&12\catcode `\#12\catcode `\^12\catcode `\_12\catcode `\%12\relax}%
\providecommand \@@startlink[1]{}%
\providecommand \@@endlink[0]{}%
\providecommand \url  [0]{\begingroup\@sanitize@url \@url }%
\providecommand \@url [1]{\endgroup\@href {#1}{\urlprefix }}%
\providecommand \urlprefix  [0]{URL }%
\providecommand \Eprint [0]{\href }%
\providecommand \doibase [0]{http://dx.doi.org/}%
\providecommand \selectlanguage [0]{\@gobble}%
\providecommand \bibinfo  [0]{\@secondoftwo}%
\providecommand \bibfield  [0]{\@secondoftwo}%
\providecommand \translation [1]{[#1]}%
\providecommand \BibitemOpen [0]{}%
\providecommand \bibitemStop [0]{}%
\providecommand \bibitemNoStop [0]{.\EOS\space}%
\providecommand \EOS [0]{\spacefactor3000\relax}%
\providecommand \BibitemShut  [1]{\csname bibitem#1\endcsname}%
\let\auto@bib@innerbib\@empty
\bibitem [{\citenamefont {Vedral}(2013)}]{Vedral:2013bk}%
  \BibitemOpen
  \bibfield  {author} {\bibinfo {author} {\bibfnamefont {V.}~\bibnamefont
  {Vedral}},\ }\href {http://books.google.com/books?id=HuABkgEACAAJ} {\emph
  {\bibinfo {title} {Introduction to Quantum Information Science}}},\ Oxford
  Graduate Texts\ (\bibinfo  {publisher} {OUP Oxford},\ \bibinfo {year}
  {2013})\BibitemShut {NoStop}%
\bibitem [{\citenamefont {Jozsa}\ and\ \citenamefont
  {Linden}(2003)}]{Jozsa2003}%
  \BibitemOpen
  \bibfield  {author} {\bibinfo {author} {\bibfnamefont {R.}~\bibnamefont
  {Jozsa}}\ and\ \bibinfo {author} {\bibfnamefont {N.}~\bibnamefont {Linden}},\
  }\href {\doibase 10.1098/rspa.2002.1097} {\bibfield  {journal} {\bibinfo
  {journal} {Proc. Royal Soc. A-Math Phys.}\ }\textbf {\bibinfo {volume}
  {459}},\ \bibinfo {pages} {2011} (\bibinfo {year} {2003})}\BibitemShut
  {NoStop}%
\bibitem [{\citenamefont {Ekert}(1991)}]{Ekert:1991xy}%
  \BibitemOpen
  \bibfield  {author} {\bibinfo {author} {\bibfnamefont {A.~K.}\ \bibnamefont
  {Ekert}},\ }\href {\doibase 10.1103/PhysRevLett.67.661} {\bibfield  {journal}
  {\bibinfo  {journal} {Phys. Rev. Lett.}\ }\textbf {\bibinfo {volume} {67}},\
  \bibinfo {pages} {661} (\bibinfo {year} {1991})}\BibitemShut {NoStop}%
\bibitem [{\citenamefont {Bennett}\ \emph {et~al.}(1993)\citenamefont
  {Bennett}, \citenamefont {Brassard}, \citenamefont {Cr\'epeau}, \citenamefont
  {Jozsa}, \citenamefont {Peres},\ and\ \citenamefont
  {Wootters}}]{Bennett:1993ta}%
  \BibitemOpen
  \bibfield  {author} {\bibinfo {author} {\bibfnamefont {C.~H.}\ \bibnamefont
  {Bennett}}, \bibinfo {author} {\bibfnamefont {G.}~\bibnamefont {Brassard}},
  \bibinfo {author} {\bibfnamefont {C.}~\bibnamefont {Cr\'epeau}}, \bibinfo
  {author} {\bibfnamefont {R.}~\bibnamefont {Jozsa}}, \bibinfo {author}
  {\bibfnamefont {A.}~\bibnamefont {Peres}}, \ and\ \bibinfo {author}
  {\bibfnamefont {W.~K.}\ \bibnamefont {Wootters}},\ }\href {\doibase
  10.1103/PhysRevLett.70.1895} {\bibfield  {journal} {\bibinfo  {journal}
  {Phys. Rev. Lett.}\ }\textbf {\bibinfo {volume} {70}},\ \bibinfo {pages}
  {1895} (\bibinfo {year} {1993})}\BibitemShut {NoStop}%
\bibitem [{\citenamefont {Horodecki}\ \emph {et~al.}(2009)\citenamefont
  {Horodecki}, \citenamefont {Horodecki},\ and\ \citenamefont
  {Horodecki}}]{Horodecki:2009gb}%
  \BibitemOpen
  \bibfield  {author} {\bibinfo {author} {\bibfnamefont {R.}~\bibnamefont
  {Horodecki}}, \bibinfo {author} {\bibfnamefont {M.}~\bibnamefont
  {Horodecki}}, \ and\ \bibinfo {author} {\bibfnamefont {K.}~\bibnamefont
  {Horodecki}},\ }\href@noop {} {\bibfield  {journal} {\bibinfo  {journal}
  {Rev. Mod. Phys.}\ }\textbf {\bibinfo {volume} {81}},\ \bibinfo {pages} {865}
  (\bibinfo {year} {2009})}\BibitemShut {NoStop}%
\bibitem [{\citenamefont {Amico}\ \emph {et~al.}(2008)\citenamefont {Amico},
  \citenamefont {Fazio}, \citenamefont {Osterloh},\ and\ \citenamefont
  {Vedral}}]{Amico2008}%
  \BibitemOpen
  \bibfield  {author} {\bibinfo {author} {\bibfnamefont {L.}~\bibnamefont
  {Amico}}, \bibinfo {author} {\bibfnamefont {R.}~\bibnamefont {Fazio}},
  \bibinfo {author} {\bibfnamefont {A.}~\bibnamefont {Osterloh}}, \ and\
  \bibinfo {author} {\bibfnamefont {V.}~\bibnamefont {Vedral}},\ }\href
  {\doibase 10.1103/RevModPhys.80.517} {\bibfield  {journal} {\bibinfo
  {journal} {Rev. Mod. Phys.}\ }\textbf {\bibinfo {volume} {80}},\ \bibinfo
  {pages} {517} (\bibinfo {year} {2008})}\BibitemShut {NoStop}%
\bibitem [{\citenamefont {Kitaev}\ and\ \citenamefont
  {Preskill}(2006)}]{Kitaev:2006dn}%
  \BibitemOpen
  \bibfield  {author} {\bibinfo {author} {\bibfnamefont {A.}~\bibnamefont
  {Kitaev}}\ and\ \bibinfo {author} {\bibfnamefont {J.}~\bibnamefont
  {Preskill}},\ }\href@noop {} {\bibfield  {journal} {\bibinfo  {journal}
  {Phys. Rev. Lett.}\ }\textbf {\bibinfo {volume} {96}},\ \bibinfo {pages}
  {110404} (\bibinfo {year} {2006})}\BibitemShut {NoStop}%
\bibitem [{\citenamefont {Levin}\ and\ \citenamefont
  {Wen}(2006)}]{Levin:2006ij}%
  \BibitemOpen
  \bibfield  {author} {\bibinfo {author} {\bibfnamefont {M.}~\bibnamefont
  {Levin}}\ and\ \bibinfo {author} {\bibfnamefont {X.-G.}\ \bibnamefont
  {Wen}},\ }\href@noop {} {\bibfield  {journal} {\bibinfo  {journal} {Phys.
  Rev. Lett.}\ }\textbf {\bibinfo {volume} {96}},\ \bibinfo {pages} {110405}
  (\bibinfo {year} {2006})}\BibitemShut {NoStop}%
\bibitem [{\citenamefont {Wolf}\ \emph {et~al.}(2008)\citenamefont {Wolf},
  \citenamefont {Verstraete}, \citenamefont {Hastings},\ and\ \citenamefont
  {Cirac}}]{Wolf:2008ht}%
  \BibitemOpen
  \bibfield  {author} {\bibinfo {author} {\bibfnamefont {M.~M.}\ \bibnamefont
  {Wolf}}, \bibinfo {author} {\bibfnamefont {F.}~\bibnamefont {Verstraete}},
  \bibinfo {author} {\bibfnamefont {M.~B.}\ \bibnamefont {Hastings}}, \ and\
  \bibinfo {author} {\bibfnamefont {J.~I.}\ \bibnamefont {Cirac}},\ }\href@noop
  {} {\bibfield  {journal} {\bibinfo  {journal} {Phys. Rev. Lett.}\ }\textbf
  {\bibinfo {volume} {100}},\ \bibinfo {pages} {070502} (\bibinfo {year}
  {2008})}\BibitemShut {NoStop}%
\bibitem [{\citenamefont {White}(1992)}]{White92}%
  \BibitemOpen
  \bibfield  {author} {\bibinfo {author} {\bibfnamefont {S.~R.}\ \bibnamefont
  {White}},\ }\href {\doibase 10.1103/PhysRevLett.69.2863} {\bibfield
  {journal} {\bibinfo  {journal} {Phys. Rev. Lett.}\ }\textbf {\bibinfo
  {volume} {69}},\ \bibinfo {pages} {2863} (\bibinfo {year}
  {1992})}\BibitemShut {NoStop}%
\bibitem [{\citenamefont {Schollw\"ock}(2005)}]{Scholl05}%
  \BibitemOpen
  \bibfield  {author} {\bibinfo {author} {\bibfnamefont {U.}~\bibnamefont
  {Schollw\"ock}},\ }\href {\doibase 10.1103/RevModPhys.77.259} {\bibfield
  {journal} {\bibinfo  {journal} {Rev. Mod. Phys.}\ }\textbf {\bibinfo {volume}
  {77}},\ \bibinfo {pages} {259} (\bibinfo {year} {2005})}\BibitemShut
  {NoStop}%
\bibitem [{\citenamefont {Hastings}\ \emph {et~al.}(2010)\citenamefont
  {Hastings}, \citenamefont {Gonz\'{a}lez}, \citenamefont {Kallin},\ and\
  \citenamefont {Melko}}]{Hastings2010}%
  \BibitemOpen
  \bibfield  {author} {\bibinfo {author} {\bibfnamefont {M.~B.}\ \bibnamefont
  {Hastings}}, \bibinfo {author} {\bibfnamefont {I.}~\bibnamefont
  {Gonz\'{a}lez}}, \bibinfo {author} {\bibfnamefont {A.~B.}\ \bibnamefont
  {Kallin}}, \ and\ \bibinfo {author} {\bibfnamefont {R.~G.}\ \bibnamefont
  {Melko}},\ }\href {\doibase 10.1103/PhysRevLett.104.157201} {\bibfield
  {journal} {\bibinfo  {journal} {Phys. Rev. Lett.}\ }\textbf {\bibinfo
  {volume} {104}},\ \bibinfo {pages} {157201} (\bibinfo {year}
  {2010})}\BibitemShut {NoStop}%
\bibitem [{\citenamefont {Melko}\ \emph {et~al.}(2010)\citenamefont {Melko},
  \citenamefont {Kallin},\ and\ \citenamefont {Hastings}}]{Melko2010}%
  \BibitemOpen
  \bibfield  {author} {\bibinfo {author} {\bibfnamefont {R.~G.}\ \bibnamefont
  {Melko}}, \bibinfo {author} {\bibfnamefont {A.~B.}\ \bibnamefont {Kallin}}, \
  and\ \bibinfo {author} {\bibfnamefont {M.~B.}\ \bibnamefont {Hastings}},\
  }\href {\doibase 10.1103/PhysRevB.82.100409} {\bibfield  {journal} {\bibinfo
  {journal} {Phys. Rev. B}\ }\textbf {\bibinfo {volume} {82}},\ \bibinfo
  {pages} {100409} (\bibinfo {year} {2010})}\BibitemShut {NoStop}%
\bibitem [{\citenamefont {Tubman}\ and\ \citenamefont
  {McMinis}(2012)}]{Tubman2012}%
  \BibitemOpen
  \bibfield  {author} {\bibinfo {author} {\bibfnamefont {N.~M.}\ \bibnamefont
  {Tubman}}\ and\ \bibinfo {author} {\bibfnamefont {J.}~\bibnamefont
  {McMinis}},\ }\href {http://arxiv.org/abs/1204.4731v2} {\  (\bibinfo {year}
  {2012})},\ \Eprint {http://arxiv.org/abs/1204.4731} {arXiv:1204.4731
  [cond-mat.str-el]} \BibitemShut {NoStop}%
\bibitem [{\citenamefont {Zhang}\ \emph {et~al.}(2012)\citenamefont {Zhang},
  \citenamefont {Grover}, \citenamefont {Turner}, \citenamefont {Oshikawa},\
  and\ \citenamefont {Vishwanath}}]{Zhang2012}%
  \BibitemOpen
  \bibfield  {author} {\bibinfo {author} {\bibfnamefont {Y.}~\bibnamefont
  {Zhang}}, \bibinfo {author} {\bibfnamefont {T.}~\bibnamefont {Grover}},
  \bibinfo {author} {\bibfnamefont {A.}~\bibnamefont {Turner}}, \bibinfo
  {author} {\bibfnamefont {M.}~\bibnamefont {Oshikawa}}, \ and\ \bibinfo
  {author} {\bibfnamefont {A.}~\bibnamefont {Vishwanath}},\ }\href {\doibase
  10.1103/PhysRevB.85.235151} {\bibfield  {journal} {\bibinfo  {journal} {Phys.
  Rev. B}\ }\textbf {\bibinfo {volume} {85}},\ \bibinfo {pages} {235151}
  (\bibinfo {year} {2012})}\BibitemShut {NoStop}%
\bibitem [{\citenamefont {Inglis}\ and\ \citenamefont
  {Melko}(2013{\natexlab{a}})}]{Inglis2013b}%
  \BibitemOpen
  \bibfield  {author} {\bibinfo {author} {\bibfnamefont {S.}~\bibnamefont
  {Inglis}}\ and\ \bibinfo {author} {\bibfnamefont {R.~G.}\ \bibnamefont
  {Melko}},\ }\href {\doibase 10.1088/1367-2630/15/7/073048} {\bibfield
  {journal} {\bibinfo  {journal} {New J. Phys.}\ }\textbf {\bibinfo {volume}
  {15}},\ \bibinfo {pages} {073048} (\bibinfo {year}
  {2013}{\natexlab{a}})}\BibitemShut {NoStop}%
\bibitem [{\citenamefont {McMinis}\ and\ \citenamefont
  {Tubman}(2013)}]{McMinis:2013dp}%
  \BibitemOpen
  \bibfield  {author} {\bibinfo {author} {\bibfnamefont {J.}~\bibnamefont
  {McMinis}}\ and\ \bibinfo {author} {\bibfnamefont {N.~M.}\ \bibnamefont
  {Tubman}},\ }\href@noop {} {\bibfield  {journal} {\bibinfo  {journal} {Phys.
  Rev. B}\ }\textbf {\bibinfo {volume} {87}},\ \bibinfo {pages} {081108}
  (\bibinfo {year} {2013})}\BibitemShut {NoStop}%
\bibitem [{\citenamefont {Selem}\ \emph {et~al.}(2013)\citenamefont {Selem},
  \citenamefont {Herdman},\ and\ \citenamefont {Whaley}}]{Selem2013}%
  \BibitemOpen
  \bibfield  {author} {\bibinfo {author} {\bibfnamefont {A.}~\bibnamefont
  {Selem}}, \bibinfo {author} {\bibfnamefont {C.~M.}\ \bibnamefont {Herdman}},
  \ and\ \bibinfo {author} {\bibfnamefont {K.~B.}\ \bibnamefont {Whaley}},\
  }\href {\doibase 10.1103/PhysRevB.87.125105} {\bibfield  {journal} {\bibinfo
  {journal} {Phys. Rev. B}\ }\textbf {\bibinfo {volume} {87}},\ \bibinfo
  {pages} {125105} (\bibinfo {year} {2013})}\BibitemShut {NoStop}%
\bibitem [{\citenamefont {Pei}\ \emph {et~al.}(2014)\citenamefont {Pei},
  \citenamefont {Han}, \citenamefont {Liao},\ and\ \citenamefont
  {Li}}]{Pei2014}%
  \BibitemOpen
  \bibfield  {author} {\bibinfo {author} {\bibfnamefont {J.}~\bibnamefont
  {Pei}}, \bibinfo {author} {\bibfnamefont {S.}~\bibnamefont {Han}}, \bibinfo
  {author} {\bibfnamefont {H.}~\bibnamefont {Liao}}, \ and\ \bibinfo {author}
  {\bibfnamefont {T.}~\bibnamefont {Li}},\ }\href {\doibase
  10.1088/0953-8984/26/3/035601} {\bibfield  {journal} {\bibinfo  {journal} {J.
  Phys.: Condens. Matter}\ }\textbf {\bibinfo {volume} {26}},\ \bibinfo {pages}
  {035601} (\bibinfo {year} {2014})}\BibitemShut {NoStop}%
\bibitem [{\citenamefont {Herdman}\ \emph {et~al.}(2014)\citenamefont
  {Herdman}, \citenamefont {Roy}, \citenamefont {Melko},\ and\ \citenamefont
  {{Del Maestro}}}]{algorithm}%
  \BibitemOpen
  \bibfield  {author} {\bibinfo {author} {\bibfnamefont {C.~M.}\ \bibnamefont
  {Herdman}}, \bibinfo {author} {\bibfnamefont {P.-N.}\ \bibnamefont {Roy}},
  \bibinfo {author} {\bibfnamefont {R.~G.}\ \bibnamefont {Melko}}, \ and\
  \bibinfo {author} {\bibfnamefont {A.}~\bibnamefont {{Del Maestro}}},\ }\href
  {\doibase 10.1103/PhysRevB.89.140501} {\bibfield  {journal} {\bibinfo
  {journal} {Phys. Rev. B}\ }\textbf {\bibinfo {volume} {89}},\ \bibinfo
  {pages} {140501} (\bibinfo {year} {2014})}\BibitemShut {NoStop}%
\bibitem [{\citenamefont {{Tubman}}\ and\ \citenamefont {{ChangMo
  Yang}}(2014)}]{Tubman2014}%
  \BibitemOpen
  \bibfield  {author} {\bibinfo {author} {\bibfnamefont {N.~M.}\ \bibnamefont
  {{Tubman}}}\ and\ \bibinfo {author} {\bibfnamefont {D.}~\bibnamefont
  {{ChangMo Yang}}},\ }\href@noop {} {\  (\bibinfo {year} {2014})},\ \Eprint
  {http://arxiv.org/abs/1402.0503} {arXiv:1402.0503 [cond-mat.str-el]}
  \BibitemShut {NoStop}%
\bibitem [{\citenamefont {Humeniuk}\ and\ \citenamefont
  {Roscilde}(2012)}]{Tommaso}%
  \BibitemOpen
  \bibfield  {author} {\bibinfo {author} {\bibfnamefont {S.}~\bibnamefont
  {Humeniuk}}\ and\ \bibinfo {author} {\bibfnamefont {T.}~\bibnamefont
  {Roscilde}},\ }\href@noop {} {\bibfield  {journal} {\bibinfo  {journal}
  {Phys. Rev. B}\ }\textbf {\bibinfo {volume} {86}},\ \bibinfo {pages} {235116}
  (\bibinfo {year} {2012})}\BibitemShut {NoStop}%
\bibitem [{\citenamefont {Inglis}\ and\ \citenamefont
  {Melko}(2013{\natexlab{b}})}]{Inglis:2013iv}%
  \BibitemOpen
  \bibfield  {author} {\bibinfo {author} {\bibfnamefont {S.}~\bibnamefont
  {Inglis}}\ and\ \bibinfo {author} {\bibfnamefont {R.~G.}\ \bibnamefont
  {Melko}},\ }\href@noop {} {\bibfield  {journal} {\bibinfo  {journal} {Phys.
  Rev. E}\ }\textbf {\bibinfo {volume} {87}},\ \bibinfo {pages} {013306}
  (\bibinfo {year} {2013}{\natexlab{b}})}\BibitemShut {NoStop}%
\bibitem [{\citenamefont {{Broecker}}\ and\ \citenamefont
  {{Trebst}}(2014)}]{Broecker}%
  \BibitemOpen
  \bibfield  {author} {\bibinfo {author} {\bibfnamefont {P.}~\bibnamefont
  {{Broecker}}}\ and\ \bibinfo {author} {\bibfnamefont {S.}~\bibnamefont
  {{Trebst}}},\ }\href@noop {} {\  (\bibinfo {year} {2014})},\ \Eprint
  {http://arxiv.org/abs/1404.3027} {arXiv:1404.3027 [cond-mat.str-el]}
  \BibitemShut {NoStop}%
\bibitem [{\citenamefont {Chung}\ \emph
  {et~al.}(2013{\natexlab{a}})\citenamefont {Chung}, \citenamefont {Bonnes},
  \citenamefont {Chen},\ and\ \citenamefont {L\"{a}uchli}}]{Chung2013}%
  \BibitemOpen
  \bibfield  {author} {\bibinfo {author} {\bibfnamefont {C.-M.}\ \bibnamefont
  {Chung}}, \bibinfo {author} {\bibfnamefont {L.}~\bibnamefont {Bonnes}},
  \bibinfo {author} {\bibfnamefont {P.}~\bibnamefont {Chen}}, \ and\ \bibinfo
  {author} {\bibfnamefont {A.~M.}\ \bibnamefont {L\"{a}uchli}},\ }\href
  {http://arxiv.org/abs/1305.6536} {\  (\bibinfo {year}
  {2013}{\natexlab{a}})},\ \Eprint {http://arxiv.org/abs/1305.6536}
  {arXiv:1305.6536 [cond-mat.quant-gas]} \BibitemShut {NoStop}%
\bibitem [{\citenamefont {Chung}\ \emph
  {et~al.}(2013{\natexlab{b}})\citenamefont {Chung}, \citenamefont {Alba},
  \citenamefont {Bonnes}, \citenamefont {Chen},\ and\ \citenamefont
  {L\"{a}uchli}}]{Chung2013a}%
  \BibitemOpen
  \bibfield  {author} {\bibinfo {author} {\bibfnamefont {C.-M.}\ \bibnamefont
  {Chung}}, \bibinfo {author} {\bibfnamefont {V.}~\bibnamefont {Alba}},
  \bibinfo {author} {\bibfnamefont {L.}~\bibnamefont {Bonnes}}, \bibinfo
  {author} {\bibfnamefont {P.}~\bibnamefont {Chen}}, \ and\ \bibinfo {author}
  {\bibfnamefont {A.~M.}\ \bibnamefont {L\"{a}uchli}},\ }\href
  {http://arxiv.org/abs/1312.1168} {\  (\bibinfo {year}
  {2013}{\natexlab{b}})},\ \Eprint {http://arxiv.org/abs/1312.1168}
  {arXiv:1312.1168 [cond-mat.str-el]} \BibitemShut {NoStop}%
\bibitem [{\citenamefont {Bombelli}\ \emph {et~al.}(1986)\citenamefont
  {Bombelli}, \citenamefont {Koul}, \citenamefont {Lee},\ and\ \citenamefont
  {Sorkin}}]{Sorkin}%
  \BibitemOpen
  \bibfield  {author} {\bibinfo {author} {\bibfnamefont {L.}~\bibnamefont
  {Bombelli}}, \bibinfo {author} {\bibfnamefont {R.~K.}\ \bibnamefont {Koul}},
  \bibinfo {author} {\bibfnamefont {J.}~\bibnamefont {Lee}}, \ and\ \bibinfo
  {author} {\bibfnamefont {R.~D.}\ \bibnamefont {Sorkin}},\ }\href {\doibase
  10.1103/PhysRevD.34.373} {\bibfield  {journal} {\bibinfo  {journal} {Phys.
  Rev. D}\ }\textbf {\bibinfo {volume} {34}},\ \bibinfo {pages} {373} (\bibinfo
  {year} {1986})}\BibitemShut {NoStop}%
\bibitem [{\citenamefont {Srednicki}(1993)}]{Shredder}%
  \BibitemOpen
  \bibfield  {author} {\bibinfo {author} {\bibfnamefont {M.}~\bibnamefont
  {Srednicki}},\ }\href {\doibase 10.1103/PhysRevLett.71.666} {\bibfield
  {journal} {\bibinfo  {journal} {Phys. Rev. Lett.}\ }\textbf {\bibinfo
  {volume} {71}},\ \bibinfo {pages} {666} (\bibinfo {year} {1993})}\BibitemShut
  {NoStop}%
\bibitem [{\citenamefont {Eisert}\ \emph {et~al.}(2010)\citenamefont {Eisert},
  \citenamefont {Cramer},\ and\ \citenamefont {Plenio}}]{Eisert2010}%
  \BibitemOpen
  \bibfield  {author} {\bibinfo {author} {\bibfnamefont {J.}~\bibnamefont
  {Eisert}}, \bibinfo {author} {\bibfnamefont {M.}~\bibnamefont {Cramer}}, \
  and\ \bibinfo {author} {\bibfnamefont {M.~B.}\ \bibnamefont {Plenio}},\
  }\href {\doibase 10.1103/RevModPhys.82.277} {\bibfield  {journal} {\bibinfo
  {journal} {Rev. Mod. Phys.}\ }\textbf {\bibinfo {volume} {82}},\ \bibinfo
  {pages} {277} (\bibinfo {year} {2010})}\BibitemShut {NoStop}%
\bibitem [{\citenamefont {Callan}\ and\ \citenamefont
  {Wilczek}(1994)}]{Callan1}%
  \BibitemOpen
  \bibfield  {author} {\bibinfo {author} {\bibfnamefont {C.}~\bibnamefont
  {Callan}}\ and\ \bibinfo {author} {\bibfnamefont {F.}~\bibnamefont
  {Wilczek}},\ }\href@noop {} {\bibfield  {journal} {\bibinfo  {journal} {Phys.
  Lett. B}\ }\textbf {\bibinfo {volume} {333}},\ \bibinfo {pages} {55 }
  (\bibinfo {year} {1994})}\BibitemShut {NoStop}%
\bibitem [{\citenamefont {Holzhey}\ \emph {et~al.}(1994)\citenamefont
  {Holzhey}, \citenamefont {Larsen},\ and\ \citenamefont {Wilczek}}]{EE1d1}%
  \BibitemOpen
  \bibfield  {author} {\bibinfo {author} {\bibfnamefont {C.}~\bibnamefont
  {Holzhey}}, \bibinfo {author} {\bibfnamefont {F.}~\bibnamefont {Larsen}}, \
  and\ \bibinfo {author} {\bibfnamefont {F.}~\bibnamefont {Wilczek}},\ }\href
  {\doibase 10.1016/0550-3213(94)90402-2} {\bibfield  {journal} {\bibinfo
  {journal} {Nucl. Phys. B}\ }\textbf {\bibinfo {volume} {424}},\ \bibinfo
  {pages} {443} (\bibinfo {year} {1994})}\BibitemShut {NoStop}%
\bibitem [{\citenamefont {Vidal}\ \emph {et~al.}(2003)\citenamefont {Vidal},
  \citenamefont {Latorre}, \citenamefont {Rico},\ and\ \citenamefont
  {Kitaev}}]{EE1d2}%
  \BibitemOpen
  \bibfield  {author} {\bibinfo {author} {\bibfnamefont {G.}~\bibnamefont
  {Vidal}}, \bibinfo {author} {\bibfnamefont {J.~I.}\ \bibnamefont {Latorre}},
  \bibinfo {author} {\bibfnamefont {E.}~\bibnamefont {Rico}}, \ and\ \bibinfo
  {author} {\bibfnamefont {A.}~\bibnamefont {Kitaev}},\ }\href@noop {}
  {\bibfield  {journal} {\bibinfo  {journal} {Phys. Rev. Lett.}\ }\textbf
  {\bibinfo {volume} {90}},\ \bibinfo {pages} {227902} (\bibinfo {year}
  {2003})}\BibitemShut {NoStop}%
\bibitem [{\citenamefont {Calabrese}\ and\ \citenamefont
  {Cardy}(2004)}]{EE1d3}%
  \BibitemOpen
  \bibfield  {author} {\bibinfo {author} {\bibfnamefont {P.}~\bibnamefont
  {Calabrese}}\ and\ \bibinfo {author} {\bibfnamefont {J.}~\bibnamefont
  {Cardy}},\ }\href@noop {} {\bibfield  {journal} {\bibinfo  {journal} {J.
  Stat. Mech.}\ ,\ \bibinfo {pages} {P06002}} (\bibinfo {year}
  {2004})}\BibitemShut {NoStop}%
\bibitem [{\citenamefont {Blote}\ \emph {et~al.}(1986)\citenamefont {Blote},
  \citenamefont {Cardy},\ and\ \citenamefont {Nightingale}}]{Cardyc}%
  \BibitemOpen
  \bibfield  {author} {\bibinfo {author} {\bibfnamefont {H.~W.~J.}\
  \bibnamefont {Blote}}, \bibinfo {author} {\bibfnamefont {J.~L.}\ \bibnamefont
  {Cardy}}, \ and\ \bibinfo {author} {\bibfnamefont {M.~P.}\ \bibnamefont
  {Nightingale}},\ }\href@noop {} {\bibfield  {journal} {\bibinfo  {journal}
  {Phys. Rev. Lett.}\ }\textbf {\bibinfo {volume} {56}},\ \bibinfo {pages}
  {742} (\bibinfo {year} {1986})}\BibitemShut {NoStop}%
\bibitem [{\citenamefont {Affleck}(1986)}]{Affleckc}%
  \BibitemOpen
  \bibfield  {author} {\bibinfo {author} {\bibfnamefont {I.}~\bibnamefont
  {Affleck}},\ }\href@noop {} {\bibfield  {journal} {\bibinfo  {journal} {Phys.
  Rev. Lett.}\ }\textbf {\bibinfo {volume} {56}},\ \bibinfo {pages} {746}
  (\bibinfo {year} {1986})}\BibitemShut {NoStop}%
\bibitem [{\citenamefont {Kallin}\ \emph {et~al.}(2013)\citenamefont {Kallin},
  \citenamefont {Hyatt}, \citenamefont {Singh},\ and\ \citenamefont
  {Melko}}]{Kallin_NLCE}%
  \BibitemOpen
  \bibfield  {author} {\bibinfo {author} {\bibfnamefont {A.~B.}\ \bibnamefont
  {Kallin}}, \bibinfo {author} {\bibfnamefont {K.}~\bibnamefont {Hyatt}},
  \bibinfo {author} {\bibfnamefont {R.~R.~P.}\ \bibnamefont {Singh}}, \ and\
  \bibinfo {author} {\bibfnamefont {R.~G.}\ \bibnamefont {Melko}},\ }\href
  {\doibase 10.1103/PhysRevLett.110.135702} {\bibfield  {journal} {\bibinfo
  {journal} {Phys. Rev. Lett.}\ }\textbf {\bibinfo {volume} {110}},\ \bibinfo
  {pages} {135702} (\bibinfo {year} {2013})}\BibitemShut {NoStop}%
\bibitem [{\citenamefont {Luitz}\ \emph {et~al.}(2014)\citenamefont {Luitz},
  \citenamefont {Alet},\ and\ \citenamefont {Laflorencie}}]{Luitz}%
  \BibitemOpen
  \bibfield  {author} {\bibinfo {author} {\bibfnamefont {D.~J.}\ \bibnamefont
  {Luitz}}, \bibinfo {author} {\bibfnamefont {F.}~\bibnamefont {Alet}}, \ and\
  \bibinfo {author} {\bibfnamefont {N.}~\bibnamefont {Laflorencie}},\
  }\href@noop {} {\bibfield  {journal} {\bibinfo  {journal} {Phys. Rev. Lett.}\
  }\textbf {\bibinfo {volume} {112}},\ \bibinfo {pages} {057203} (\bibinfo
  {year} {2014})}\BibitemShut {NoStop}%
\bibitem [{\citenamefont {Fradkin}\ and\ \citenamefont
  {Moore}(2006)}]{FradkinMoore}%
  \BibitemOpen
  \bibfield  {author} {\bibinfo {author} {\bibfnamefont {E.}~\bibnamefont
  {Fradkin}}\ and\ \bibinfo {author} {\bibfnamefont {J.~E.}\ \bibnamefont
  {Moore}},\ }\href {\doibase 10.1103/PhysRevLett.97.050404} {\bibfield
  {journal} {\bibinfo  {journal} {Phys. Rev. Lett.}\ }\textbf {\bibinfo
  {volume} {97}},\ \bibinfo {pages} {050404} (\bibinfo {year}
  {2006})}\BibitemShut {NoStop}%
\bibitem [{\citenamefont {Metlitski}\ \emph {et~al.}(2009)\citenamefont
  {Metlitski}, \citenamefont {Fuertes},\ and\ \citenamefont {Sachdev}}]{Max}%
  \BibitemOpen
  \bibfield  {author} {\bibinfo {author} {\bibfnamefont {M.~A.}\ \bibnamefont
  {Metlitski}}, \bibinfo {author} {\bibfnamefont {C.~A.}\ \bibnamefont
  {Fuertes}}, \ and\ \bibinfo {author} {\bibfnamefont {S.}~\bibnamefont
  {Sachdev}},\ }\href@noop {} {\bibfield  {journal} {\bibinfo  {journal} {Phys.
  Rev. B}\ }\textbf {\bibinfo {volume} {80}},\ \bibinfo {pages} {115122}
  (\bibinfo {year} {2009})}\BibitemShut {NoStop}%
\bibitem [{\citenamefont {Casini}\ and\ \citenamefont
  {Huerta}(2009)}]{Casini_FieldTheory}%
  \BibitemOpen
  \bibfield  {author} {\bibinfo {author} {\bibfnamefont {H.}~\bibnamefont
  {Casini}}\ and\ \bibinfo {author} {\bibfnamefont {M.}~\bibnamefont
  {Huerta}},\ }\href@noop {} {\bibfield  {journal} {\bibinfo  {journal} {J.
  Phys. A: Math. Theor.}\ }\textbf {\bibinfo {volume} {42}},\ \bibinfo {pages}
  {504007} (\bibinfo {year} {2009})}\BibitemShut {NoStop}%
\bibitem [{\citenamefont {Nishioka}\ \emph {et~al.}(2009)\citenamefont
  {Nishioka}, \citenamefont {Ryu},\ and\ \citenamefont {Takayanagi}}]{ryu_2}%
  \BibitemOpen
  \bibfield  {author} {\bibinfo {author} {\bibfnamefont {T.}~\bibnamefont
  {Nishioka}}, \bibinfo {author} {\bibfnamefont {S.}~\bibnamefont {Ryu}}, \
  and\ \bibinfo {author} {\bibfnamefont {T.}~\bibnamefont {Takayanagi}},\
  }\href {http://stacks.iop.org/1751-8121/42/i=50/a=504008} {\bibfield
  {journal} {\bibinfo  {journal} {J. Phys. A}\ }\textbf {\bibinfo {volume}
  {42}},\ \bibinfo {pages} {504008} (\bibinfo {year} {2009})}\BibitemShut
  {NoStop}%
\bibitem [{\citenamefont {{Myers}}\ and\ \citenamefont
  {{Singh}}(2012)}]{Robcorner}%
  \BibitemOpen
  \bibfield  {author} {\bibinfo {author} {\bibfnamefont {R.~C.}\ \bibnamefont
  {{Myers}}}\ and\ \bibinfo {author} {\bibfnamefont {A.}~\bibnamefont
  {{Singh}}},\ }\href {\doibase 10.1007/JHEP09(2012)013} {\bibfield  {journal}
  {\bibinfo  {journal} {J. High Energy Phys.}\ }\textbf {\bibinfo {volume}
  {9}},\ \bibinfo {pages} {13} (\bibinfo {year} {2012})}\BibitemShut {NoStop}%
\bibitem [{\citenamefont {Zamolodchikov}(1986)}]{Zamo}%
  \BibitemOpen
  \bibfield  {author} {\bibinfo {author} {\bibfnamefont {A.}~\bibnamefont
  {Zamolodchikov}},\ }\href@noop {} {\bibfield  {journal} {\bibinfo  {journal}
  {JETP Lett.}\ }\textbf {\bibinfo {volume} {43}},\ \bibinfo {pages} {731}
  (\bibinfo {year} {1986})}\BibitemShut {NoStop}%
\bibitem [{\citenamefont {Cardy}(1988)}]{Cardy_Cth}%
  \BibitemOpen
  \bibfield  {author} {\bibinfo {author} {\bibfnamefont {J.~L.}\ \bibnamefont
  {Cardy}},\ }\href@noop {} {\bibfield  {journal} {\bibinfo  {journal} {Phys.
  Lett. B}\ }\textbf {\bibinfo {volume} {215}},\ \bibinfo {pages} {749}
  (\bibinfo {year} {1988})}\BibitemShut {NoStop}%
\bibitem [{\citenamefont {Casini}\ and\ \citenamefont
  {Huerta}(2012)}]{Casini12}%
  \BibitemOpen
  \bibfield  {author} {\bibinfo {author} {\bibfnamefont {H.}~\bibnamefont
  {Casini}}\ and\ \bibinfo {author} {\bibfnamefont {M.}~\bibnamefont
  {Huerta}},\ }\href@noop {} {\bibfield  {journal} {\bibinfo  {journal} {Phys.
  Rev. D}\ }\textbf {\bibinfo {volume} {85}},\ \bibinfo {pages} {125016}
  (\bibinfo {year} {2012})}\BibitemShut {NoStop}%
\bibitem [{\citenamefont {Grover}(2014)}]{Grover_C}%
  \BibitemOpen
  \bibfield  {author} {\bibinfo {author} {\bibfnamefont {T.}~\bibnamefont
  {Grover}},\ }\href {\doibase 10.1103/PhysRevLett.112.151601} {\bibfield
  {journal} {\bibinfo  {journal} {Phys. Rev. Lett.}\ }\textbf {\bibinfo
  {volume} {112}},\ \bibinfo {pages} {151601} (\bibinfo {year}
  {2014})}\BibitemShut {NoStop}%
\bibitem [{\citenamefont {Simon}(2002)}]{Simon2002}%
  \BibitemOpen
  \bibfield  {author} {\bibinfo {author} {\bibfnamefont {C.}~\bibnamefont
  {Simon}},\ }\href {\doibase 10.1103/PhysRevA.66.052323} {\bibfield  {journal}
  {\bibinfo  {journal} {Phys. Rev. A}\ }\textbf {\bibinfo {volume} {66}},\
  \bibinfo {pages} {052323} (\bibinfo {year} {2002})}\BibitemShut {NoStop}%
\bibitem [{\citenamefont {Vedral}(2003)}]{Vedral2003}%
  \BibitemOpen
  \bibfield  {author} {\bibinfo {author} {\bibfnamefont {V.}~\bibnamefont
  {Vedral}},\ }\href@noop {} {\bibfield  {journal} {\bibinfo  {journal} {Cent.
  Eur. J. Phys.}\ }\textbf {\bibinfo {volume} {2}},\ \bibinfo {pages} {289}
  (\bibinfo {year} {2003})}\BibitemShut {NoStop}%
\bibitem [{\citenamefont {Hines}\ \emph {et~al.}(2003)\citenamefont {Hines},
  \citenamefont {McKenzie},\ and\ \citenamefont {Milburn}}]{Hines2003}%
  \BibitemOpen
  \bibfield  {author} {\bibinfo {author} {\bibfnamefont {A.~P.}\ \bibnamefont
  {Hines}}, \bibinfo {author} {\bibfnamefont {R.~H.}\ \bibnamefont {McKenzie}},
  \ and\ \bibinfo {author} {\bibfnamefont {G.~J.}\ \bibnamefont {Milburn}},\
  }\href {\doibase 10.1103/PhysRevA.67.013609} {\bibfield  {journal} {\bibinfo
  {journal} {Phys. Rev. A}\ }\textbf {\bibinfo {volume} {67}},\ \bibinfo
  {pages} {013609} (\bibinfo {year} {2003})}\BibitemShut {NoStop}%
\bibitem [{\citenamefont {Heaney}\ \emph {et~al.}(2007)\citenamefont {Heaney},
  \citenamefont {Anders}, \citenamefont {Kaszlikowski},\ and\ \citenamefont
  {Vedral}}]{Heaney2007}%
  \BibitemOpen
  \bibfield  {author} {\bibinfo {author} {\bibfnamefont {L.}~\bibnamefont
  {Heaney}}, \bibinfo {author} {\bibfnamefont {J.}~\bibnamefont {Anders}},
  \bibinfo {author} {\bibfnamefont {D.}~\bibnamefont {Kaszlikowski}}, \ and\
  \bibinfo {author} {\bibfnamefont {V.}~\bibnamefont {Vedral}},\ }\href
  {\doibase 10.1103/PhysRevA.76.053605} {\bibfield  {journal} {\bibinfo
  {journal} {Phys. Rev. A}\ }\textbf {\bibinfo {volume} {76}},\ \bibinfo
  {pages} {053605} (\bibinfo {year} {2007})}\BibitemShut {NoStop}%
\bibitem [{\citenamefont {{Heaney}}(2007)}]{Heaney2007a}%
  \BibitemOpen
  \bibfield  {author} {\bibinfo {author} {\bibfnamefont {L.}~\bibnamefont
  {{Heaney}}},\ }\href@noop {} {\  (\bibinfo {year} {2007})},\ \Eprint
  {http://arxiv.org/abs/0711.0942} {arXiv:0711.0942 [quant-ph]} \BibitemShut
  {NoStop}%
\bibitem [{\citenamefont {Kaszlikowski}\ and\ \citenamefont
  {Wiesniak}(2007)}]{Kaszlikowski2007}%
  \BibitemOpen
  \bibfield  {author} {\bibinfo {author} {\bibfnamefont {D.}~\bibnamefont
  {Kaszlikowski}}\ and\ \bibinfo {author} {\bibfnamefont {M.}~\bibnamefont
  {Wiesniak}},\ }\href@noop {} {\  (\bibinfo {year} {2007})},\ \Eprint
  {http://arxiv.org/abs/0712.0990} {arXiv:0712.0990 [quant-ph]} \BibitemShut
  {NoStop}%
\bibitem [{\citenamefont {{Vedral}}(2007)}]{Vedral2008}%
  \BibitemOpen
  \bibfield  {author} {\bibinfo {author} {\bibfnamefont {V.}~\bibnamefont
  {{Vedral}}},\ }\href@noop {} {\  (\bibinfo {year} {2007})},\ \Eprint
  {http://arxiv.org/abs/0703129} {arXiv:0703129 [quant-ph]} \BibitemShut
  {NoStop}%
\bibitem [{\citenamefont {Heaney}\ and\ \citenamefont
  {Vedral}(2009)}]{Heaney2009}%
  \BibitemOpen
  \bibfield  {author} {\bibinfo {author} {\bibfnamefont {L.}~\bibnamefont
  {Heaney}}\ and\ \bibinfo {author} {\bibfnamefont {V.}~\bibnamefont
  {Vedral}},\ }\href {\doibase 10.1103/PhysRevLett.103.200502} {\bibfield
  {journal} {\bibinfo  {journal} {Phys. Rev. Lett.}\ }\textbf {\bibinfo
  {volume} {103}},\ \bibinfo {pages} {200502} (\bibinfo {year}
  {2009})}\BibitemShut {NoStop}%
\bibitem [{\citenamefont {Goold}\ \emph {et~al.}(2009)\citenamefont {Goold},
  \citenamefont {Heaney}, \citenamefont {Busch},\ and\ \citenamefont
  {Vedral}}]{Goold2009}%
  \BibitemOpen
  \bibfield  {author} {\bibinfo {author} {\bibfnamefont {J.}~\bibnamefont
  {Goold}}, \bibinfo {author} {\bibfnamefont {L.}~\bibnamefont {Heaney}},
  \bibinfo {author} {\bibfnamefont {T.}~\bibnamefont {Busch}}, \ and\ \bibinfo
  {author} {\bibfnamefont {V.}~\bibnamefont {Vedral}},\ }\href {\doibase
  10.1103/PhysRevA.80.022338} {\bibfield  {journal} {\bibinfo  {journal} {Phys.
  Rev. A}\ }\textbf {\bibinfo {volume} {80}},\ \bibinfo {pages} {022338}
  (\bibinfo {year} {2009})}\BibitemShut {NoStop}%
\bibitem [{\citenamefont {Ding}\ and\ \citenamefont {Yang}(2009)}]{Ding2009}%
  \BibitemOpen
  \bibfield  {author} {\bibinfo {author} {\bibfnamefont {W.}~\bibnamefont
  {Ding}}\ and\ \bibinfo {author} {\bibfnamefont {K.}~\bibnamefont {Yang}},\
  }\href {\doibase 10.1103/PhysRevA.80.012329} {\bibfield  {journal} {\bibinfo
  {journal} {Phys. Rev. A}\ }\textbf {\bibinfo {volume} {80}},\ \bibinfo
  {pages} {012329} (\bibinfo {year} {2009})}\BibitemShut {NoStop}%
\bibitem [{\citenamefont {Gagatsos}\ \emph {et~al.}(2012)\citenamefont
  {Gagatsos}, \citenamefont {Karanikas},\ and\ \citenamefont
  {Kordas}}]{Gagatsos2012}%
  \BibitemOpen
  \bibfield  {author} {\bibinfo {author} {\bibfnamefont {C.~N.}\ \bibnamefont
  {Gagatsos}}, \bibinfo {author} {\bibfnamefont {A.~I.}\ \bibnamefont
  {Karanikas}}, \ and\ \bibinfo {author} {\bibfnamefont {G.~I.}\ \bibnamefont
  {Kordas}},\ }\href@noop {} {\  (\bibinfo {year} {2012})},\ \Eprint
  {http://arxiv.org/abs/1207.0303} {arXiv:1207.0303 [quant-ph]} \BibitemShut
  {NoStop}%
\bibitem [{\citenamefont {Calabrese}\ \emph
  {et~al.}(2011{\natexlab{a}})\citenamefont {Calabrese}, \citenamefont
  {Mintchev},\ and\ \citenamefont {Vicari}}]{Calabrese2011a}%
  \BibitemOpen
  \bibfield  {author} {\bibinfo {author} {\bibfnamefont {P.}~\bibnamefont
  {Calabrese}}, \bibinfo {author} {\bibfnamefont {M.}~\bibnamefont {Mintchev}},
  \ and\ \bibinfo {author} {\bibfnamefont {E.}~\bibnamefont {Vicari}},\ }\href
  {\doibase 10.1088/1742-5468/2011/09/P09028} {\bibfield  {journal} {\bibinfo
  {journal} {J. Stat. Mech.: Theory E}\ }\textbf {\bibinfo {volume} {2011}},\
  \bibinfo {pages} {P09028} (\bibinfo {year} {2011}{\natexlab{a}})}\BibitemShut
  {NoStop}%
\bibitem [{\citenamefont {Calabrese}\ \emph
  {et~al.}(2011{\natexlab{b}})\citenamefont {Calabrese}, \citenamefont
  {Mintchev},\ and\ \citenamefont {Vicari}}]{Calabrese2011b}%
  \BibitemOpen
  \bibfield  {author} {\bibinfo {author} {\bibfnamefont {P.}~\bibnamefont
  {Calabrese}}, \bibinfo {author} {\bibfnamefont {M.}~\bibnamefont {Mintchev}},
  \ and\ \bibinfo {author} {\bibfnamefont {E.}~\bibnamefont {Vicari}},\ }\href
  {\doibase 10.1103/PhysRevLett.107.020601} {\bibfield  {journal} {\bibinfo
  {journal} {Phys. Rev. Lett.}\ }\textbf {\bibinfo {volume} {107}},\ \bibinfo
  {pages} {020601} (\bibinfo {year} {2011}{\natexlab{b}})}\BibitemShut
  {NoStop}%
\bibitem [{\citenamefont {Burnett}\ \emph {et~al.}(2001)\citenamefont
  {Burnett}, \citenamefont {Choi}, \citenamefont {Davis}, \citenamefont
  {Dunningham},\ and\ \citenamefont {Morgan}}]{Burnett2001}%
  \BibitemOpen
  \bibfield  {author} {\bibinfo {author} {\bibfnamefont {K.}~\bibnamefont
  {Burnett}}, \bibinfo {author} {\bibfnamefont {S.}~\bibnamefont {Choi}},
  \bibinfo {author} {\bibfnamefont {M.}~\bibnamefont {Davis}}, \bibinfo
  {author} {\bibfnamefont {J.~A.}\ \bibnamefont {Dunningham}}, \ and\ \bibinfo
  {author} {\bibfnamefont {S.~A.}\ \bibnamefont {Morgan}},\ }\href {\doibase
  http://dx.doi.org/10.1016/S1296-2147(01)01185-4} {\bibfield  {journal}
  {\bibinfo  {journal} {Comptes Rendus de l'Acad\'{e}mie des Sciences - Series
  IV - Physics}\ }\textbf {\bibinfo {volume} {2}},\ \bibinfo {pages} {399}
  (\bibinfo {year} {2001})}\BibitemShut {NoStop}%
\bibitem [{\citenamefont {Dunningham}\ \emph {et~al.}(2002)\citenamefont
  {Dunningham}, \citenamefont {Bose}, \citenamefont {Henderson}, \citenamefont
  {Vedral},\ and\ \citenamefont {Burnett}}]{Dunningham2002a}%
  \BibitemOpen
  \bibfield  {author} {\bibinfo {author} {\bibfnamefont {J.~A.}\ \bibnamefont
  {Dunningham}}, \bibinfo {author} {\bibfnamefont {S.}~\bibnamefont {Bose}},
  \bibinfo {author} {\bibfnamefont {L.}~\bibnamefont {Henderson}}, \bibinfo
  {author} {\bibfnamefont {V.}~\bibnamefont {Vedral}}, \ and\ \bibinfo {author}
  {\bibfnamefont {K.}~\bibnamefont {Burnett}},\ }\href {\doibase
  10.1103/PhysRevA.65.064302} {\bibfield  {journal} {\bibinfo  {journal} {Phys.
  Rev. A}\ }\textbf {\bibinfo {volume} {65}},\ \bibinfo {pages} {064302}
  (\bibinfo {year} {2002})}\BibitemShut {NoStop}%
\bibitem [{\citenamefont {Eisert}\ \emph {et~al.}(2002)\citenamefont {Eisert},
  \citenamefont {Simon},\ and\ \citenamefont {Plenio}}]{Eisert:2002bx}%
  \BibitemOpen
  \bibfield  {author} {\bibinfo {author} {\bibfnamefont {J.}~\bibnamefont
  {Eisert}}, \bibinfo {author} {\bibfnamefont {C.}~\bibnamefont {Simon}}, \
  and\ \bibinfo {author} {\bibfnamefont {M.~B.}\ \bibnamefont {Plenio}},\
  }\href@noop {} {\bibfield  {journal} {\bibinfo  {journal} {J. Phys. A: Math.
  Gen.}\ }\textbf {\bibinfo {volume} {35}},\ \bibinfo {pages} {3911} (\bibinfo
  {year} {2002})}\BibitemShut {NoStop}%
\bibitem [{\citenamefont {Adesso}\ and\ \citenamefont
  {Illuminati}(2007)}]{Adesso:2007eq}%
  \BibitemOpen
  \bibfield  {author} {\bibinfo {author} {\bibfnamefont {G.}~\bibnamefont
  {Adesso}}\ and\ \bibinfo {author} {\bibfnamefont {F.}~\bibnamefont
  {Illuminati}},\ }\href@noop {} {\bibfield  {journal} {\bibinfo  {journal} {J.
  Phys. A: Math. Theor.}\ }\textbf {\bibinfo {volume} {40}},\ \bibinfo {pages}
  {7821} (\bibinfo {year} {2007})}\BibitemShut {NoStop}%
\bibitem [{\citenamefont {Wehrl}(1978)}]{Wehrl:1978ku}%
  \BibitemOpen
  \bibfield  {author} {\bibinfo {author} {\bibfnamefont {A.}~\bibnamefont
  {Wehrl}},\ }\href@noop {} {\bibfield  {journal} {\bibinfo  {journal} {Rev.
  Mod. Phys.}\ }\textbf {\bibinfo {volume} {50}},\ \bibinfo {pages} {221}
  (\bibinfo {year} {1978})}\BibitemShut {NoStop}%
\bibitem [{\citenamefont {Zanardi}(2002)}]{Zanardi2002}%
  \BibitemOpen
  \bibfield  {author} {\bibinfo {author} {\bibfnamefont {P.}~\bibnamefont
  {Zanardi}},\ }\href {\doibase 10.1103/PhysRevA.65.042101} {\bibfield
  {journal} {\bibinfo  {journal} {Phys. Rev. A}\ }\textbf {\bibinfo {volume}
  {65}},\ \bibinfo {pages} {042101} (\bibinfo {year} {2002})}\BibitemShut
  {NoStop}%
\bibitem [{\citenamefont {Shi}(2003)}]{Shi2003}%
  \BibitemOpen
  \bibfield  {author} {\bibinfo {author} {\bibfnamefont {Y.}~\bibnamefont
  {Shi}},\ }\href {\doibase 10.1103/PhysRevA.67.024301} {\bibfield  {journal}
  {\bibinfo  {journal} {Phys. Rev. A}\ }\textbf {\bibinfo {volume} {67}},\
  \bibinfo {pages} {024301} (\bibinfo {year} {2003})}\BibitemShut {NoStop}%
\bibitem [{\citenamefont {Fang}\ and\ \citenamefont {Chang}(2003)}]{Fang2003}%
  \BibitemOpen
  \bibfield  {author} {\bibinfo {author} {\bibfnamefont {A.}~\bibnamefont
  {Fang}}\ and\ \bibinfo {author} {\bibfnamefont {Y.}~\bibnamefont {Chang}},\
  }\href {\doibase 10.1016/S0375-9601(03)00546-2} {\bibfield  {journal}
  {\bibinfo  {journal} {Phys. Lett. A}\ }\textbf {\bibinfo {volume} {311}},\
  \bibinfo {pages} {443} (\bibinfo {year} {2003})}\BibitemShut {NoStop}%
\bibitem [{\citenamefont {Zozulya}\ \emph {et~al.}(2008)\citenamefont
  {Zozulya}, \citenamefont {Haque},\ and\ \citenamefont
  {Schoutens}}]{Zozulya2008}%
  \BibitemOpen
  \bibfield  {author} {\bibinfo {author} {\bibfnamefont {O.~S.}\ \bibnamefont
  {Zozulya}}, \bibinfo {author} {\bibfnamefont {M.}~\bibnamefont {Haque}}, \
  and\ \bibinfo {author} {\bibfnamefont {K.}~\bibnamefont {Schoutens}},\ }\href
  {\doibase 10.1103/PhysRevA.78.042326} {\bibfield  {journal} {\bibinfo
  {journal} {Phys. Rev. A}\ }\textbf {\bibinfo {volume} {78}},\ \bibinfo
  {pages} {042326} (\bibinfo {year} {2008})}\BibitemShut {NoStop}%
\bibitem [{\citenamefont {Eckert}\ \emph {et~al.}(2002)\citenamefont {Eckert},
  \citenamefont {Schliemann}, \citenamefont {Bru\ss},\ and\ \citenamefont
  {Lewenstein}}]{Eckert2002}%
  \BibitemOpen
  \bibfield  {author} {\bibinfo {author} {\bibfnamefont {K.}~\bibnamefont
  {Eckert}}, \bibinfo {author} {\bibfnamefont {J.}~\bibnamefont {Schliemann}},
  \bibinfo {author} {\bibfnamefont {D.}~\bibnamefont {Bru\ss}}, \ and\ \bibinfo
  {author} {\bibfnamefont {M.}~\bibnamefont {Lewenstein}},\ }\href {\doibase
  10.1006/aphy.2002.6268} {\bibfield  {journal} {\bibinfo  {journal} {Ann.
  Phys.}\ }\textbf {\bibinfo {volume} {299}},\ \bibinfo {pages} {88} (\bibinfo
  {year} {2002})}\BibitemShut {NoStop}%
\bibitem [{\citenamefont {Haque}\ \emph {et~al.}(2009)\citenamefont {Haque},
  \citenamefont {Zozulya},\ and\ \citenamefont {Schoutens}}]{Haque2009}%
  \BibitemOpen
  \bibfield  {author} {\bibinfo {author} {\bibfnamefont {M.}~\bibnamefont
  {Haque}}, \bibinfo {author} {\bibfnamefont {O.~S.}\ \bibnamefont {Zozulya}},
  \ and\ \bibinfo {author} {\bibfnamefont {K.}~\bibnamefont {Schoutens}},\
  }\href {\doibase 10.1088/1751-8113/42/50/504012} {\bibfield  {journal}
  {\bibinfo  {journal} {J. Phys. A: Math Theory}\ }\textbf {\bibinfo {volume}
  {42}},\ \bibinfo {pages} {504012} (\bibinfo {year} {2009})}\BibitemShut
  {NoStop}%
\bibitem [{\citenamefont {Haque}\ \emph {et~al.}(2007)\citenamefont {Haque},
  \citenamefont {Zozulya},\ and\ \citenamefont {Schoutens}}]{Haque:2007il}%
  \BibitemOpen
  \bibfield  {author} {\bibinfo {author} {\bibfnamefont {M.}~\bibnamefont
  {Haque}}, \bibinfo {author} {\bibfnamefont {O.}~\bibnamefont {Zozulya}}, \
  and\ \bibinfo {author} {\bibfnamefont {K.}~\bibnamefont {Schoutens}},\ }\href
  {\doibase 10.1103/PhysRevLett.98.060401} {\bibfield  {journal} {\bibinfo
  {journal} {Phys. Rev. Lett.}\ }\textbf {\bibinfo {volume} {98}},\ \bibinfo
  {pages} {060401} (\bibinfo {year} {2007})}\BibitemShut {NoStop}%
\bibitem [{\citenamefont {Zozulya}\ \emph {et~al.}(2007)\citenamefont
  {Zozulya}, \citenamefont {Haque}, \citenamefont {Schoutens},\ and\
  \citenamefont {Rezayi}}]{Zozulya2007a}%
  \BibitemOpen
  \bibfield  {author} {\bibinfo {author} {\bibfnamefont {O.~S.}\ \bibnamefont
  {Zozulya}}, \bibinfo {author} {\bibfnamefont {M.}~\bibnamefont {Haque}},
  \bibinfo {author} {\bibfnamefont {K.}~\bibnamefont {Schoutens}}, \ and\
  \bibinfo {author} {\bibfnamefont {E.~H.}\ \bibnamefont {Rezayi}},\ }\href
  {\doibase 10.1103/PhysRevB.76.125310} {\bibfield  {journal} {\bibinfo
  {journal} {Phys. Rev. B}\ }\textbf {\bibinfo {volume} {76}},\ \bibinfo
  {pages} {125310} (\bibinfo {year} {2007})}\BibitemShut {NoStop}%
\bibitem [{\citenamefont {Dunningham}\ \emph {et~al.}(2005)\citenamefont
  {Dunningham}, \citenamefont {Rau},\ and\ \citenamefont
  {Burnett}}]{Dunningham2005}%
  \BibitemOpen
  \bibfield  {author} {\bibinfo {author} {\bibfnamefont {J.}~\bibnamefont
  {Dunningham}}, \bibinfo {author} {\bibfnamefont {A.}~\bibnamefont {Rau}}, \
  and\ \bibinfo {author} {\bibfnamefont {K.}~\bibnamefont {Burnett}},\ }\href
  {\doibase 10.1126/science.1109545} {\bibfield  {journal} {\bibinfo  {journal}
  {Science}\ }\textbf {\bibinfo {volume} {307}},\ \bibinfo {pages} {872}
  (\bibinfo {year} {2005})}\BibitemShut {NoStop}%
\bibitem [{\citenamefont {Balachandran}\ \emph {et~al.}(2013)\citenamefont
  {Balachandran}, \citenamefont {Govindarajan}, \citenamefont {de~Queiroz},\
  and\ \citenamefont {Reyes-Lega}}]{Balachandran2013}%
  \BibitemOpen
  \bibfield  {author} {\bibinfo {author} {\bibfnamefont {A.~P.}\ \bibnamefont
  {Balachandran}}, \bibinfo {author} {\bibfnamefont {T.~R.}\ \bibnamefont
  {Govindarajan}}, \bibinfo {author} {\bibfnamefont {A.~R.}\ \bibnamefont
  {de~Queiroz}}, \ and\ \bibinfo {author} {\bibfnamefont {A.~F.}\ \bibnamefont
  {Reyes-Lega}},\ }\href {\doibase 10.1103/PhysRevLett.110.080503} {\bibfield
  {journal} {\bibinfo  {journal} {Phys. Rev. Lett.}\ }\textbf {\bibinfo
  {volume} {110}},\ \bibinfo {pages} {080503} (\bibinfo {year}
  {2013})}\BibitemShut {NoStop}%
\bibitem [{\citenamefont {Benatti}\ \emph {et~al.}(2012)\citenamefont
  {Benatti}, \citenamefont {Floreanini},\ and\ \citenamefont
  {Marzolino}}]{Benatti2012a}%
  \BibitemOpen
  \bibfield  {author} {\bibinfo {author} {\bibfnamefont {F.}~\bibnamefont
  {Benatti}}, \bibinfo {author} {\bibfnamefont {R.}~\bibnamefont {Floreanini}},
  \ and\ \bibinfo {author} {\bibfnamefont {U.}~\bibnamefont {Marzolino}},\
  }\href {\doibase 10.1103/PhysRevA.85.042329} {\bibfield  {journal} {\bibinfo
  {journal} {Phys. Rev. A}\ }\textbf {\bibinfo {volume} {85}},\ \bibinfo
  {pages} {042329} (\bibinfo {year} {2012})}\BibitemShut {NoStop}%
\bibitem [{\citenamefont {Wiseman}\ and\ \citenamefont
  {Vaccaro}(2003)}]{Wiseman2003}%
  \BibitemOpen
  \bibfield  {author} {\bibinfo {author} {\bibfnamefont {H.~M.}\ \bibnamefont
  {Wiseman}}\ and\ \bibinfo {author} {\bibfnamefont {J.~A.}\ \bibnamefont
  {Vaccaro}},\ }\href {\doibase 10.1103/PhysRevLett.91.097902} {\bibfield
  {journal} {\bibinfo  {journal} {Phys. Rev. Lett.}\ }\textbf {\bibinfo
  {volume} {91}},\ \bibinfo {pages} {097902} (\bibinfo {year}
  {2003})}\BibitemShut {NoStop}%
\bibitem [{\citenamefont {Wiseman}\ \emph {et~al.}(2003)\citenamefont
  {Wiseman}, \citenamefont {Bartlett},\ and\ \citenamefont
  {Vaccaro}}]{Wiseman2003fb}%
  \BibitemOpen
  \bibfield  {author} {\bibinfo {author} {\bibfnamefont {H.~M.}\ \bibnamefont
  {Wiseman}}, \bibinfo {author} {\bibfnamefont {S.~D.}\ \bibnamefont
  {Bartlett}}, \ and\ \bibinfo {author} {\bibfnamefont {J.~A.}\ \bibnamefont
  {Vaccaro}},\ }\href@noop {} {\  (\bibinfo {year} {2003})},\ \Eprint
  {http://arxiv.org/abs/0309046} {arXiv:0309046 [quant-ph]} \BibitemShut
  {NoStop}%
\bibitem [{\citenamefont {Boschi}\ \emph {et~al.}(1998)\citenamefont {Boschi},
  \citenamefont {Branca}, \citenamefont {De~Martini}, \citenamefont {Hardy},\
  and\ \citenamefont {Popescu}}]{Boschi:1998qt}%
  \BibitemOpen
  \bibfield  {author} {\bibinfo {author} {\bibfnamefont {D.}~\bibnamefont
  {Boschi}}, \bibinfo {author} {\bibfnamefont {S.}~\bibnamefont {Branca}},
  \bibinfo {author} {\bibfnamefont {F.}~\bibnamefont {De~Martini}}, \bibinfo
  {author} {\bibfnamefont {L.}~\bibnamefont {Hardy}}, \ and\ \bibinfo {author}
  {\bibfnamefont {S.}~\bibnamefont {Popescu}},\ }\href {\doibase
  10.1103/PhysRevLett.80.1121} {\bibfield  {journal} {\bibinfo  {journal}
  {Phys. Rev. Lett.}\ }\textbf {\bibinfo {volume} {80}},\ \bibinfo {pages}
  {1121} (\bibinfo {year} {1998})}\BibitemShut {NoStop}%
\bibitem [{\citenamefont {Killoran}\ \emph {et~al.}(2014)\citenamefont
  {Killoran}, \citenamefont {Cramer},\ and\ \citenamefont
  {Plenio}}]{Killoran:2014gu}%
  \BibitemOpen
  \bibfield  {author} {\bibinfo {author} {\bibfnamefont {N.}~\bibnamefont
  {Killoran}}, \bibinfo {author} {\bibfnamefont {M.}~\bibnamefont {Cramer}}, \
  and\ \bibinfo {author} {\bibfnamefont {M.~B.}\ \bibnamefont {Plenio}},\
  }\href@noop {} {\bibfield  {journal} {\bibinfo  {journal} {Phys. Rev. Lett.}\
  }\textbf {\bibinfo {volume} {112}},\ \bibinfo {pages} {150501} (\bibinfo
  {year} {2014})}\BibitemShut {NoStop}%
\bibitem [{\citenamefont {Byrnes}\ \emph {et~al.}(2012)\citenamefont {Byrnes},
  \citenamefont {Wen},\ and\ \citenamefont {Yamamoto}}]{Byrnes:2012mq}%
  \BibitemOpen
  \bibfield  {author} {\bibinfo {author} {\bibfnamefont {T.}~\bibnamefont
  {Byrnes}}, \bibinfo {author} {\bibfnamefont {K.}~\bibnamefont {Wen}}, \ and\
  \bibinfo {author} {\bibfnamefont {Y.}~\bibnamefont {Yamamoto}},\ }\href
  {\doibase 10.1103/PhysRevA.85.040306} {\bibfield  {journal} {\bibinfo
  {journal} {Phys. Rev. A}\ }\textbf {\bibinfo {volume} {85}},\ \bibinfo
  {pages} {040306} (\bibinfo {year} {2012})}\BibitemShut {NoStop}%
\bibitem [{\citenamefont {Aharonov}\ and\ \citenamefont
  {Susskind}(1967)}]{Aharonov1967}%
  \BibitemOpen
  \bibfield  {author} {\bibinfo {author} {\bibfnamefont {Y.}~\bibnamefont
  {Aharonov}}\ and\ \bibinfo {author} {\bibfnamefont {L.}~\bibnamefont
  {Susskind}},\ }\href {\doibase 10.1103/PhysRev.155.1428} {\bibfield
  {journal} {\bibinfo  {journal} {Physical Review}\ }\textbf {\bibinfo {volume}
  {155}},\ \bibinfo {pages} {1428} (\bibinfo {year} {1967})}\BibitemShut
  {NoStop}%
\bibitem [{\citenamefont {Dowling}\ \emph {et~al.}(2006)\citenamefont
  {Dowling}, \citenamefont {Doherty},\ and\ \citenamefont
  {Wiseman}}]{Dowling2006}%
  \BibitemOpen
  \bibfield  {author} {\bibinfo {author} {\bibfnamefont {M.~R.}\ \bibnamefont
  {Dowling}}, \bibinfo {author} {\bibfnamefont {A.~C.}\ \bibnamefont
  {Doherty}}, \ and\ \bibinfo {author} {\bibfnamefont {H.~M.}\ \bibnamefont
  {Wiseman}},\ }\href {\doibase 10.1103/PhysRevA.73.052323} {\bibfield
  {journal} {\bibinfo  {journal} {Phys. Rev. A}\ }\textbf {\bibinfo {volume}
  {73}},\ \bibinfo {pages} {052323} (\bibinfo {year} {2006})}\BibitemShut
  {NoStop}%
\bibitem [{\citenamefont {Ceperley}(1995)}]{Ceperley1995}%
  \BibitemOpen
  \bibfield  {author} {\bibinfo {author} {\bibfnamefont {D.~M.}\ \bibnamefont
  {Ceperley}},\ }\href {\doibase 10.1103/RevModPhys.67.279} {\bibfield
  {journal} {\bibinfo  {journal} {Rev. Mod. Phys.}\ }\textbf {\bibinfo {volume}
  {67}},\ \bibinfo {pages} {279} (\bibinfo {year} {1995})}\BibitemShut
  {NoStop}%
\bibitem [{\citenamefont {Sarsa}\ \emph {et~al.}(2000)\citenamefont {Sarsa},
  \citenamefont {Schmidt},\ and\ \citenamefont {Magro}}]{Sarsa2000}%
  \BibitemOpen
  \bibfield  {author} {\bibinfo {author} {\bibfnamefont {A.}~\bibnamefont
  {Sarsa}}, \bibinfo {author} {\bibfnamefont {K.~E.}\ \bibnamefont {Schmidt}},
  \ and\ \bibinfo {author} {\bibfnamefont {W.~R.}\ \bibnamefont {Magro}},\
  }\href {\doibase 10.1063/1.481926} {\bibfield  {journal} {\bibinfo  {journal}
  {J. Chem. Phys.}\ }\textbf {\bibinfo {volume} {113}},\ \bibinfo {pages}
  {1366} (\bibinfo {year} {2000})}\BibitemShut {NoStop}%
\bibitem [{\citenamefont {Cuervo}\ \emph {et~al.}(2005)\citenamefont {Cuervo},
  \citenamefont {Roy},\ and\ \citenamefont {Boninsegni}}]{Cuervo2005a}%
  \BibitemOpen
  \bibfield  {author} {\bibinfo {author} {\bibfnamefont {J.~E.}\ \bibnamefont
  {Cuervo}}, \bibinfo {author} {\bibfnamefont {P.-N.}\ \bibnamefont {Roy}}, \
  and\ \bibinfo {author} {\bibfnamefont {M.}~\bibnamefont {Boninsegni}},\
  }\href {\doibase 10.1063/1.1872775} {\bibfield  {journal} {\bibinfo
  {journal} {J. Chem. Phys.}\ }\textbf {\bibinfo {volume} {122}},\ \bibinfo
  {pages} {114504} (\bibinfo {year} {2005})}\BibitemShut {NoStop}%
\bibitem [{\citenamefont {Boninsegni}\ and\ \citenamefont
  {Moroni}(2012)}]{Boninsegni2012}%
  \BibitemOpen
  \bibfield  {author} {\bibinfo {author} {\bibfnamefont {M.}~\bibnamefont
  {Boninsegni}}\ and\ \bibinfo {author} {\bibfnamefont {S.}~\bibnamefont
  {Moroni}},\ }\href {\doibase 10.1103/PhysRevE.86.056712} {\bibfield
  {journal} {\bibinfo  {journal} {Phys. Rev. E}\ }\textbf {\bibinfo {volume}
  {86}},\ \bibinfo {pages} {056712} (\bibinfo {year} {2012})}\BibitemShut
  {NoStop}%
\bibitem [{Note1()}]{Note1}%
  \BibitemOpen
  \bibinfo {note} {For $S_2$, $\Pi _{2}^A$ is called the ``SWAP'' operator in
  the literature for spatial entanglement. However, we do not use this notation
  here to avoid confusion with a {\protect \it swap} update used in a
  continuous space worm algorithm.}\BibitemShut {Stop}%
\bibitem [{\citenamefont {Cardy}(2011)}]{Cardy}%
  \BibitemOpen
  \bibfield  {author} {\bibinfo {author} {\bibfnamefont {J.}~\bibnamefont
  {Cardy}},\ }\href {\doibase 10.1103/PhysRevLett.106.150404} {\bibfield
  {journal} {\bibinfo  {journal} {Phys. Rev. Lett.}\ }\textbf {\bibinfo
  {volume} {106}},\ \bibinfo {pages} {150404} (\bibinfo {year}
  {2011})}\BibitemShut {NoStop}%
\bibitem [{\citenamefont {Jang}\ \emph {et~al.}(2001)\citenamefont {Jang},
  \citenamefont {Jang},\ and\ \citenamefont {Voth}}]{Jang2001}%
  \BibitemOpen
  \bibfield  {author} {\bibinfo {author} {\bibfnamefont {S.}~\bibnamefont
  {Jang}}, \bibinfo {author} {\bibfnamefont {S.}~\bibnamefont {Jang}}, \ and\
  \bibinfo {author} {\bibfnamefont {G.~A.}\ \bibnamefont {Voth}},\ }\href
  {\doibase 10.1063/1.1410117} {\bibfield  {journal} {\bibinfo  {journal} {J.
  Chem. Phys.}\ }\textbf {\bibinfo {volume} {115}},\ \bibinfo {pages} {7832}
  (\bibinfo {year} {2001})}\BibitemShut {NoStop}%
\bibitem [{\citenamefont {Peschel}(2012)}]{Peschel2012}%
  \BibitemOpen
  \bibfield  {author} {\bibinfo {author} {\bibfnamefont {I.}~\bibnamefont
  {Peschel}},\ }\href {\doibase 10.1007/s13538-012-0074-1} {\bibfield
  {journal} {\bibinfo  {journal} {Braz. J. Phys.}\ }\textbf {\bibinfo {volume}
  {42}},\ \bibinfo {pages} {267} (\bibinfo {year} {2012})}\BibitemShut
  {NoStop}%
\bibitem [{\citenamefont {Han}\ \emph {et~al.}(1999)\citenamefont {Han},
  \citenamefont {Kim},\ and\ \citenamefont {Noz}}]{Han1999}%
  \BibitemOpen
  \bibfield  {author} {\bibinfo {author} {\bibfnamefont {D.}~\bibnamefont
  {Han}}, \bibinfo {author} {\bibfnamefont {Y.~S.}\ \bibnamefont {Kim}}, \ and\
  \bibinfo {author} {\bibfnamefont {M.~E.}\ \bibnamefont {Noz}},\ }\href
  {\doibase 10.1119/1.19192} {\bibfield  {journal} {\bibinfo  {journal} {Am. J.
  of Phys.}\ }\textbf {\bibinfo {volume} {67}},\ \bibinfo {pages} {61}
  (\bibinfo {year} {1999})}\BibitemShut {NoStop}%
\bibitem [{\citenamefont {{Benavides-Riveros}}\ \emph
  {et~al.}(2014)\citenamefont {{Benavides-Riveros}}, \citenamefont
  {{Toranzo}},\ and\ \citenamefont {{Dehesa}}}]{Benavides-Riveros2014}%
  \BibitemOpen
  \bibfield  {author} {\bibinfo {author} {\bibfnamefont {C.~L.}\ \bibnamefont
  {{Benavides-Riveros}}}, \bibinfo {author} {\bibfnamefont {I.~V.}\
  \bibnamefont {{Toranzo}}}, \ and\ \bibinfo {author} {\bibfnamefont {J.~S.}\
  \bibnamefont {{Dehesa}}},\ }\href@noop {} {\  (\bibinfo {year} {2014})},\
  \Eprint {http://arxiv.org/abs/1404.4447} {arXiv:1404.4447 [quant-ph]}
  \BibitemShut {NoStop}%
\bibitem [{\citenamefont {Kallin}\ \emph {et~al.}(2011)\citenamefont {Kallin},
  \citenamefont {Hastings}, \citenamefont {Melko},\ and\ \citenamefont
  {Singh}}]{Kallin2011}%
  \BibitemOpen
  \bibfield  {author} {\bibinfo {author} {\bibfnamefont {A.~B.}\ \bibnamefont
  {Kallin}}, \bibinfo {author} {\bibfnamefont {M.~B.}\ \bibnamefont
  {Hastings}}, \bibinfo {author} {\bibfnamefont {R.~G.}\ \bibnamefont {Melko}},
  \ and\ \bibinfo {author} {\bibfnamefont {R.~R.~P.}\ \bibnamefont {Singh}},\
  }\href {\doibase 10.1103/PhysRevB.84.165134} {\bibfield  {journal} {\bibinfo
  {journal} {Phys. Rev. B}\ }\textbf {\bibinfo {volume} {84}},\ \bibinfo
  {pages} {165134} (\bibinfo {year} {2011})}\BibitemShut {NoStop}%
\bibitem [{\citenamefont {Boninsegni}\ \emph
  {et~al.}(2006{\natexlab{a}})\citenamefont {Boninsegni}, \citenamefont
  {Prokof'ev},\ and\ \citenamefont {Svistunov}}]{Boninsegni:2006ed}%
  \BibitemOpen
  \bibfield  {author} {\bibinfo {author} {\bibfnamefont {M.}~\bibnamefont
  {Boninsegni}}, \bibinfo {author} {\bibfnamefont {N.}~\bibnamefont
  {Prokof'ev}}, \ and\ \bibinfo {author} {\bibfnamefont {B.}~\bibnamefont
  {Svistunov}},\ }\href {\doibase 10.1103/PhysRevLett.96.070601} {\bibfield
  {journal} {\bibinfo  {journal} {Phys. Rev. Lett.}\ }\textbf {\bibinfo
  {volume} {96}},\ \bibinfo {pages} {070601} (\bibinfo {year}
  {2006}{\natexlab{a}})}\BibitemShut {NoStop}%
\bibitem [{\citenamefont {Boninsegni}\ \emph
  {et~al.}(2006{\natexlab{b}})\citenamefont {Boninsegni}, \citenamefont
  {Prokofev},\ and\ \citenamefont {Svistunov}}]{Boninsegni:2006gc}%
  \BibitemOpen
  \bibfield  {author} {\bibinfo {author} {\bibfnamefont {M.}~\bibnamefont
  {Boninsegni}}, \bibinfo {author} {\bibfnamefont {N.~V.}\ \bibnamefont
  {Prokofev}}, \ and\ \bibinfo {author} {\bibfnamefont {B.~V.}\ \bibnamefont
  {Svistunov}},\ }\href@noop {} {\bibfield  {journal} {\bibinfo  {journal}
  {Phys. Rev. E}\ }\textbf {\bibinfo {volume} {74}},\ \bibinfo {pages} {036701}
  (\bibinfo {year} {2006}{\natexlab{b}})}\BibitemShut {NoStop}%
\bibitem [{\citenamefont {Prokof'ev}\ and\ \citenamefont
  {Svistunov}(1998)}]{PhysRevLett.81.2514}%
  \BibitemOpen
  \bibfield  {author} {\bibinfo {author} {\bibfnamefont {N.~V.}\ \bibnamefont
  {Prokof'ev}}\ and\ \bibinfo {author} {\bibfnamefont {B.~V.}\ \bibnamefont
  {Svistunov}},\ }\href {\doibase 10.1103/PhysRevLett.81.2514} {\bibfield
  {journal} {\bibinfo  {journal} {Phys. Rev. Lett.}\ }\textbf {\bibinfo
  {volume} {81}},\ \bibinfo {pages} {2514} (\bibinfo {year}
  {1998})}\BibitemShut {NoStop}%
\bibitem [{\citenamefont {Singh}\ \emph {et~al.}(2011)\citenamefont {Singh},
  \citenamefont {Hastings}, \citenamefont {Kallin},\ and\ \citenamefont
  {Melko}}]{Singh:2011jx}%
  \BibitemOpen
  \bibfield  {author} {\bibinfo {author} {\bibfnamefont {R.~R.~P.}\
  \bibnamefont {Singh}}, \bibinfo {author} {\bibfnamefont {M.~B.}\ \bibnamefont
  {Hastings}}, \bibinfo {author} {\bibfnamefont {A.~B.}\ \bibnamefont
  {Kallin}}, \ and\ \bibinfo {author} {\bibfnamefont {R.~G.}\ \bibnamefont
  {Melko}},\ }\href@noop {} {\bibfield  {journal} {\bibinfo  {journal} {Phys.
  Rev. Lett.}\ }\textbf {\bibinfo {volume} {106}},\ \bibinfo {pages} {135701}
  (\bibinfo {year} {2011})}\BibitemShut {NoStop}%
\bibitem [{\citenamefont {S\"orensen}\ \emph {et~al.}(2001)\citenamefont
  {S\"orensen}, \citenamefont {Duan}, \citenamefont {Cirac},\ and\
  \citenamefont {Zoller}}]{Sorensen2001}%
  \BibitemOpen
  \bibfield  {author} {\bibinfo {author} {\bibfnamefont {A.}~\bibnamefont
  {S\"orensen}}, \bibinfo {author} {\bibfnamefont {L.~M.}\ \bibnamefont
  {Duan}}, \bibinfo {author} {\bibfnamefont {J.~I.}\ \bibnamefont {Cirac}}, \
  and\ \bibinfo {author} {\bibfnamefont {P.}~\bibnamefont {Zoller}},\ }\href
  {\doibase 10.1038/35051038} {\bibfield  {journal} {\bibinfo  {journal}
  {Nature}\ }\textbf {\bibinfo {volume} {409}},\ \bibinfo {pages} {63}
  (\bibinfo {year} {2001})}\BibitemShut {NoStop}%
\bibitem [{\citenamefont {Est{\`e}ve}\ \emph {et~al.}(2008)\citenamefont
  {Est{\`e}ve}, \citenamefont {Gross}, \citenamefont {Weller}, \citenamefont
  {Giovanazzi},\ and\ \citenamefont {Oberthaler}}]{Esteve:2008ij}%
  \BibitemOpen
  \bibfield  {author} {\bibinfo {author} {\bibfnamefont {J.}~\bibnamefont
  {Est{\`e}ve}}, \bibinfo {author} {\bibfnamefont {C.}~\bibnamefont {Gross}},
  \bibinfo {author} {\bibfnamefont {A.}~\bibnamefont {Weller}}, \bibinfo
  {author} {\bibfnamefont {S.}~\bibnamefont {Giovanazzi}}, \ and\ \bibinfo
  {author} {\bibfnamefont {M.~K.}\ \bibnamefont {Oberthaler}},\ }\href@noop {}
  {\bibfield  {journal} {\bibinfo  {journal} {Nature}\ }\textbf {\bibinfo
  {volume} {455}},\ \bibinfo {pages} {1216} (\bibinfo {year}
  {2008})}\BibitemShut {NoStop}%
\bibitem [{\citenamefont {Riedel}\ \emph {et~al.}(2010)\citenamefont {Riedel},
  \citenamefont {B{\"o}hi}, \citenamefont {Li}, \citenamefont {H{\"a}nsch},
  \citenamefont {Sinatra},\ and\ \citenamefont {Treutlein}}]{Riedel:2010gea}%
  \BibitemOpen
  \bibfield  {author} {\bibinfo {author} {\bibfnamefont {M.~F.}\ \bibnamefont
  {Riedel}}, \bibinfo {author} {\bibfnamefont {P.}~\bibnamefont {B{\"o}hi}},
  \bibinfo {author} {\bibfnamefont {Y.}~\bibnamefont {Li}}, \bibinfo {author}
  {\bibfnamefont {T.~W.}\ \bibnamefont {H{\"a}nsch}}, \bibinfo {author}
  {\bibfnamefont {A.}~\bibnamefont {Sinatra}}, \ and\ \bibinfo {author}
  {\bibfnamefont {P.}~\bibnamefont {Treutlein}},\ }\href@noop {} {\bibfield
  {journal} {\bibinfo  {journal} {Nature}\ }\textbf {\bibinfo {volume} {464}},\
  \bibinfo {pages} {1170} (\bibinfo {year} {2010})}\BibitemShut {NoStop}%
\bibitem [{\citenamefont {Benatti}\ \emph {et~al.}(2011)\citenamefont
  {Benatti}, \citenamefont {Floreanini},\ and\ \citenamefont
  {Marzolino}}]{Benatti:2011hf}%
  \BibitemOpen
  \bibfield  {author} {\bibinfo {author} {\bibfnamefont {F.}~\bibnamefont
  {Benatti}}, \bibinfo {author} {\bibfnamefont {R.}~\bibnamefont {Floreanini}},
  \ and\ \bibinfo {author} {\bibfnamefont {U.}~\bibnamefont {Marzolino}},\
  }\href@noop {} {\bibfield  {journal} {\bibinfo  {journal} {J. Phys. B: Atom.,
  Mol. Opt.}\ }\textbf {\bibinfo {volume} {44}},\ \bibinfo {pages} {091001}
  (\bibinfo {year} {2011})}\BibitemShut {NoStop}%
\bibitem [{\citenamefont {He}\ \emph {et~al.}(2011)\citenamefont {He},
  \citenamefont {Reid}, \citenamefont {Vaughan}, \citenamefont {Gross},
  \citenamefont {Oberthaler},\ and\ \citenamefont {Drummond}}]{He:2011jf}%
  \BibitemOpen
  \bibfield  {author} {\bibinfo {author} {\bibfnamefont {Q.~Y.}\ \bibnamefont
  {He}}, \bibinfo {author} {\bibfnamefont {M.~D.}\ \bibnamefont {Reid}},
  \bibinfo {author} {\bibfnamefont {T.~G.}\ \bibnamefont {Vaughan}}, \bibinfo
  {author} {\bibfnamefont {C.}~\bibnamefont {Gross}}, \bibinfo {author}
  {\bibfnamefont {M.}~\bibnamefont {Oberthaler}}, \ and\ \bibinfo {author}
  {\bibfnamefont {P.~D.}\ \bibnamefont {Drummond}},\ }\href@noop {} {\bibfield
  {journal} {\bibinfo  {journal} {Phys. Rev. Lett.}\ }\textbf {\bibinfo
  {volume} {106}},\ \bibinfo {pages} {120405} (\bibinfo {year}
  {2011})}\BibitemShut {NoStop}%
\bibitem [{\citenamefont {Hyllus}\ \emph {et~al.}(2012)\citenamefont {Hyllus},
  \citenamefont {Pezz{\'e}}, \citenamefont {Smerzi},\ and\ \citenamefont
  {T{\'o}th}}]{Hyllus:2012dl}%
  \BibitemOpen
  \bibfield  {author} {\bibinfo {author} {\bibfnamefont {P.}~\bibnamefont
  {Hyllus}}, \bibinfo {author} {\bibfnamefont {L.}~\bibnamefont {Pezz{\'e}}},
  \bibinfo {author} {\bibfnamefont {A.}~\bibnamefont {Smerzi}}, \ and\ \bibinfo
  {author} {\bibfnamefont {G.}~\bibnamefont {T{\'o}th}},\ }\href@noop {}
  {\bibfield  {journal} {\bibinfo  {journal} {Phys. Rev. A}\ }\textbf {\bibinfo
  {volume} {86}},\ \bibinfo {pages} {012337} (\bibinfo {year}
  {2012})}\BibitemShut {NoStop}%
\bibitem [{\citenamefont {L{\"u}cke}\ \emph {et~al.}(2014)\citenamefont
  {L{\"u}cke}, \citenamefont {Peise}, \citenamefont {Vitagliano}, \citenamefont
  {Arlt}, \citenamefont {Santos}, \citenamefont {T{\'o}th},\ and\ \citenamefont
  {Klempt}}]{Lucke:2014hz}%
  \BibitemOpen
  \bibfield  {author} {\bibinfo {author} {\bibfnamefont {B.}~\bibnamefont
  {L{\"u}cke}}, \bibinfo {author} {\bibfnamefont {J.}~\bibnamefont {Peise}},
  \bibinfo {author} {\bibfnamefont {G.}~\bibnamefont {Vitagliano}}, \bibinfo
  {author} {\bibfnamefont {J.}~\bibnamefont {Arlt}}, \bibinfo {author}
  {\bibfnamefont {L.}~\bibnamefont {Santos}}, \bibinfo {author} {\bibfnamefont
  {G.}~\bibnamefont {T{\'o}th}}, \ and\ \bibinfo {author} {\bibfnamefont
  {C.}~\bibnamefont {Klempt}},\ }\href@noop {} {\bibfield  {journal} {\bibinfo
  {journal} {Phys. Rev. Lett.}\ }\textbf {\bibinfo {volume} {112}},\ \bibinfo
  {pages} {155304} (\bibinfo {year} {2014})}\BibitemShut {NoStop}%
\bibitem [{\citenamefont {Dalton}\ \emph {et~al.}(2014)\citenamefont {Dalton},
  \citenamefont {Heaney}, \citenamefont {Goold}, \citenamefont {Garraway},\
  and\ \citenamefont {Busch}}]{Dalton:2014dv}%
  \BibitemOpen
  \bibfield  {author} {\bibinfo {author} {\bibfnamefont {B.~J.}\ \bibnamefont
  {Dalton}}, \bibinfo {author} {\bibfnamefont {L.}~\bibnamefont {Heaney}},
  \bibinfo {author} {\bibfnamefont {J.}~\bibnamefont {Goold}}, \bibinfo
  {author} {\bibfnamefont {B.~M.}\ \bibnamefont {Garraway}}, \ and\ \bibinfo
  {author} {\bibfnamefont {T.}~\bibnamefont {Busch}},\ }\href@noop {}
  {\bibfield  {journal} {\bibinfo  {journal} {New J Phys.}\ }\textbf {\bibinfo
  {volume} {16}},\ \bibinfo {pages} {013026} (\bibinfo {year}
  {2014})}\BibitemShut {NoStop}%
\bibitem [{\citenamefont {Li}\ \emph {et~al.}(2010)\citenamefont {Li},
  \citenamefont {{Le Roy}}, \citenamefont {Roy},\ and\ \citenamefont
  {McKellar}}]{Li2010}%
  \BibitemOpen
  \bibfield  {author} {\bibinfo {author} {\bibfnamefont {H.}~\bibnamefont
  {Li}}, \bibinfo {author} {\bibfnamefont {R.~J.}\ \bibnamefont {{Le Roy}}},
  \bibinfo {author} {\bibfnamefont {P.-N.}\ \bibnamefont {Roy}}, \ and\
  \bibinfo {author} {\bibfnamefont {A.~R.~W.}\ \bibnamefont {McKellar}},\
  }\href {\doibase 10.1103/PhysRevLett.105.133401} {\bibfield  {journal}
  {\bibinfo  {journal} {Phys. Rev. Lett.}\ }\textbf {\bibinfo {volume} {105}},\
  \bibinfo {pages} {133401} (\bibinfo {year} {2010})}\BibitemShut {NoStop}%
\bibitem [{\citenamefont {Raston}\ \emph {et~al.}(2012)\citenamefont {Raston},
  \citenamefont {J{\"a}ger}, \citenamefont {Li}, \citenamefont {{Le Roy}},\
  and\ \citenamefont {Roy}}]{raston_coh2_superfluid}%
  \BibitemOpen
  \bibfield  {author} {\bibinfo {author} {\bibfnamefont {P.~L.}\ \bibnamefont
  {Raston}}, \bibinfo {author} {\bibfnamefont {W.}~\bibnamefont {J{\"a}ger}},
  \bibinfo {author} {\bibfnamefont {H.}~\bibnamefont {Li}}, \bibinfo {author}
  {\bibfnamefont {R.~J.}\ \bibnamefont {{Le Roy}}}, \ and\ \bibinfo {author}
  {\bibfnamefont {P.-N.}\ \bibnamefont {Roy}},\ }\href@noop {} {\bibfield
  {journal} {\bibinfo  {journal} {Phys. Rev. Lett.}\ }\textbf {\bibinfo
  {volume} {108}},\ \bibinfo {pages} {253402} (\bibinfo {year}
  {2012})}\BibitemShut {NoStop}%
\bibitem [{\citenamefont {Zeng}\ and\ \citenamefont
  {Roy}(2014)}]{zeng2014microscopic}%
  \BibitemOpen
  \bibfield  {author} {\bibinfo {author} {\bibfnamefont {T.}~\bibnamefont
  {Zeng}}\ and\ \bibinfo {author} {\bibfnamefont {P.-N.}\ \bibnamefont {Roy}},\
  }\href@noop {} {\bibfield  {journal} {\bibinfo  {journal} {Rep. Prog. Phys.}\
  }\textbf {\bibinfo {volume} {77}},\ \bibinfo {pages} {046601} (\bibinfo
  {year} {2014})}\BibitemShut {NoStop}%
\bibitem [{\citenamefont {Verstraete}\ and\ \citenamefont
  {Cirac}(2010)}]{Verstraete2010}%
  \BibitemOpen
  \bibfield  {author} {\bibinfo {author} {\bibfnamefont {F.}~\bibnamefont
  {Verstraete}}\ and\ \bibinfo {author} {\bibfnamefont {J.~I.}\ \bibnamefont
  {Cirac}},\ }\href {\doibase 10.1103/PhysRevLett.104.190405} {\bibfield
  {journal} {\bibinfo  {journal} {Phys. Rev. Lett.}\ }\textbf {\bibinfo
  {volume} {104}},\ \bibinfo {pages} {190405} (\bibinfo {year}
  {2010})}\BibitemShut {NoStop}%
\end{thebibliography}%

\end{document}